\documentclass{article}

\usepackage{PRIMEarxiv}

\usepackage[utf8]{inputenc} 
\usepackage[T1]{fontenc}    
\usepackage{xcolor}         
\usepackage{hyperref}       
\usepackage{url}            
\hypersetup{
    colorlinks=true,
    urlcolor=blue,
    linkcolor=blue,
    citecolor=blue
}
\usepackage{booktabs}       
\usepackage{amsfonts}       
\usepackage{nicefrac}       
\usepackage{microtype}      
\usepackage{lipsum}
\usepackage{fancyhdr}       
\usepackage{graphicx}       
\graphicspath{{Figure/}{./}}
\usepackage{algorithm}
\usepackage{algpseudocode}
\usepackage{amsmath,amssymb,mathtools,bm}
\usepackage{tabularx}
\usepackage{multirow}
\usepackage{multicol}
\usepackage{upgreek}
\usepackage{natbib}
\usepackage{cleveref}
\usepackage{environ}
\usepackage{makecell}
\usepackage{threeparttable}

\newsavebox\CBox
\newcommand*\textBF[1]{\sbox\CBox{#1}\resizebox{\wd\CBox}{\ht\CBox}{\textbf{#1}}}

\newcommand{\sig}[2]{%
  \mbox{#1}%
  \rlap{$^{{\fontsize{2pt}{2pt}\selectfont #2}}$}%
}

\usepackage{xcolor}
\makeatletter
\DeclareRobustCommand{\blue}{%
  \@ifnextchar\bgroup{\blue@arg}{\color{black}}%
}
\newcommand{\blue@arg}[1]{{\color{black}#1}}
\AtBeginDocument{%
  \@ifpackageloaded{hyperref}{%
    \pdfstringdefDisableCommands{%
      \def\BluePDF{\futurelet\BlueNext\BluePDFi}%
      \def\BluePDFi{%
        \ifx\BlueNext\bgroup
          \expandafter\BluePDFarg
        \else
          \relax
        \fi
      }%
      \def\BluePDFarg#1{#1}%
      \let\blue\BluePDF
    }%
  }{}%
}
\makeatother

\title{Cyclic 2.5D Perceptual Loss for Cross-Modal 3D Medical Image Synthesis: T1w MRI to Tau PET}

\author{
  Junho Moon$^{1}$, Symac Kim$^{2}$, Haejun Chung$^{1,2,3}$, Ikbeom Jang$^{4,5,6}$\\
  for the Alzheimer's Disease Neuroimaging Initiative\\[0.75ex]
  {\normalfont\small $^{1}$Department of Artificial Intelligence Semiconductor Engineering, Hanyang University, Seoul, South Korea}\\
  {\normalfont\small $^{2}$Department of Artificial Intelligence, Hanyang University, Seoul, South Korea}\\
  {\normalfont\small $^{3}$Department of Electronic Engineering, Hanyang University, Seoul, South Korea}\\
  {\normalfont\small $^{4}$Division of Computer Engineering, Hankuk University of Foreign Studies, Yongin, South Korea}\\
  {\normalfont\small $^{5}$Division of AI Data Convergence, Hankuk University of Foreign Studies, Yongin, South Korea}\\
  {\normalfont\small $^{6}$Division of Language \& AI, Hankuk University of Foreign Studies, Seoul, South Korea}\\[0.75ex]
  {\normalfont\small Correspondence: \texttt{ijang@hufs.ac.kr}; \texttt{haejun@hanyang.ac.kr}}
}

\begin{document}
\maketitle

\renewcommand\thefootnote{}
\footnotetext{A journal version of this paper is available at \url{https://doi.org/10.1002/hbm.70508}.}
\renewcommand\thefootnote{\fnsymbol{footnote}}
\setcounter{footnote}{0}

\begin{abstract}
Positron emission tomography (PET) provides an in vivo molecular marker for various diseases, including Alzheimer’s disease and related dementias (ADRD). PET has become increasingly integrated into diagnostic decision-making, disease staging, and clinical trial enrichment. However, its widespread use remains constrained by high costs, government regulations, and the invasiveness of radiotracer injection. Modern diagnostic frameworks emphasize the importance of multimodal biomarker assessment, such as the ``amyloid/tau/neurodegeneration'' (A/T/N) framework for Alzheimer's disease; however, they are constrained by these barriers. Medical image synthesis or translation offers a potential solution by enabling the reconstruction of unavailable modalities. The clinical utility of PET depends on accurately capturing regional uptake patterns rather than exact voxel-wise intensities, motivating the use of perceptual loss functions to assess higher-level semantic features in generative models. While 2D, 3D, and 2.5D perceptual losses are utilized in 3D synthesis, each encounters challenges, including limited volumetric context, the scarcity of pre-trained 3D models, and difficulty balancing optimization across anatomical planes. In this work, we address cross-modal synthesis of tau PET from structural magnetic resonance imaging (MRI), generating 3D pseudo-[\textsuperscript{18}F]flortaucipir standardized uptake value ratio (SUVR) maps from 3D T1-weighted MR images. We propose a cyclic 2.5D perceptual loss that cyclically optimizes the axial, coronal, and sagittal planes over training phases, thereby enhancing volumetric consistency. Furthermore, we standardize PET SUVRs by scanner manufacturer, reducing inter-manufacturer variability and better preserving high‑uptake regions. We evaluate the proposed approach on cohorts spanning the ADRD spectrum using data from the Alzheimer’s Disease Neuroimaging Initiative and the Standardized Centralized Alzheimer’s Disease and Related Dementias Neuroimaging cohort. Our approach is broadly applicable across various generative frameworks and achieves high quantitative and qualitative performance on diverse architectures, including U-Net, UNETR, SwinUNETR, CycleGAN, and Pix2Pix. Notably, it achieves better agreement between synthesized SUVRs and measured PET scans in key brain regions relevant to Alzheimer-type tau pathology. The code is publicly available at
\url{https://github.com/labhai/Cyclic-2.5D-Perceptual-Loss}.
\end{abstract}


\section{Introduction}\label{sec:intro}
\subsection{Clinical Motivation}
\blue{Alzheimer’s disease (AD), the most common cause of dementia within the broader Alzheimer’s disease and related dementias (ADRD) spectrum \citep{montine2014recommendations}, is a major contributor to cognitive decline worldwide \citep{knopman2021alzheimer}.} Despite extensive research efforts, there is currently no cure, and only limited disease-modifying therapies are available \citep{winblad2016defeating,wang2022human}. Recently, anti-amyloid monoclonal antibodies such as lecanemab \citep{van2023lecanemab} and donanemab \citep{sims2023donanemab} have demonstrated modest efficacy in slowing clinical progression in early AD. However, these treatments are not curative and are associated with important safety considerations and barriers to access. Early and accurate diagnosis is therefore critical to optimize prognosis, preserve cognitive function \citep{rasmussen2019alzheimer}, and reduce both direct and indirect costs of care \citep{wong2020economic}. The urgency of this need is emphasized by demographic projections, with the number of individuals living with dementia expected to rise from approximately 57 million in 2019 to more than 150 million by 2050 \citep{nichols2022estimation}.

Positron emission tomography (PET) has emerged as a powerful imaging modality capable of in vivo visualization of diverse molecular markers across multiple diseases, including amyloid beta (A) and tau (T) in AD \citep{chandra2019applications}. This is particularly important given that the defining neuropathological hallmarks of AD are amyloid beta (A$\beta$) plaques and neurofibrillary tangles (NFTs) composed of hyperphosphorylated tau protein \citep{maass2017comparison,knopman2021alzheimer}. The National Institute on Aging--Alzheimer's Association (NIA--AA) ``amyloid/tau/neurodegeneration'' (A/T/N) research framework \blue{\citep{jack2018nia}} has increasingly been adopted in clinical practice to support diagnosis and therapeutic decision making \citep{altomare2019applying,rosenberg2022beta}. Current guidelines emphasize that amyloid and tau PET provide spatially resolved information on neuropathology and should be used when results are expected to alter diagnosis or management \citep{rabinovici2025updated}. \blue{Recently revised clinical criteria from the Alzheimer’s Association (AA) workgroup further note that biomarker results can establish the presence of AD, and that tau PET contributes prognostic value and enhances diagnostic confidence \citep{jack2024revised}}.

Particularly, tau PET provides a unique window into the accumulation of abnormally phosphorylated tau proteins that form NFTs among these modalities \citep{pooler2013propagation,leuzy2020diagnostic}. In early AD, tau PET reveals NFT deposition in the entorhinal cortex, followed by spread to the hippocampus and subsequently the neocortex as the disease advances \citep{pooler2013propagation,cho2016tau}. These topographies closely recapitulate Braak staging, from transentorhinal to widespread neocortical involvement \citep{braak1991neuropathological,braak1995}, and strongly correlate with cognitive decline \citep{bejanin2017tau,phillips2018tau,dang2023tau}. Consequently, tau PET has emerged as a powerful tool for diagnosis and staging, offering greater etiologic specificity and closer coupling to contemporaneous clinical severity than structural MRI alone \citep{ossenkoppele2021accuracy,leuzy2020diagnostic}.

Despite its diagnostic utility, tau PET imaging presents several limitations that hinder its routine implementation in research and clinical settings. Compared to ``A'' and ``N'' biomarkers, it is often the most difficult to obtain in routine clinical practice due to both practical and logistical challenges. The procedure requires intravenous administration of radioactive tracers \citep{gambhir2002molecular,guo2021characterization}, rendering it invasive, and each scan entails substantial costs, often several thousand USD \citep{berger2003cost,dietlein2000cost,buck2010economic}. Access is largely restricted to specialized centers, limiting availability for many patients \citep{saif2010role,slough2016clinical}. Incorporating tau PET into diagnostic workflows---often already including MRI, FDG PET, and amyloid PET---adds procedural complexity and patient burden, including cumulative exposure to multiple radiopharmaceuticals \citep{lee2024synthesizing}. Typical effective doses for [\textsuperscript{18}F]flortaucipir (AV-1451) are approximately 5--10 mSv per study, with additional exposure if combined PET/CT is performed \citep{choi2016human}. Furthermore, quantification is challenged by off-target binding and inter-site variability in acquisition and reconstruction, motivating ongoing efforts to harmonize these processes \citep{marquie2017lessons,baker2019effect,akamatsu2023review}. While tau PET is widely regarded as a valuable biomarker for AD, its limited accessibility emphasizes the need for methodological alternatives that balance diagnostic utility with real-world feasibility.

Medical image synthesis or translation presents a promising approach to addressing the limitations of tau PET imaging. These approaches enable the generation of synthetic images for modalities that may be unavailable by leveraging information from existing ones. In clinical practice, multimodal analysis is essential, as different imaging modalities provide complementary diagnostic information \citep{yu2020medical,nie2018medical}. Integrating these modalities enhances diagnostic accuracy and facilitates more precise lesion characterization \citep{du2016overview,ortner2019amyloid,zhu2019mri}. From a machine learning perspective, missing modalities reduce the size of training datasets and limit model performance in computer-aided diagnosis \citep{pan2020spatially}. In this context, cross-modal synthesis has emerged as an active area of research, showing encouraging results across FDG and amyloid PET synthesis, and more recently in tau PET surrogates \citep{dayarathna2023deep,zhang2022bpgan,zhang2023mrabeta,jang2023taupetgen}. Thus, the ability to approximate absent modalities is of considerable interest both clinically and in research, with particular relevance for improving the accessibility of tau PET imaging.

The present study provides a proof-of-concept demonstration that tau PET-like images---specifically [\textsuperscript{18}F]flortaucipir standardized uptake value ratio (SUVR) maps---can be approximated from structural T1-weighted (T1w) MRI scans. This approach is supported by extensive neurobiological evidence linking structural alterations captured by MRI, such as cortical atrophy and regional volume loss, to tau burden measured by PET imaging \citep{sepulcre2016vivo,xia2017association,mak2018vivo,la2020prospective,rahmani2023t1}. While deep learning-based image-to-image translation has already demonstrated feasibility in generating PET images from MRI \citep{shin2020gandalf,zhang2022bpgan,zhang2023mrabeta,jang2023taupetgen,lee2024synthesizing}, to our knowledge, few studies have focused specifically on approximating tau PET patterns from T1w MRI alone. Accordingly, the goal of this study is to establish feasibility rather than propose an immediately deployable clinical substitute.

In PET imaging, the diagnostic emphasis lies on the regional distribution of radiotracer uptake, which reflects underlying molecular pathology, rather than on high-resolution structural detail. This emphasis on uptake patterns motivates the incorporation of perceptual loss \citep{johnson2016perceptual} into the generative modeling framework. Unlike conventional pixel-wise losses, perceptual loss prioritizes feature-level similarity, reducing overly smooth reconstructions and improving the fidelity of synthesized uptake patterns \citep{johnson2016perceptual,armanious2020medgan}. For 3D medical environments, perceptual loss can be implemented in 2D, 3D, or 2.5D forms; however, each approach has inherent limitations. The 2D approach \citep{johnson2016perceptual} may fail to capture inter-slice continuity and spatial coherence, which are essential for accurately representing three-dimensional anatomical structures. Conversely, the 3D approach \citep{zhang2022soup} is limited by the scarcity of pre-trained 3D models and may struggle to comprehensively encode spatial information across the axial, coronal, and sagittal planes---information critical clinically. The 2.5D approach \citep{monai_perceptualloss_docs}, while partially addressing these issues, faces challenges in balancing plane-wise perceptual losses effectively.

\begin{figure*}[t]
  \centering
  \includegraphics[width=\linewidth]{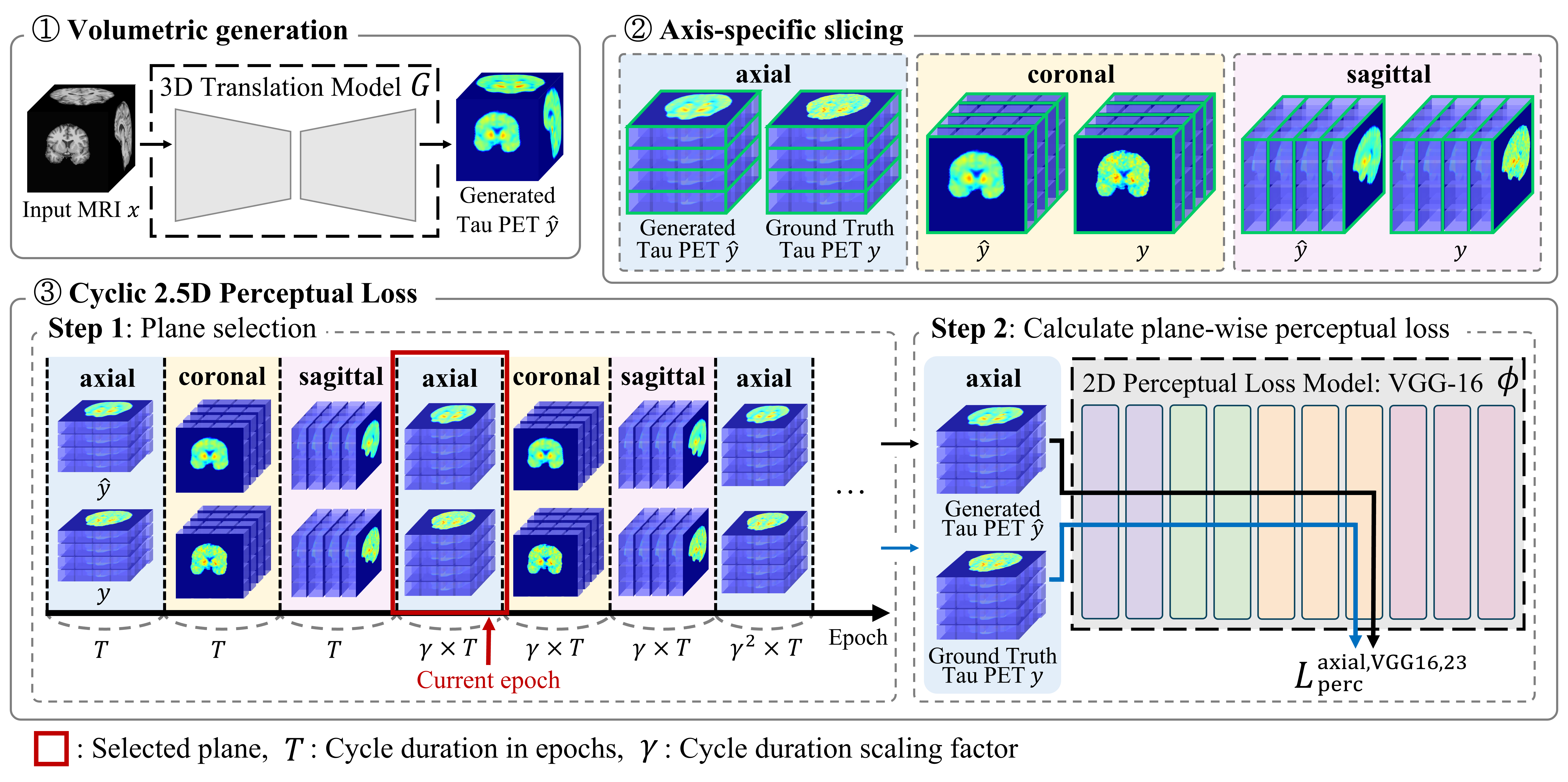}
  \caption{Overview of our proposed cyclic 2.5D perceptual loss. Following the generation of a 3D tau PET image from the given 3D T1w MR image, both the generated tau PET image $\hat{y}$ and the ground truth tau PET image $y$ are sliced into axial, coronal, or sagittal planes based on the current epoch. The sliced image pairs are then processed through the first 23 layers of a pre-trained VGG-16 model, and the corresponding feature maps are extracted. For each slice pair, the mean squared error between the paired feature maps is computed and averaged.}
  \label{fig:cyclic_perceptual}
\end{figure*}

To tackle these problems, we introduce a cyclic 2.5D perceptual loss for 3D medical image synthesis, as illustrated in \Cref{fig:cyclic_perceptual}. This method sequentially calculates 2D perceptual losses for the axial, coronal, and sagittal planes, allocating an equal number of epochs to each plane per cycle. As training progresses, the cycles repeat with fewer epochs. By incorporating mean squared error (MSE) and structural similarity index measure (SSIM) losses, this strategy preserves both semantic and structural similarity in the generated images. Additionally, we employ a by-manufacturer standardization approach to address variations in SUVR distributions across PET images from different scanner manufacturers. This step aims to retain better focal high-SUVR regions, which might otherwise be diminished by conventional min--max normalization. \blue{We evaluate the proposed approach across cohorts spanning the ADRD spectrum using data from the Alzheimer’s Disease Neuroimaging Initiative (ADNI) and the Standardized Centralized Alzheimer’s Disease and Related Dementias Neuroimaging (SCAN) cohort.} Our approach is applied to deep generative models, including a 3D U-Net \citep{cciccek20163d} and its variants such as UNETR \citep{hatamizadeh2022unetr} and SwinUNETR \citep{hatamizadeh2021swin}, as well as generative adversarial network (GAN)-based methods like CycleGAN \citep{zhu2017unpaired} and Pix2Pix \citep{pix2pix}, demonstrating significant performance improvements in both 2D cross-sectional slices and 3D volumetric reconstructions.

In this context, this study makes the following key contributions to neuroimaging for \blue{ADRD} and cross-modal medical image synthesis:
\begin{itemize}
\item We present a proof of concept that T1w MRI alone can approximate regional [\textsuperscript{18}F]flortaucipir tau PET SUVR, supporting its potential as a structural proxy for regional tau deposition patterns.
\item We introduce a cyclic 2.5D perceptual loss that sequentially optimizes axial, coronal, and sagittal planes over training cycles, enhancing 3D volumetric consistency and balanced plane-wise feature learning.
\item We propose a by‑manufacturer SUVR standardization intended to reduce inter-manufacturer variability while better preserving regions of elevated tau uptake than conventional min--max normalization.
\item The proposed approach yielded synthesized PET SUVRs closer to the actual PET images at both voxel and ROI levels compared to existing perceptual losses, with particularly significant improvements in many key tau pathology regions of Alzheimer's disease.
\end{itemize}

\subsection{Related Work}
\label{sec:backgrounds}

\textbf{Tau PET}. PET detects coincident pairs of 511-keV annihilation photons generated following positron emission, enabling quantitative in vivo estimation of radiotracer activity concentrations \citep{vaquero2015positron}. Tau PET specifically allows visualization and quantification of aggregated tau for the evaluation of tauopathies, including AD, as well as selected non-AD tauopathies such as progressive supranuclear palsy and corticobasal degeneration, through the use of dedicated radiotracers \citep{leuzy2019tau}. First-generation tau PET tracers include [\textsuperscript{18}F]THK5317, [\textsuperscript{18}F]THK5351, [\textsuperscript{18}F]flortaucipir, and [\textsuperscript{11}C]PBB3 \citep{petersen2022overview,leuzy2019tau}. Among these, [\textsuperscript{18}F]flortaucipir is the first to receive U.S. FDA approval (marketed as flortaucipir F 18). This tracer exhibits high affinity for paired-helical-filament tau, minimal cross-reactivity with A$\beta$, favorable brain penetration, and suitable kinetic and metabolic characteristics \citep{petersen2022overview,mattay2020brain}. While [\textsuperscript{18}F]flortaucipir demonstrates high sensitivity for AD, its performance in non-AD tauopathies remains limited \citep{petersen2022overview}. Alternative diagnostic approaches for AD include cerebrospinal fluid biomarkers \citep{shaw2018appropriate} and emerging blood-based plasma biomarkers \citep{barthelemy2024highly}. However, unlike fluid-based measures, tau PET enables direct in vivo visualization of NFTs in the brain, supporting assessment of disease phenotypes, staging, and the spatial distribution of tau pathology.

\textbf{Perceptual Loss}. \label{sec:2Dperceq} \citet{johnson2016perceptual} argued that traditional per-pixel loss functions are insufficient for capturing perceptual differences in images. To address these limitations, they introduced feature reconstruction loss and style reconstruction loss. Feature reconstruction loss, commonly referred to as perceptual loss \citep{liu2021generic}, measures the similarity between feature maps of a pre-trained model, such as VGG \citep{simonyan2014very}, for the ground truth and the generated images. This loss is computed using the average squared Euclidean distance, formally expressed as
\begin{equation}
L_{\text{perc}}^{\phi, j}(\hat{y}, y) = \frac{1}{C_j H_j W_j} ||\phi_j(\hat{y})-\phi_j(y)||_2^2,
\label{eq:2Dperceq}
\end{equation}
where $y$ represents the ground truth image, $\hat{y}$ denotes the generated image, $\phi$ refers to the pre-trained model, $j$ is the layer index in $\phi$, and $C_j$, $H_j$, $W_j$ are the number of channels, height, and width of the feature map, respectively. For three-dimensional medical imaging, the 2D perceptual loss can be adapted by applying it to individual slices along a specific anatomical axis (e.g., axial, coronal, or sagittal). In this slice-wise formulation, each slice of the 3D volume is processed independently, and perceptual similarity is assessed using a 2D pre-trained model. However, this approach does not enforce inter-slice continuity or volumetric coherence, potentially overlooking clinically relevant contextual information inherent in the full 3D structure.

Beyond this slice-wise 2D formulation, volumetric data can also be addressed using 3D or 2.5D perceptual loss strategies, which aim to capture richer spatial context. The 3D approach \citep{zhang2022soup} extends perceptual loss to volumetric domains by leveraging a pre-trained 3D feature extractor, thereby explicitly incorporating the depth dimension $D_j$. The 3D perceptual loss is given by:
\begin{equation}
L_{\text{3Dperc}}^{\phi, j}(\hat{y}, y) = \frac{1}{C_j H_j W_j D_j} ||\phi_j(\hat{y})-\phi_j(y)||_2^2.
\label{eq:3Dperc}
\end{equation}
While the 3D formulation enables direct modeling of volumetric structure, its practical adoption is hindered by the scarcity of robust, generalizable pre-trained 3D models. Moreover, it may struggle to represent plane-specific structures consistently across axial, coronal, and sagittal views.

An alternative is the 2.5D approach \citep{monai_perceptualloss_docs}, which approximates volumetric perceptual consistency by aggregating 2D perceptual losses computed along three orthogonal anatomical planes. This method utilizes a pre-trained 2D model, such as VGG \citep{ha20243d}, to independently extract feature representations from axial, coronal, and sagittal slices. The losses for each plane are then combined to approximate the structural context of the full volume:
\begin{equation}
L_{\text{2.5Dperc}}^{\phi, j}(\hat{y}, y) = L_{\text{perc}}^{\text{axial},\phi, j}(\hat{y}, y)+L_{\text{perc}}^{\text{coronal},\phi, j}(\hat{y}, y)+L_{\text{perc}}^{\text{sagittal},\phi, j}(\hat{y}, y),
\label{eq:25Dperc}
\end{equation}
where $L_{\text{perc}}^{\text{axial},\phi, j}(\hat{y}, y)$, $L_{\text{perc}}^{\text{coronal},\phi, j}(\hat{y}, y)$, and $L_{\text{perc}}^{\text{sagittal},\phi, j}(\hat{y}, y)$ are the 2D perceptual losses computed along the axial, coronal, and sagittal planes, respectively. Although this strategy offers a straightforward means of leveraging established 2D feature extractors to incorporate multi-view information, the direct summation of independent plane-wise losses can lead to imbalanced optimization across views and fails to address cross-plane inconsistencies, thereby limiting volumetric coherence explicitly. Compared to these strategies, our proposed cyclic 2.5D perceptual loss mitigates the limitations of conventional perceptual losses by sequentially optimizing axial, coronal, and sagittal losses across training epochs. This cyclic scheduling balances learning across planes while promoting consistency in volumetric representation.

\textbf{Cross-Modality Brain Image Synthesis with Deep Learning}. 
Cross-modality image synthesis aims to learn the relationship between input and target images to generate realistic images in a different modality or contrast. In brain imaging, this technique is primarily used to synthesize images in different modalities (e.g., MRI, PET) or contrasts (e.g., T1w, diffusion-weighted imaging) from available scans. Clinically, it offers benefits, such as reducing scanner time and imaging costs, while deep learning enables high-quality synthesis despite the complex nonlinear mappings between modalities \citep{dayarathna2023deep}. Several backbone models have been employed for cross-modality synthesis, including CycleGAN (e.g., \citealt{yang2018unpaired,abu2021paired}), U-Net \citep{ronneberger2015u} (e.g., \citealt{sikka2018mri}), and Denoising Diffusion Probabilistic Models \citep{ho2020denoising} (e.g., \citealt{pan2023cycle,xie2023synthesizing}). Some approaches incorporate perceptual loss or similar loss functions as part of their composite loss formulation to enhance perceptual fidelity \citep{armanious2019unsupervised, dar2019image}.

Various approaches have been developed for translating medical images into PET images. GANDALF \citep{shin2020gandalf} introduced a conditional generative adversarial network with discriminator-adaptive loss fine-tuning to synthesize [\textsuperscript{18}F]AV-45 and [\textsuperscript{18}F]FDG PET images from T1w MRI scans while simultaneously classifying AD. BPGAN \citep{zhang2022bpgan} demonstrated T1w MRI to [\textsuperscript{18}F]FDG PET image translation using a conditional variational autoencoder GAN and a conditional latent regressor GAN with multiple loss functions, including SSIM loss. \citet{jang2023taupetgen} proposed TauPETGen, a latent-diffusion approach that generates [\textsuperscript{18}F]MK-6240 PET from text-reported MMSE scores and/or MRI. \citet{lee2024synthesizing} evaluated a convolutional neural network (CNN) for synthesizing tau PET from T1w MRI alone as well as from FDG PET and amyloid PET, and reported that although MRI-only input can feasibly generate tau PET, it yields substantially lower accuracy compared to PET-conditioned models. \citet{wang2024joint} proposed a joint learning framework comprising ShareGAN for the cross-modal synthesis of T1w MRI and [\textsuperscript{18}F]FDG PET, as well as a diagnosis network for AD detection; by jointly training these components, the framework improved diagnostic performance. \citet{li2024pasta} proposed PASTA, a conditional-diffusion dual-arm architecture that synthesized [\textsuperscript{18}F]FDG PET from T1w MRI and incorporated adaptive normalization layers to integrate multimodal conditions, enabling the generation of pathology-aware PET images. In contrast to this literature, our study focuses on the more constrained yet clinically practical setting of synthesizing [\textsuperscript{18}F]flortaucipir tau PET solely from T1w MRI.

\begin{center}
\begin{table*}[!h]
\caption{\blue{Demographic, clinical, and biomarker characteristics of participants in the ADNI cohort. For the total, training, validation, and test datasets, information is provided on gender, age (mean ± standard deviation), race, and clinical type (AD, LMCI, MCI, and EMCI). The ages are computed using the MRI acquisition date.
\blue{A$\beta$/Tau group counts are reported for participants with available amyloid status (Total n=494/516; Train n=348/360; Validation n=74/78; Test n=72/78).}\label{tab:dataset}}}
\resizebox{\textwidth}{!}{%
\begin{tabular}{@{\extracolsep\fill}cccccc@{}}
\toprule
\textBF{Data Split} &
\makecell{\textBF{Gender} \\ \textBF{(Male/Female)}} &
\makecell{\textBF{Age} \\ \textBF{(Min-Max)}} &
\makecell{\textBF{Race} \\ \textBF{(White/Black/Others)}} &
\makecell{\textBF{Diagnosis} \\ \textBF{(AD/LMCI/MCI/EMCI)}} &
\makecell{\textBF{\blue{A$\beta$/Tau}} \\ \textBF{\blue{(A$\beta$--T--/A$\beta$--T+/A$\beta$+T--/A$\beta$+T+})}} \\
\midrule
Total (516) & 291/225 &  74.2±7.9 (55.3-94.0) & 462/31/23 & 97/50/267/102 & \blue{169/29/76/220} \\
Train (360) & 206/154 & 73.9±8.0 (55.3-94.0) & 321/22/17 & 66/37/187/70 & \blue{120/19/52/157} \\
Validation (78) & 41/37 & 75.5±7.5 (59.4-91.6) & 69/4/5 & 16/8/39/15 & \blue{24/4/15/31} \\
Test (78) & 44/34 & 74.7±7.5 (57.4-92.0) & 72/5/1 & 15/5/41/17 & \blue{25/6/9/32} \\
\bottomrule
\end{tabular}
}
\end{table*}
\end{center}

\section{Method}
\blue{\subsection{Datasets and Study Design}}
\label{ssec:datacollection}

\blue{\subsubsection{ADNI}}

We compile a dataset of 516 subject-level pairs of 3D T1w MRI volumes and tau PET scans from the ADNI, phase 3 (ADNI-3). For every subject, the interval between the MRI and PET acquisitions is less than one year. \blue{\Cref{tab:dataset} summarizes the participant demographics, the distribution of clinical diagnoses, and the composition of the A$\beta$/Tau biomarker strata.} Clinically, the subjects are categorized as AD, late mild cognitive impairment (LMCI), mild cognitive impairment (MCI), or early mild cognitive impairment (EMCI). \blue{We further categorize the subset of participants with available amyloid status (n=494) into four A$\beta$/Tau biomarker groups (A$\beta$--T--, A$\beta$--T+, A$\beta$+T--, and A$\beta$+T+) for stratified analyses; the biomarker labeling procedure is described in the Biomarker stratification subsection.}

\blue{
For the primary experiments, we randomly partition the cohort at the subject level into training, validation, and test subsets containing 360, 78, and 78 MRI--PET pairs, respectively. In addition, we construct a site-held-out split based on the ADNI site code embedded in the participant identifier: all MRI–PET pairs from a given site code are assigned to a single subset, and no site code is present in more than one subset. This strategy reduces the risk of information leakage arising from site-specific acquisition characteristics and enables evaluation of generalization to unseen sites. To support a direct comparison with the primary split, the site-held-out split uses the same subset sizes of 360, 78, and 78 pairs, and includes 50 training sites, 4 validation sites, and 4 test sites. \Cref{tab:dataset} reports participant characteristics for the primary subject-level split, whereas the site-held-out split is reserved exclusively for the robustness analysis. Here, ``site'' refers to the ADNI site code in the participant identifier rather than the number of ADNI imaging centers. Detailed acquisition procedures for the T1w MRI and tau PET scans are described in the following section.
}

\textbf{MRI acquisition}.\label{sec:adni_mri} Structural T1w MRI data are acquired on ADNI‑qualified 3T scanners at 57 \blue{ADNI imaging centers}, including Siemens, GE, and Philips systems. All scans follow the standard ADNI sagittal 3D inversion-recovery spoiled gradient-echo protocol, with the following nominal parameters: TR $\approx$ 2,300 ms; TE $\approx$ min full echo (e.g., 2.8 ms); TI $\approx$ 900 ms; flip angle = 8--9\textdegree; matrix = 256 $\times$ 240; isotropic voxel size = 1.0 mm; and acquisition time $\approx$ 6 min with parallel imaging. 

\textbf{PET acquisition}. Tau imaging is performed using [\textsuperscript{18}F]flortaucipir. Participants receive an intravenous injection of 370 MBq (±10 \%), followed by scanning between 75 and 105 minutes post-injection, acquired as six consecutive 5-minute frames. For attenuation correction, PET/CT systems employ low-dose CT, whereas PET-only systems use post-emission transmission scans with rotating rod sources. Both configurations include Siemens, GE, and Philips scanners. The emission data are acquired in the 3D list mode and reconstructed locally using vendor-recommended iterative methods with standard corrections. 
Details of subsequent post-reconstruction processing to achieve uniform resolution across sites are provided in the following subsection.

\blue{
\textbf{Biomarker stratification}.\label{ssec:biomarker} We obtain amyloid PET region-of-interest (ROI) summary measures processed by the ADNI UC Berkeley PET Core using their standardized pipeline (uniform 6-mm resolution). Amyloid positivity (A$\beta$+) is determined using the ADNI-provided binary amyloid status, derived from tracer-specific cortical summary SUVR thresholds normalized to the whole cerebellum (SUVR $\ge$ 1.11 for florbetapir and $\ge$ 1.08 for florbetaben; \citealt{royse2021validation,pichet2022amyloid}). When available, we use the amyloid PET scan acquired at the same study visit as the tau PET scan; otherwise, we select the amyloid scan closest in time to the tau scan within a 2-year window. Tau positivity (T+) is defined as [${}^{18}$F]flortaucipir SUVR $\ge$ 1.27 in the meta-temporal ROI \citep{ossenkoppele2018discriminative,thijssen2021association}. The meta-temporal ROI includes the bilateral entorhinal, amygdala, fusiform, inferior temporal, and middle temporal cortices, and SUVR values are normalized to the inferior cerebellar gray matter. Biomarker labels are used solely for stratified reporting and are not provided as model inputs.
}

\blue{
\subsubsection{SCAN}
\label{ssec:dataset_scan}
We curate an independent cohort from the SCAN initiative comprising 96 subject-level paired 3D T1w MR and tau PET scans. MRI--PET pairs are selected such that the absolute inter-scan interval is less than one year. The SCAN subset used in this study includes only cognitively impaired participants and excludes cognitively unimpaired controls. Syndromic clinical categories comprise EMCI (n=29), LMCI (n=24), cognitive impairment not meeting criteria for mild cognitive impairment (n=19), and dementia (n=24). Dementia cases include clinically diagnosed AD dementia (n=9) as well as dementia attributed to non-AD or uncertain etiologies (n=15).

To explicitly characterize etiologic heterogeneity beyond AD, we further stratify participants into etiologic diagnostic groupings indicating vascular contributions to cognitive impairment and dementia (VCID; n=23), frontotemporal dementia (FTD) spectrum disorders (n=16), and Lewy body disease (LBD; n=10). These etiologic groupings are not mutually exclusive, and mixed etiologies are present (FTD/VCID: n=2; FTD/LBD: n=2). Only a minority of participants have AD as a primary or contributing etiology (n=14), making this cohort a stringent setting for evaluating generalizability beyond AD-enriched research cohorts and for probing the limited specificity of structural atrophy patterns to AD molecular pathology.

We partition the SCAN cohort into training, validation, and test subsets containing 67, 15, and 14 MRI--PET pairs, respectively, using a site-held-out strategy to reduce information leakage from site-specific acquisition characteristics. Specifically, all MRI--PET pairs acquired at a given imaging site are assigned to a single subset, and no site is shared across subsets. The resulting split comprises 5 training sites, 3 validation sites, and 3 test sites.

\textbf{MRI acquisition}. T1w MRI data are acquired at Alzheimer’s Disease Research Centers' imaging sites using SCAN-qualified scanners, including Siemens, GE, and Philips systems. The SCAN MR acquisition scheme is intentionally aligned with the ADNI protocol and includes, at minimum, an accelerated sagittal 3D T1w sequence (MPRAGE/IRSPGR) and a sagittal 3D FLAIR sequence; in the present study, only the T1w scan is used. The accelerated sagittal 3D T1w acquisition is harmonized to ADNI with respect to geometry and acceleration (e.g., isotropic voxel size $\approx$ 1.0 mm, $\sim$2$\times$ parallel imaging, and an acquisition time of $\sim$6 min). Vendor-specific timing parameters may vary across implementations and are specified via SCAN MRI Core--provided electronic protocol files to minimize manual parameter entry and ensure protocol fidelity.

\textbf{PET acquisition}. Tau PET is performed using [\textsuperscript{18}F]flortaucipir. The target injected activity is 370 MBq ($\pm$10\%), with a minimum injectable dose of 225 MBq. Emission acquisition is designed to fully encompass the target 80--100 min post-injection window and is reconstructed as consecutive 5-min frames. Attenuation correction is performed using scanner-specific procedures (PET/CT, PET/MR, or PET-only). For PET/CT systems, a low-dose CT scan is acquired with low effective mAs (typically $\sim$30), starting approximately 5 min prior to the PET emission scan. For PET-only systems, a 5--6 min transmission scan is acquired for attenuation correction (pre-emission for dynamic scans; post-emission for static scans using rod sources), and standard segmentation and re-projection routines are applied. PET images are reconstructed using scanner- and site-specific parameters provided in the SCAN PET manual appendices, with standard correction procedures applied.

}

\subsection{Data Preprocessing}
\label{ssec:datapreprocessing}
\begin{figure*}[tb]
  \centering
  \includegraphics[width=\linewidth]{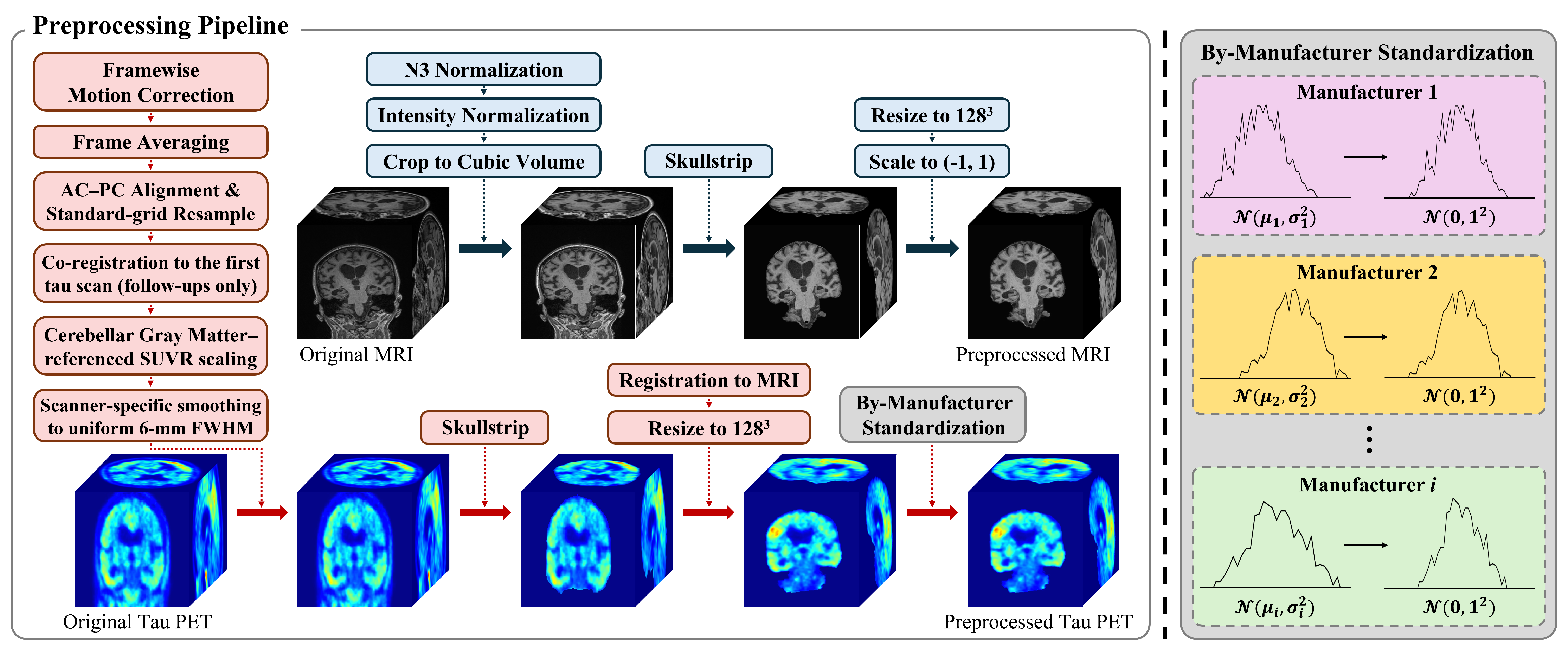}
  \caption{End-to-end preprocessing workflow for T1w MR and tau PET images, including the by-manufacturer standardization of PET SUVR. \blue{Tau PET processing involves framewise rigid-body motion correction and temporal averaging, alignment to the AC--PC plane with resampling to a standardized voxel grid, longitudinal within-participant co-registration to the first flortaucipir scan, SUVR normalization using a cerebellar gray matter reference region, and scanner-specific 3D Gaussian smoothing to harmonize effective spatial resolution.} T1w MRI processing includes non-parametric non-uniform intensity correction, intensity normalization, and conversion to a cubic volume. Skull and other non-brain tissues are removed from both modalities. Tau PET scans are subsequently co-registered to the corresponding T1w MRIs. Finally, T1w voxel intensities are rescaled to the range $(-1, 1)$, whereas tau PET SUVRs are z-standardized (mean = 0, standard deviation = 1) within each scanner vendor.
  }
  \label{fig:preprocessing_pipeline}
\end{figure*}

\blue{
\Cref{fig:preprocessing_pipeline} summarizes our data preprocessing workflow, spanning vendor- and consortium-provided processing through our in-house preparation for model training. Steps performed using \texttt{FreeSurfer} (version 7.4.1; \citealt{fischl2012freesurfer}) are explicitly indicated; all subsequent processing steps are implemented in our custom scripts. The procedures described below are applied identically to both the ADNI and SCAN datasets.
}

\textbf{MRI preprocessing}. Preprocessing starts with non-parametric non-uniform intensity correction \citep{sled1998nonparametric} using the \texttt{mri\_nu\_correct.mni} command, followed by initial intensity normalization \citep{dale1999cortical} via \texttt{mri\_normalize}. After normalization, volumes are cropped to a cubic volume ($L \times L \times L$). Axial slices dominated by background voxels at the superior and inferior extremes are discarded to avoid introducing empty regions during resizing. Skull stripping and removal of extracranial tissue are performed with SynthStrip \citep{hoopes2022synthstrip} (\texttt{mri\_synthstrip}). The resulting brain-only images are resampled to $128 \times 128 \times 128$ voxels and linearly scaled to the interval $(-1, 1)$ with min--max normalization.

\textbf{PET preprocessing}. \blue{For both datasets, we analyze the most fully preprocessed [\textsuperscript{18}F]flortaucipir images (Step 4; ADNI: ``Co-registered, Averaged, Standardized Image and Voxel Size, Uniform 6 mm Res''; SCAN: ``T80 Coreg, Avg, Rigid Reg to Std Img/Vox Size, 80--100*, 6 mm Res''). These images are generated using a standardized, centrally implemented post-processing pipeline that harmonizes data across scanner platforms. Briefly, for each acquisition, individual 5-min frames are rigidly co-registered to the first extracted frame for motion correction and then averaged to produce a single static image. For each participant, the first tau PET scan is rigidly reoriented to anterior commissure--posterior commissure (AC--PC) alignment and resampled to a $160 \times 160 \times 96$ matrix with 1.5 mm isotropic voxels; subsequent flortaucipir scans are co-registered to the first scan to maintain a consistent AC--PC orientation across time points. The images are intensity-normalized using an atlas-space cerebellar gray matter reference region to generate SUVR images. To harmonize spatial resolution, scanner-specific smoothing kernels are applied to achieve an effective full-width at half-maximum (FWHM) of 6 mm.} Building on these uniform-resolution SUVR images, we remove non-brain tissue using \texttt{mri\_synthstrip}, rigidly co-register each tau PET volume to its corresponding T1w MRI using \texttt{mri\_coreg} and \texttt{mri\_vol2vol}, and resample the co-registered volumes to $128 \times 128 \times 128$ voxels to match the processed MRI dimensions. All subsequent analyses are performed on SUVR images normalized to cerebellar gray matter.

\blue{
We observe residual manufacturer-dependent differences in the overall scale and dispersion of cortical SUVR values even after the ``Step 4'' pipeline, which applies scanner-specific smoothing to harmonize effective spatial resolution and generates cerebellar gray matter--referenced SUVR images. To mitigate the remaining intensity-scale heterogeneity prior to model training, we perform by-manufacturer $z$-standardization of SUVR values. We deliberately avoid global min--max rescaling because it can undesirably compress the upper tail of the SUVR distribution. Importantly, this manufacturer-wise standardization serves as a pragmatic intensity normalization step applied after PET processing to facilitate model training; it is not intended as a PET harmonization procedure \citep{joshi2009reducing} aimed at correcting scanner-physics effects or site-related biological variability.

We perform by-manufacturer $z$-standardization of SUVR values to mitigate inter-scanner intensity variation, yielding distributions with mean $0$ and standard deviation $1$ within each manufacturer. Let $m \in \mathcal{M}$ denote the PET scanner manufacturer; in our dataset, $\mathcal{M}=\{\text{Siemens}, \text{GE}, \text{Philips}\}$. For a PET volume $y$ acquired on a scanner from manufacturer $m$, we apply the following by-manufacturer standardization:
\begin{equation}
\tilde{y} = \mathcal{S}_m(y) = \frac{y-\mu_m}{\sigma_m+\varepsilon_{\text{std}}},
\label{eq:vendor_std}
\end{equation}
where $\mu_m$ and $\sigma_m$ denote the mean and standard deviation estimated from the training subset for manufacturer $m$, and $\varepsilon_{\text{std}}>0$ is a small constant included for numerical stability. For each dataset configuration, we estimate $\mu_m$ and $\sigma_m$ exclusively on the corresponding training subset and apply them unchanged to the validation and test subsets to prevent information leakage. Because \Cref{eq:vendor_std} defines an affine transformation within each manufacturer, it preserves the within-manufacturer rank ordering of SUVR values while reducing between-manufacturer differences in intensity scale. All loss terms in \Cref{ssec:cyclic} are computed using the standardized volumes $\tilde{y}$ and $\hat{\tilde{y}}$. For notational convenience, we subsequently omit tildes and adopt the assignments $y \leftarrow \tilde{y}$ and $\hat{y} \leftarrow \hat{\tilde{y}}$ in the loss definitions.
}

\subsection{3D T1w MRI-to-Tau PET Translation}
\label{ssec:cyclic} 

Given a T1w MRI modality $X$ paired with the corresponding tau PET modality $Y$, our objective is to learn a mapping $G: X \rightarrow Y$ such that $G(x) \approx y$ for each $x \in X$ to its paired $y \in Y$. A 3D image translation model $G$ is trained to minimize the difference between the generated tau PET image $\hat{y}$ and the ground truth tau PET image $y$ by optimizing the objective function described below, with a 3D T1w MRI $x$ as input. The proposed loss function is not specific to a particular model but can be applied to various image-to-image translation models, such as U-Net and GANs.

\blue{
\textbf{Cyclic 2.5D Perceptual Loss.}
We introduce a cyclic 2.5D perceptual loss, denoted by
$L^{\phi,j}_{\mathrm{cyc2.5D}}(\hat{y},y; e)$, which computes a plane-wise perceptual loss while selecting the slicing plane in an epoch-dependent cyclic manner, as illustrated in  \Cref{fig:cyclic_perceptual}. Let $y \in \mathbb{R}^{D \times H \times W}$ denote the ground-truth PET volume and let $\hat{y}=G(x)$ denote the generated PET volume. For each plane
$p \in \{\mathrm{axial}, \mathrm{coronal}, \mathrm{sagittal}\}$, we write $\mathrm{slice}_p(\cdot,i)$ for the $i$-th 2D slice extracted along plane $p$, and we let $N_p$ be the number of slices along that plane. We apply min--max normalization to each extracted slice prior to evaluation by a pre-trained 2D perceptual network:
\begin{equation}
\mathcal{N}(u) = \frac{u-\min(u)}{\max(u)-\min(u)+\varepsilon_{mm}},
\label{eq:minmax_norm}
\end{equation}
where $\varepsilon_{mm}>0$ prevents division by zero. Let $\phi$ denote the pre-trained 2D perceptual model, and let $\phi_j(\cdot)$ denote the feature representation produced by its first $j$ layers. We define the plane-wise, slice-averaged perceptual loss as
\begin{equation}
L^{p,\phi,j}_{\mathrm{perc}}(\hat{y},y)
= \frac{1}{N_p}\sum_{i=1}^{N_p}
\mathrm{MSE}\!\left(
\phi_j\!\left(\mathcal{N}(\mathrm{slice}_p(\hat{y},i))\right),
\phi_j\!\left(\mathcal{N}(\mathrm{slice}_p(y,i))\right)
\right),
\label{eq:plane_perc}
\end{equation}
where $\mathrm{MSE}(a,b)=\frac{1}{|a|}\|a-b\|_2^2$ averages the squared error over all feature-map elements.

In the proposed cyclic formulation, exactly one slicing plane is activated at each epoch, in contrast to the standard 2.5D perceptual loss in \Cref{eq:25Dperc}, which aggregates perceptual losses from all planes simultaneously. Let $e \in \{0,1,2,\dots\}$ denote the epoch index. We define a sequence of cycle durations $\{T_k\}_{k\ge 0}$ by setting $T_0=T$ and updating
\begin{equation}
T_{k+1} = \max\!\left(1,\mathrm{round}(\gamma T_k)\right),
\label{eq:Tk_update}
\end{equation}
where $\gamma$ controls the decay of the cycle length and the $\max$ operator enforces a minimum duration of one epoch. We further define the cumulative start index of cycle $k$ as $s_0=0$ and $s_{k+1}=s_k+3T_k$, so that each cycle consists of three consecutive segments of length $T_k$, one per plane. For a given epoch $e$, let $k(e)$ be the unique integer satisfying $s_{k(e)} \le e < s_{k(e)+1}$. We then compute the within-cycle segment index
\begin{equation}
q(e)=\left\lfloor \frac{e-s_{k(e)}}{T_{k(e)}}\right\rfloor \in \{0,1,2\},
\end{equation}
which identifies which of the three plane segments is active at epoch $e$. The selected plane is defined as
\begin{equation}
\pi(e)=
\begin{cases}
\mathrm{axial}, & q(e)=0,\\
\mathrm{coronal}, & q(e)=1,\\
\mathrm{sagittal}, & q(e)=2.
\end{cases}
\label{eq:plane_schedule}
\end{equation}

Using this schedule, we define the cyclic 2.5D perceptual loss at epoch $e$ as an epoch-dependent weighted sum of plane-specific perceptual losses:
\begin{equation}
L^{\phi,j}_{\mathrm{cyc2.5D}}(\hat{y},y; e)
=
\sum_{p \in \mathcal{P}} w_p(e)\, L^{p,\phi,j}_{\mathrm{perc}}(\hat{y},y),
\quad
w_p(e) = \mathbb{I}\!\left[p = \pi(e)\right],
\label{eq:cyc25d}
\end{equation}
where $\mathcal{P}=\{\mathrm{axial},\mathrm{coronal},\mathrm{sagittal}\}$ denotes the set of slicing planes, $L^{p,\phi,j}_{\mathrm{perc}}(\hat{y},y)$ is the perceptual loss evaluated on plane $p$, and $\pi(e)\in\mathcal{P}$ specifies the plane selected at epoch $e$. The indicator function $\mathbb{I}[\cdot]$ equals $1$ if its argument is true and $0$ otherwise; consequently, $L^{\phi,j}_{\mathrm{cyc2.5D}}(\hat{y},y; e)$ reduces to $L^{\pi(e),\phi,j}_{\mathrm{perc}}(\hat{y},y)$ at each epoch. 
}

\textbf{Structural Preservation Loss}.
We utilize SSIM as a loss function to preserve structural similarity between the generated PET image $\hat{y}$ and the ground truth PET image $y$. SSIM is a metric that evaluates statistical similarity based on luminance and contrast \citep{wang2004image}. Given a predetermined window size, a Gaussian filter with a standard deviation of $\sigma$ is applied, and the window scans across both images, $\hat{y}$ and $y$. For each window pair, the means $\mu_{\hat{y}}$, $\mu_y$, variances $\sigma_{\hat{y}}$, $\sigma_y$, and covariance $\sigma_{\hat{y} y}$ are calculated. SSIM is computed as:
\begin{equation}
\text{SSIM}(\hat{y}, y) = \frac{(2\mu_{\hat{y}}\mu_y + C_1)(2\sigma_{\hat{y}y} + C_2)}{(\mu_{\hat{y}}^2 + \mu_y^2 + C_1)(\sigma_{\hat{y}}^2 + \sigma_y^2 + C_2)},
\label{eq:SSIM}
\end{equation}
where $C_1 = (k_1 \cdot L)^2$, $C_2 = (k_2 \cdot L)^2$, $L$ represents the data range, and $k_1$ and $k_2$ are set to 0.01 and 0.03, respectively. After the window completes its traversal over the image, the average SSIM value is computed. The window has the same dimensionality as the images. When used as a loss function, the SSIM loss is defined as:
\begin{equation}
L_{\text{SSIM}}(\hat{y}, y) = 1 - \text{SSIM}(\hat{y}, y).
\label{eq:SSIMLoss}
\end{equation}

\textbf{Voxel-level Fidelity Loss}.
In addition, a voxel-wise loss function $L_{\text{voxel}}$ is defined based on the MSE to retain low-level voxel information, as follows:
\begin{equation}
L_{\text{voxel}}(\hat{y}, y) = \frac{1}{D \times H \times W} \sum_{i=1}^D \sum_{j=1}^H \sum_{k=1}^W (\hat{y}_{ijk}-y_{ijk})^2,
\label{eq:MSE}
\end{equation}
where $D$, $H$, and $W$ represent the depth, height, and width of the images, respectively, and $i$, $j$, $k$ are the voxel indices along these dimensions.

Our final combined training objective $L_{\text{combined}}$ is formulated as:
\blue{
\begin{equation}
L_{\mathrm{combined}}(e)
= L_{\mathrm{voxel}}(\hat{y},y)
+ L_{\mathrm{SSIM}}(\hat{y},y)
+ \lambda\, L^{\phi,j}_{\mathrm{cyc2.5D}}(\hat{y},y;e).
\label{eq:combinedPerc}
\end{equation}
}
where $\lambda$ is a weighting factor for the cyclic 2.5D perceptual loss. The weights of $L_{\text{voxel}}$ and $L_{\text{SSIM}}$ are fixed to 1 to streamline the hyperparameter selection process.

\subsection{Experiment}
\label{ssec:experiment}
\textbf{Training Details}.
All primary experiments, except the benchmark, use an image translation model $G$ implemented as a 3D U-Net with channel depths of 64, 128, 256, 512, and 1024. Instance normalization \citep{ulyanov2016instance} replaces batch normalization \citep{ioffe2015batch} to improve stability when training with small batch sizes. Models are optimized for up to 1,000 epochs. Early stopping halts training when the validation loss fails to improve for 30 consecutive epochs, counting only epochs completed after the second axial--coronal--sagittal cycle. Optimization uses Adam \citep{kingma2015adam} with a cosine‑annealing learning‑rate scheduler \citep{loshchilov2017sgdr}; the maximum learning rate is $5\times10^{-4}$ and the annealing period is fixed at 120 epochs. A dropout layer with a probability of 0.2 is added to reduce overfitting. On-the-fly data augmentation is applied independently at every epoch to increase robustness. Each paired MRI–PET volume could undergo 3D elastic deformation (magnitude 50--100, Gaussian kernel $\sigma =$ 4--7, trilinear interpolation), random affine transformation (rotations within ±15\textdegree and scaling within ±10 \%), and random flips about each anatomical axis; every augmentation is sampled with 0.5 probability and applied identically to both modalities. Gaussian noise with mean 0 and standard deviation uniformly sampled from 0 to 0.1 is injected into the MRI channel to simulate acquisition artifacts. \blue{The primary experiments are performed on the ADNI dataset, and the identical 3D U-Net architecture and training protocol are applied to the SCAN dataset for independent-cohort replication.} Experiments are conducted on NVIDIA H100 GPUs (80 GB VRAM) and NVIDIA RTX 6000 Ada GPUs (48 GB VRAM); the hardware is kept constant for experiments on the same topic to minimize confounds. The code is publicly available at
\url{https://github.com/labhai/Cyclic-2.5D-Perceptual-Loss}.

\textbf{Evaluation Metrics}. We employ the SSIM, peak signal-to-noise ratio (PSNR), and mean absolute error (MAE) as metrics to evaluate the quality of the synthesized PET images quantitatively. The analysis includes both 3D average test values and 2D slice-wise evaluations for the SSIM, PSNR, and MAE. 
\blue{All quantitative results are reported as mean $\pm$ standard deviation across subjects. For the volumetric evaluation, each metric is computed directly on the entire 3D PET volume for each subject, producing a single scalar score per subject. These volumetric scores are reported as ``Overall (3D)'' in the quantitative tables.}
For the 2D slice-wise evaluations, assessments are conducted on axial, coronal, and sagittal slices, with results averaged across slices 20 to 90. 
\blue{These plane-wise results are reported as ``Axial,'' ``Coronal,'' and ``Sagittal,'' respectively. Because the plane-wise analysis is restricted to a subset of slices that excludes peripheral regions largely dominated by background, the ``Overall (3D)'' scores are not necessarily equal to the average of the three plane-wise scores.}

The PSNR quantifies the signal-to-noise ratio relative to the peak intensity and is defined as follows:
\begin{equation}
\text{PSNR}(\hat{y}, y) = 10\log_{10}(\frac{R^2}{\text{MSE}(\hat{y}, y)})
\end{equation}
where $R$ is the maximum SUVR value of the ground truth PET image $y$. The MAE measures the mean of absolute differences between the generated PET image $\hat{y}$ and the ground truth PET image $y$, expressed as:
\begin{equation}
\text{MAE}(\hat{y}, y) = \frac{1}{N} \sum_{i=1}^N |\hat{y}_{i}-y_{i}|,
\end{equation}
where $N$ denotes the total number of voxels. 

The SSIM is selected as the primary evaluation metric due to its ability to assess structural similarity, providing a more accurate representation of perceptual quality as perceived by the human visual system. Unlike metrics such as PSNR or MAE, which rely on pixel-level differences, SSIM evaluates the structural integrity of images. PSNR and MAE are sensitive to pixel value discrepancies and can be impacted by misregistration or noise introduced during the co-registration of MR and PET images. This sensitivity may lead to biased evaluations of image quality. Consequently, SSIM is deemed a more suitable metric for evaluating the performance of tau PET synthesis and is adopted as the primary criterion for assessment.

\textbf{Cyclic 2.5D Perceptual Loss Hyperparameter Selection}.
We conduct hyperparameter selection for the cyclic 2.5D perceptual loss using the validation dataset. The primary objective of the initial experiments is to identify the optimal cycle duration $T$, which determines the number of training epochs devoted to a single anatomical plane before switching to the next cycle. A greedy search strategy is employed, starting from $T=60$ and incrementally increasing by 30. As shown in \Cref{tab:cycle_duration}, the SSIM scores improve progressively with increasing $T$ from 60 to 90 and further to 120. However, we observe a decline at $T=150$, indicating diminishing returns. Consequently, $T=120$ is selected as the optimal duration, as it yields the best performance before any degradation occurs. During this search, the cycle duration scaling factor $\gamma$ and the perceptual loss weight $\lambda$ are both fixed at 1. For consistency across experiments, we employ a VGG11-based perceptual loss model, pre-trained on ImageNet-1K \citep{imagenetrussakovsky2015imagenet}, which extracts feature maps after the fourth ReLU layer.

\begin{table*}[!t]
\centering
\caption{Greedy search of the cycle duration $T$ for the cyclic 2.5D perceptual loss. Starting from $T=60$, the value is incrementally increased in steps of 30, and model performance on the validation dataset is monitored. Experiments are performed using a 3D U-Net model trained on datasets preprocessed with by-manufacturer PET SUVR standardization. \blue{The values are reported as mean ± standard deviation across subjects. ``Overall (3D)'' refers to the 3D SSIM or PSNR computed on the entire volume; ``Axial/Coronal/Sagittal'' corresponds to the slice-wise 2D SSIM or PSNR averaged over slices 20–90 in each plane. The best-performing value in each column is shown in \textbf{bold}.}\label{tab:cycle_duration}} 
\resizebox{\textwidth}{!}{%
\begin{tabular}{@{}ccccccccc@{}}
\toprule
\multirow{2}{*}{\textBF{Cycle Duration ($T$)}} & \multicolumn{4}{c}{\textBF{SSIM(\%)↑}} & \multicolumn{4}{c}{\textBF{PSNR(dB)↑}} \\ 
\cmidrule(lr){2-5} \cmidrule(lr){6-9} 
& \textBF{Overall (3D)} & \textBF{Axial} & \textBF{Coronal} & \textBF{Sagittal} & \textBF{Overall (3D)} & \textBF{Axial} & \textBF{Coronal} & \textBF{Sagittal} \\ 
\midrule
60  & 90.37±3.68 & 86.89±4.40 & 86.67±4.37 & 85.39±4.83 & 29.01±3.04 & \textBF{30.66±4.99} & 27.82±2.96 & 28.21±3.11 \\
90  & 90.38±3.72 & 86.88±4.47 & 86.68±4.42 & 85.41±4.88 & 29.00±3.14 & 30.59±5.02 & 27.78±3.06 & 28.18±3.20 \\
120 & \textBF{90.61±3.62} & \textBF{87.16±4.32} & \textBF{86.95±4.33} & \textBF{85.74±4.76} & \textBF{29.20±3.04} & 30.54±4.61 & \textBF{27.96±2.96} & \textBF{28.32±3.07} \\
150 & 90.42±3.67 & 86.90±4.44 & 86.72±4.37 & 85.49±4.78 & 28.89±3.17 & 30.48±5.05 & 27.68±3.12 & 28.06±3.23 \\
\bottomrule
\end{tabular}
}
\end{table*}
\begin{table*}[!t]
\centering
\caption{Grid search of the cycle duration scaling factor $\gamma$ for the cyclic 2.5D perceptual loss. Experiments are conducted using increased, fixed, and decreased cycle durations with the validation dataset. The results are obtained from a 3D U-Net model trained on datasets preprocessed with by-manufacturer PET SUVR standardization. \blue{The values are reported as mean ± standard deviation across subjects. ``Overall (3D)'' refers to the 3D SSIM or PSNR computed on the entire volume; ``Axial/Coronal/Sagittal'' corresponds to the slice-wise 2D SSIM or PSNR averaged over slices 20–90 in each plane. The best-performing value in each column is shown in \textbf{bold}.}\label{tab:cycle_duration_factor}} 
\resizebox{\textwidth}{!}{%
\begin{tabular}{c@{\hspace{3pt}}ccccccccc}
\toprule
\multicolumn{2}{c}{\multirow{2}{*}{\begin{tabular}[c]{@{}c@{}}\textBF{Cycle Duration}\\\textBF{Scaling Factor ($\gamma$)}\end{tabular}}} & \multicolumn{4}{c}{\textBF{SSIM(\%)↑}} & \multicolumn{4}{c}{\textBF{PSNR(dB)↑}} \\ 
\cmidrule(lr){3-6} \cmidrule(lr){7-10} 
& & \textBF{Overall (3D)} & \textBF{Axial} & \textBF{Coronal} & \textBF{Sagittal} & \textBF{Overall (3D)} & \textBF{Axial} & \textBF{Coronal} & \textBF{Sagittal} \\ 
\midrule
Increase 2.0 & ($\times$ 2.0) & 90.47±3.66 & 86.96±4.38 & 86.79±4.36 & 85.51±4.80 & 29.12±3.07 & 30.36±4.52 & 27.91±2.99 & 28.22±3.09 \\
Increase 1.5 & ($\times$ 1.5) & 90.62±3.62 & 87.16±4.35 & 86.96±4.35 & 85.74±4.76 & 29.24±3.07 & \textBF{30.55±4.56} & 27.99±2.99 & 28.34±3.10 \\
Fixed & ($\times$ 1.0) & 90.61±3.62 & 87.16±4.32 & 86.95±4.33 & 85.74±4.76 & 29.20±3.04 & 30.54±4.61 & 27.96±2.96 & 28.32±3.07 \\
Decrease 1.5 & ($\times$ 0.67) & \textBF{90.64±3.61} & \textBF{87.22±4.32} & \textBF{87.03±4.33} & \textBF{85.74±4.76} & \textBF{29.24±3.12} & 30.51±4.58 & \textBF{28.02±3.06} & \textBF{28.36±3.10} \\
Decrease 2.0 & ($\times$ 0.5) & 90.51±3.69 & 87.02±4.40 & 86.83±4.42 & 85.59±4.83 & 29.09±3.03 & 30.34±4.52 & 27.88±2.98 & 28.21±3.07 \\
\bottomrule
\end{tabular}
}
\end{table*}
\begin{table*}[!t]
\centering
\caption{Experimental results for selecting the cyclic 2.5D perceptual loss model $\phi$ and the corresponding loss weight $\lambda$. Models pre-trained on the medical image dataset RadImageNet \citep{mei2022radimagenet} and the natural image dataset ImageNet-1K \citep{imagenetrussakovsky2015imagenet} are evaluated using the validation set for performance assessment. The loss weight $\lambda$ is tested at values \{0.1, 0.3, 0.5, 1, 3, 5\} for each model, and only the best-performing configuration is reported. The complete set of experimental results is provided in Suppl. A.1. \blue{The values are reported as mean ± standard deviation across subjects.} The SSIM, PSNR, or MAE values represent performance across the 3D volume. The best-performing value in each column is shown in \textbf{bold}.\label{tab:perceptual_loss_model}} 
\resizebox{\textwidth}{!}{%
\begin{tabular}{@{}ccccccccc@{}}
\toprule
\textBF{Dataset}           & \begin{tabular}[c]{@{}c@{}}\textBF{Perceptual Loss}\\\textBF{Model ($\phi$)}\end{tabular} & \begin{tabular}[c]{@{}c@{}}\textBF{Feature Extraction}\\\textBF{Layer ($j$)}\end{tabular} & \textBF{$n_{\text{params}}$} & \begin{tabular}[c]{@{}c@{}}\textBF{Perceptual Loss}\\\textBF{Weight ($\lambda$)}\end{tabular} & \textBF{SSIM(\%)↑} & \textBF{PSNR(dB)↑} & \textBF{MAE↓} \\
\midrule
\multirow{2}{*}{RadImageNet} & ResNet-50 & 3rd Block Stack & 8.53M & 1.0 & 90.37±3.58 & 29.02±3.10 & 10.05±4.85 \\
                             & InceptionNet v3 & 8th Inception Module & 8.97M & 0.5 & 90.48±3.67 & 29.11±3.19 & 10.03±4.79 \\ 
[0.3cm]
\multirow{3}{*}{ImageNet-1K} & ResNet-50 & 3rd Block Stack & 8.53M & 0.5 & 90.46±3.68 & 29.06±3.16 & 9.99±4.97 \\
                             & InceptionNet v3 & 8th Inception Module & 8.97M & 0.5 & 90.65±3.57 & 29.30±3.09 & 9.79±4.65 \\
                             & VGG-16 & 10th ReLU & 7.63M & 0.5 & \textBF{90.69±3.59} & \textBF{29.31±3.19} & \textBF{9.73±4.83} \\
\bottomrule
\end{tabular}
}
\end{table*}

Following this, we optimize the cycle duration scaling factor $\gamma$, which governs the progression of epoch intervals between loss cycles across perceptual planes. We evaluate several strategies, including constant, increasing, and decreasing durations between cycles. As detailed in \Cref{tab:cycle_duration_factor}, a grid search reveals that progressively shortening the cycle duration by a factor of 1.5 (i.e., multiplying by 0.67) yields the most stable and generalizable performance across both global and individual planes. In contrast, fixing $\gamma=1.0$ or increasing it by a factor of 1.5 leads to overfitting, with improvements limited to specific planes. Therefore, we adopt a decreasing schedule with $\gamma=0.67$ for all subsequent experiments. These experiments utilize the same perceptual loss model and weights as previously described.

We next evaluate different pre-trained 2D models $\phi$ and corresponding weights $\lambda$ for feature extraction in computing the cyclic 2.5D perceptual loss. The results are summarized in \Cref{tab:perceptual_loss_model}. We compare models trained on medical image datasets with those trained on natural image datasets to assess the impact of domain-specific pretraining. For the medical domain, we utilize ResNet-50 \citep{resnethe2016deep} and Inception-v3 \citep{inceptionv3szegedy2016rethinking}, both pre-trained on RadImageNet \citep{mei2022radimagenet}. For the natural domain, we evaluate ResNet-50, Inception-v3, and VGG-16, all of which are pre-trained on ImageNet-1K. The total number of parameters from the input layer to the feature extraction layer is maintained at a similar level across models to ensure a fair comparison. Among all tested configurations, the VGG-16 model pre-trained on ImageNet-1K and configured to compute perceptual loss at the 23rd layer (10th ReLU) achieves the best validation performance. Accordingly, we select the VGG-16 model as the final perceptual loss model $\phi$, with the feature extraction layer $j$ fixed at 23. The optimal loss weight $\lambda$ for this configuration is determined to be 0.5. For completeness, we evaluate $\lambda \in \{0.1, 0.5, 1, 3, 5\}$ and report the best-performing setting. Full results are provided in Suppl. Section 1.

\textbf{Benchmark}.
Quantitative evaluations and visual qualitative assessments are performed to evaluate the performance of the proposed cyclic 2.5D perceptual loss for T1w MRI to tau PET translation. For ablation, the cyclic 2.5D component in the combined loss (Eq. \ref{eq:combinedPerc}) is individually replaced with the conventional 2D, 3D, or 2.5D perceptual loss. Because the structural MR volumes are acquired in the sagittal plane, the 2D perceptual loss is computed on sagittal slices to preserve anatomical correspondence. We also evaluate the impact of by-manufacturer standardization by comparing datasets generated using min--max normalization within the range of $(–1, 1)$ during the final PET normalization step. The perceptual loss weight $\lambda$ is set to 0.5 for all experiments. The perceptual loss model $\phi$ utilizes the VGG-16 pre-trained on ImageNet-1K for the 2D, 2.5D, and cyclic 2.5D approaches, using the 10th ReLU layer as the feature extraction layer $j$. In contrast, the 3D approach employs the pre-trained ResNet-50 MedicalNet model \citep{chen2019med3d}, with the feature extraction layer $j$ set to the fourth block stack. A variety of generative approaches are employed for the 3D image translation model $G$, including the 3D U-Net, U-Net-like models (UNETR and SwinUNETR), and GAN-based models (CycleGAN and Pix2Pix), with the 3D U-Net serving as the generator in the GAN-based models. The experiments involving the 3D U-Net and its variants employ the combined loss function described in \Cref{eq:combinedPerc}. Two distinct loss configurations are evaluated for the CycleGAN and Pix2Pix experiments: the original loss and a combination of the original loss with the proposed combined loss. CycleGAN's original loss consists of adversarial loss, cycle consistency loss, and identity loss. Pix2Pix's original loss comprises adversarial loss and L1 loss. The proposed loss is incorporated into CycleGAN’s original loss to create a combined loss configuration. For Pix2Pix, the L1 loss is replaced with the proposed loss. Since experiments involving the 2D, 3D, and 2.5D perceptual loss models do not require delayed early stopping, early stopping is applied after completing the first cycle of the cosine annealing scheduler across all experiments. The maximum learning rate is set to 2e-4 for the CycleGAN and Pix2Pix experiments. 

Quantitative comparisons are conducted with the state-of-the-art algorithms designed for synthesizing PET images from MR images to validate the effectiveness of the proposed approach. Specifically, the open-source implementations of PASTA \citep{li2024pasta} and ShareGAN with joint learning \citep{wang2024joint} are utilized. We employ our own dataset and preprocessing pipeline because the datasets and preprocessing scripts used in their experiments are unavailable, apart from the training algorithms. The MR and PET images are preprocessed according to the steps outlined in \Cref{ssec:datapreprocessing}, excluding resizing and normalization. Subsequently, the image pairs are resized and normalized to match the input requirements of each framework. Specifically, PASTA utilizes MRI and PET pairs normalized to the range $(0, 1)$ with dimensions ($128 \times 128 \times 128$), whereas ShareGAN applies normalization to the range $(-1, 1)$ with dimensions ($256 \times 256 \times 256$). Patient metadata beyond the MR and PET images is excluded during training with PASTA to ensure a fair comparison. For ShareGAN, joint learning is performed via a binary classification task to distinguish AD from EMCI/MCI/LMCI.

\textbf{ROI Analysis}.
An ROI analysis is performed after the quantitative and visual qualitative analysis to demonstrate the effectiveness of the proposed method in capturing tau PET brain signatures in Alzheimer's patients. The proposed method, which utilizes the combined loss incorporating the cyclic 2.5D perceptual loss, is compared to the models trained by replacing the cyclic 2.5D perceptual loss with the 2D, 3D, or 2.5D loss.

All PET scans, including the ground truth PET images and those generated by the aforementioned models, are registered to a standard anatomical space defined by the fsaverage template provided by \texttt{FreeSurfer}. Specifically, the MR images are first registered to the fsaverage surface, and the linear transformation matrix obtained during this registration is subsequently applied to align the corresponding PET images to the fsaverage surface. Since the MR and PET images are co-registered in the MRI space, voxel-wise correspondence of the atlas-aligned PET data is ensured across subjects. ROIs are defined based on the Desikan--Killiany atlas \citep{desikan2006automated} provided by \texttt{FreeSurfer}. The SUVRs within each region, as defined by the atlas-based ROI segmentation map, are averaged to derive a representative PET SUVR value for each region to extract region-specific PET signals. The mean SUVR for each ROI is calculated using the following formula:
\begin{equation}
\mu_{\text{ROI}} = \frac{1}{N_{\text{ROI}}} \sum_{i=1}^{N_{\text{ROI}}} I(v_i),
\end{equation}
where $\mu_{\text{ROI}}$ represents the mean SUVR for a specific ROI, $N_{\text{ROI}}$ denotes the total number of voxels within the ROI, and $I(v_i)$ indicates the SUVR of the $i$-th voxel within the ROI.

The ROI analysis is conducted by calculating the MSE for each cortical region defined in the Desikan--Killiany atlas. This process involves comparing the mean SUVR from the ground truth PET images with the mean SUVR from the generated PET images across different settings. The analysis is performed for all 78 subjects in the test dataset, and the results are averaged to compute the overall MSE for each ROI across all experimental conditions. Specifically, the MSE for a given ROI, denoted as $\text{MSE}_{\text{ROI}}$, is calculated as follows:
\begin{equation}
\text{MSE}_{\text{ROI}} = \frac{1}{N_{\text{subject}}} \sum_{i=1}^{N_{\text{subject}}} (\hat{\mu}_{\text{ROI}, i} - \mu_{\text{ROI}, i})^2,
\end{equation}
where $N_{\text{subject}}$ represents the total number of test subjects, $\hat{\mu}_{\text{ROI,i}}$ denotes the mean SUVR of the generated PET image for the specified ROI of the $i$-th subject, and $\mu_{\text{ROI,i}}$ corresponds to the mean SUVR of the ground truth PET image for the same ROI and subject.

\blue{
\textbf{Diagnosis- and A$\beta$/Tau-stratified evaluation.}
We conduct a stratified evaluation on the ADNI test set to examine whether the fidelity of the synthesized tau PET differs across clinical diagnostic categories and biomarker-defined disease stages. Using the same subject-level metrics as in the primary quantitative evaluation, we compute performance separately for each diagnosis group (EMCI, MCI, LMCI, and AD) and for each A$\beta$/Tau biomarker stratum (A$\beta$--T--, A$\beta$--T+, A$\beta$+T--, and A$\beta$+T+). Within each stratum, we report results for the proposed method and for model variants trained with conventional perceptual losses (2D, 3D, and 2.5D) under otherwise identical experimental settings. Diagnosis and biomarker labels serve only to define subgroups for reporting and are not provided to the model as input features.
}

\blue{
\textbf{Site-held-out evaluation.}
We also perform a site-held-out evaluation using the site-disjoint partition described in \Cref{ssec:datacollection} to assess robustness to site-specific acquisition effects and to reduce potential leakage from overlap in imaging centers. In this split, all MRI--PET pairs acquired at the same ADNI site are assigned exclusively to a single subset, ensuring that no site appears in more than one of the training, validation, or test sets. Under this site-held-out protocol, we repeat the main perceptual-loss comparison, contrasting conventional 2D, 3D, and 2.5D perceptual losses with the proposed cyclic 2.5D perceptual loss. We keep all preprocessing steps, loss weights, optimization settings, and evaluation procedures identical to those used in the primary quantitative evaluation to enable a direct comparison.
}

\blue{
\textbf{Independent-cohort replication in SCAN}. 
To assess generalizability beyond ADNI, we perform an additional evaluation on the independent SCAN cohort. This experiment serves as an independent-cohort replication to determine whether the relative advantage of the cyclic 2.5D perceptual loss persists outside ADNI. We train the 3D U-Net on the SCAN training subset and select model checkpoints using the SCAN validation subset, while keeping the architecture, optimization scheme, and hyperparameters fixed to those described in Training Details. The SCAN partition is site-disjoint by design. The by-manufacturer PET SUVR standardization parameters are estimated exclusively from the SCAN training subset and are applied without modification to the validation and test subsets. We reproduce the perceptual-loss comparison by training model variants that differ only in the perceptual term (2D, 3D, 2.5D, and cyclic 2.5D), while keeping all other loss components and weights constant.
}

\textbf{Statistical Analysis}.
Statistical analyses are conducted to evaluate the performance of the proposed cyclic 2.5D perceptual loss relative to the conventional 2D, 3D, and 2.5D perceptual losses. A paired design at the subject level is employed, ensuring that identical sample sets are used for three predefined pairwise contrasts: cyclic 2.5D versus (vs.) 2D, vs. 3D, and vs. 2.5D. Performance evaluation is based on subject-level metrics, including the global 3D averages, the plane-wise 2D averages over slices 20 to 90, and the ROI-level prediction accuracy quantified by the mean squared error ($\text{MSE}_{\text{ROI}}$). For each subject, paired differences between the cyclic 2.5D and the corresponding comparator are calculated. The distribution of these differences is assessed for normality using the Shapiro--Wilk test. When the assumption of normality is met, a paired $t$-test is applied; otherwise, inference relies on the non-parametric Wilcoxon signed-rank test to ensure robustness against deviations from normality. One-sided hypotheses are specified to reflect the directional objective of demonstrating superiority of the cyclic 2.5D, with the alternative hypothesis favoring higher SSIM and PSNR values and lower MAE and $\text{MSE}_{\text{ROI}}$ values for the proposed approach. To address multiplicity across metrics, anatomical planes, and regions, the false discovery rate is controlled via the Benjamini--Hochberg (BH) procedure, applied separately within each family of hypotheses defined by the pairwise contrast and the performance endpoint.

\begin{table*}[!t]
\centering
\caption{Quantitative comparisons of state-of-the-art MRI-to-PET translation frameworks and perceptual loss functions for tau PET synthesis. Evaluations are performed on test datasets preprocessed using distinct normalization methods. The dagger symbol ($\dagger$) denotes models trained with the proposed by-manufacturer standardization of PET SUVR. In contrast, models without this symbol are trained using min--max normalization. The label “None” indicates that training is performed solely with the baseline loss function of the corresponding framework, without incorporating perceptual loss. \blue{The values are reported as mean ± standard deviation across subjects. ``Overall (3D)'' refers to the 3D SSIM or PSNR computed on the entire volume; ``Axial/Coronal/Sagittal'' corresponds to the slice-wise 2D SSIM or PSNR averaged over slices 20–90 in each plane}; $^{*}p < 0.05$; $^{**}p < 0.01.$ The best-performing value in each column is shown in \textbf{bold}.\label{tab:benchmark}} 
\resizebox{\textwidth}{!}{%
\begin{tabular}{@{}cccccccccc@{}}
\toprule
\multirow{2}{*}{\textBF{Model}} & \multirow{2}{*}{\textBF{Perceptual Loss}} & \multicolumn{4}{c}{\textBF{SSIM(\%)↑}} & \multicolumn{4}{c}{\textBF{PSNR(dB)↑}} \\ 
\cmidrule(lr){3-6} \cmidrule(lr){7-10}
                                      &                                   & \multicolumn{1}{c}{\textBF{Overall (3D)}} & \multicolumn{1}{c}{\textBF{Axial}} & \multicolumn{1}{c}{\textBF{Coronal}} & \multicolumn{1}{c}{\textBF{Sagittal}} & \multicolumn{1}{c}{\textBF{Overall (3D)}} & \multicolumn{1}{c}{\textBF{Axial}} & \multicolumn{1}{c}{\textBF{Coronal}} & \multicolumn{1}{c}{\textBF{Sagittal}} \\ \midrule
\begin{tabular}[c]{@{}c@{}}PASTA\\{\small w/ MRI only}\end{tabular} & None & 85.32±4.49 & 80.79±5.28 & 80.26±4.81 & 79.10±5.12 & 24.98±2.88 & 25.98±3.52 & 23.55±2.87 & 24.22±3.07 \\ \midrule
\begin{tabular}[c]{@{}c@{}}ShareGAN\\{\small w/ AD vs. MCI}\end{tabular} & None & 85.49±4.59 & 79.98±6.06 & 79.41±5.85 & 78.02±6.53 & 25.01±2.57 & 26.19±4.32 & 23.57±2.48 & 24.24±2.86 \\ \midrule

\multirow{4}{*}{U-Net}
  & 2D
    & \sig{87.67±5.50}{**} & \sig{82.63±6.69}{*} & 82.66±7.04 & 81.43±7.58
    & 26.46±2.43 & 26.16±2.34 & \textBF{25.41±2.57} & 25.41±2.59 \\

  & 3D
    & \sig{87.18±5.47}{**} & \sig{82.23±6.71}{**} & \sig{82.12±7.03}{**} & \sig{80.89±7.62}{**}
    & \sig{25.51±1.99}{**} & \sig{25.01±1.80}{**} & \sig{24.48±2.12}{**} & \sig{24.44±2.16}{**} \\

  & 2.5D
    & \sig{85.33±5.35}{**} & \sig{80.40±6.48}{**} & \sig{80.18±6.80}{**} & \sig{79.03±7.37}{**}
    & \sig{23.09±1.47}{**} & \sig{22.74±1.44}{**} & \sig{22.55±1.66}{**} & \sig{22.52±1.70}{**} \\

  & Cyclic 2.5D (Ours)
    & \textBF{88.03±5.89} & \textBF{82.78±7.59} & \textBF{82.75±7.59} & \textBF{81.49±8.26}
    & \textBF{26.71±2.81} & \textBF{26.56±2.68} & 25.32±2.83 & \textBF{25.48±2.87} \\ \midrule

\multirow{4}{*}{U\textendash Net$^\dagger$}
  & 2D
    & \sig{89.95±3.53}{**} & \sig{85.95±4.35}{**} & \sig{86.05±4.22}{**} & \sig{84.80±4.48}{**}
    & \sig{28.19±2.81}{**} & \sig{29.10±3.98}{**} & \sig{26.95±2.69}{**} & \sig{27.33±2.92}{**} \\

  & 3D
    & \sig{89.77±3.80}{**} & \sig{85.76±4.60}{**} & \sig{85.77±4.54}{**} & \sig{84.57±4.80}{**}
    & \sig{28.23±2.73}{**} & \sig{29.02±3.82}{**} & \sig{26.99±2.61}{**} & \sig{27.40±2.84}{**} \\

  & 2.5D
    & \sig{90.02±3.57}{**} & \sig{86.06±4.41}{**} & \sig{86.11±4.27}{**} & \sig{84.93±4.54}{**}
    & \sig{28.23±2.77}{**} & \sig{29.18±4.01}{**} & \sig{26.98±2.66}{**} & \sig{27.41±2.91}{**} \\

  & Cyclic 2.5D (Ours)
    & \textBF{90.47±3.62} & \textBF{86.70±4.38} & \textBF{86.74±4.27} & \textBF{85.57±4.57}
    & \textBF{28.95±2.73} & \textBF{29.77±3.76} & \textBF{27.70±2.56} & \textBF{28.10±2.86} \\ \midrule
 
\multirow{4}{*}{UNETR$^\dagger$}
 & 2D & 89.84±3.54 & 85.94±4.32 & \sig{85.97±4.19}{*} & \sig{84.60±4.48}{*} & 28.32±2.45 & 29.22±3.53 & 27.10±2.30 & 27.50±2.55 \\
 & 3D & \sig{89.68±3.76}{*} & 85.80±4.46 & \sig{85.80±4.39}{*} & \sig{84.54±4.73}{**} & 28.25±2.41 & \sig{28.82±3.26}{**} & \textBF{27.16±2.22} & 27.40±2.48 \\ 
 & 2.5D & \sig{89.68±3.66}{**} & \sig{85.77±4.30}{**} & \sig{85.76±4.22}{**} & \sig{84.58±4.50}{**} & \textBF{28.34±2.29} & 29.16±3.38 & 27.11±2.03 & \textBF{27.57±2.44} \\ 
 & Cyclic 2.5D (Ours) & \textBF{89.93±3.29} & \textBF{86.03±3.94} & \textBF{86.07±3.92} & \textBF{84.84±4.21} & 28.34±2.52 & \textBF{29.32±3.67} & 27.13±2.30 & 27.57±2.65 \\ \midrule

\multirow{4}{*}{SwinUNETR$^\dagger$}
 & 2D & \sig{90.16±3.55}{*} & \sig{86.35±4.33}{**} & \sig{86.46±4.13}{*} & \sig{85.14±4.43}{**} & 28.53±2.55 & \textBF{29.49±3.68} & 27.30±2.36 & 27.79±2.72 \\
 & 3D & \sig{90.20±3.68}{*} & \sig{86.42±4.44}{*} & \sig{86.45±4.29}{**} & \sig{85.26±4.64}{**} & \sig{28.28±2.64}{*} & \sig{29.08±3.72}{*} & \sig{27.12±2.38}{**} & \sig{27.46±2.76}{*} \\ 
 & 2.5D & 90.23±3.55 & \sig{86.42±4.35}{*} & \sig{86.51±4.19}{*} & \sig{85.30±4.53}{*} & \textBF{28.60±2.58} & 29.47±3.61 & 27.37±2.41 & \textBF{27.86±2.72} \\ 
 & Cyclic 2.5D (Ours) & \textBF{90.29±3.40} & \textBF{86.57±4.11} & \textBF{86.69±4.01} & \textBF{85.36±4.38} & 28.52±2.59 & 29.47±3.59 & \textBF{27.41±2.35} & 27.72±2.71 \\ \midrule

\multirow{5}{*}{CycleGAN$^\dagger$} & None
  &  \sig{84.69±4.81}{**} & \sig{79.19±5.58}{**} & \sig{78.49±5.40}{**} & \sig{76.94±5.75}{**}
  & \sig{26.12±2.39}{**} & \sig{26.70±3.16}{**} & \sig{24.75±2.22}{**} & \sig{25.25±2.51}{**} \\
 & 2D
  & \sig{86.93±4.42}{**} & \sig{81.82±5.19}{**} & \sig{81.68±5.05}{**} & \sig{80.03±5.38}{**}
  & 27.15±2.32          & \textBF{27.83±3.33}          & 25.85±2.11          & 26.20±2.44          \\
 & 3D
  & \sig{85.63±5.03}{**} & \sig{80.58±5.67}{**} & \sig{79.84±5.65}{**} & \sig{78.39±5.98}{**}
  & \sig{26.34±2.38}{**} & \sig{26.87±3.17}{**} & \sig{24.99±2.20}{**}  & \sig{25.45±2.55}{**} \\
 & 2.5D
  & \sig{87.15±4.32}{**} & \sig{82.28±5.01}{**}  & \sig{82.16±4.89}{**}  & \sig{80.75±5.14}{**}
  & \textBF{27.18±2.45}          & 27.71±3.26          & \textBF{25.91±2.24}          & \textBF{26.27±2.51}          \\
 & Cyclic 2.5D (Ours)
  & \textBF{87.69±3.97}           & \textBF{82.82±4.78}           & \textBF{82.78±4.62}           & \textBF{81.27±4.95}
  & 27.16±2.32           & 27.62±3.21           & 25.90±2.19           & 26.24±2.43 \\ \midrule
\multirow{5}{*}{Pix2Pix$^\dagger$} & None
  & \sig{82.69±4.39}{**} & \sig{76.73±5.05}{**} & \sig{76.35±4.83}{**} & \sig{74.56±5.20}{**}
  & \sig{26.48±2.01}{**} & \sig{26.56±2.48}{**} & \sig{25.14±1.82}{**} & \sig{25.40±2.06}{**} \\
 & 2D
  & \sig{85.70±4.28}{**} & \sig{80.30±4.96}{**} & \sig{80.19±4.76}{**} & \sig{78.48±5.26}{**}
  & \sig{26.70±2.09}{**} & \sig{27.02±2.86}{**} & \sig{25.38±1.95}{**} & \sig{25.50±2.12}{**} \\
 & 3D
  & \sig{84.10±4.72}{**} & \sig{78.02±5.30}{**} & \sig{77.82±5.11}{**} & \sig{75.90±5.49}{**}
  & \sig{26.47±2.14}{**} & \sig{26.48±2.62}{**} & \sig{25.12±1.90}{**} & \sig{25.35±2.20}{**} \\
 & 2.5D
  & \sig{86.66±4.08}{**} & \sig{81.41±4.74}{**} & \sig{81.45±4.54}{**} & \sig{79.72±4.85}{**}
  & \sig{27.06±2.12}{**} & \sig{27.41±2.90}{**} & \sig{25.74±1.95}{**} & \sig{26.02±2.23}{**} \\
 & Cyclic 2.5D (Ours)
  & \textBF{87.63±3.91}           & \textBF{82.65±4.63}           & \textBF{82.67±4.45}           & \textBF{81.08±4.74}
  & \textBF{27.55±2.21}           & \textBF{28.18±3.24}           & \textBF{26.25±1.99}           & \textBF{26.66±2.34} \\ \bottomrule
\end{tabular}
}
\end{table*}

\blue{
\textbf{Assessment of by‐manufacturer SUVR standardization}.
Because the scanner manufacturer is partially confounded with acquisition site and participant composition, we explicitly evaluate whether by-manufacturer standardization attenuates manufacturer-dependent distributional differences within each diagnostic category while preserving disease-related SUVR differences between diagnostic groups within each manufacturer. We conduct a paired pre-standardization and post-standardization analysis using the ADNI cohort. Manufacturer-specific parameters ($\mu_m$ and $\sigma_m$) are estimated using only the training subset and are then applied without modification to all subjects to produce post-standardized PET volumes. Cortical voxels are defined using the Desikan--Killiany atlas.

We pool cortical voxel SUVR values by manufacturer and diagnostic group (AD and LMCI/MCI/EMCI) and generate violin plots in both the pre-standardization space (SUVR) and the post-standardization space (manufacturer-wise $z$-scored SUVR) for visualization. We annotate the 75th (p75), 90th (p90), and 99th (p99) percentiles to emphasize distributional differences, particularly in the upper tail. For statistical inference, we summarize each subject's cortical SUVR distribution using three subject-level metrics: the median, p75, and p90. We apply a non-parametric Kruskal--Wallis H (KW) test across the three manufacturers (Siemens, GE, and Philips) to assess manufacturer effects within each diagnostic group. When the omnibus test indicates significance, we conduct post hoc pairwise Mann--Whitney U (MWU) tests and control the false discovery rate using the BH procedure. 
}

\begin{figure*}[t]
  \centering
  \includegraphics[width=\linewidth]{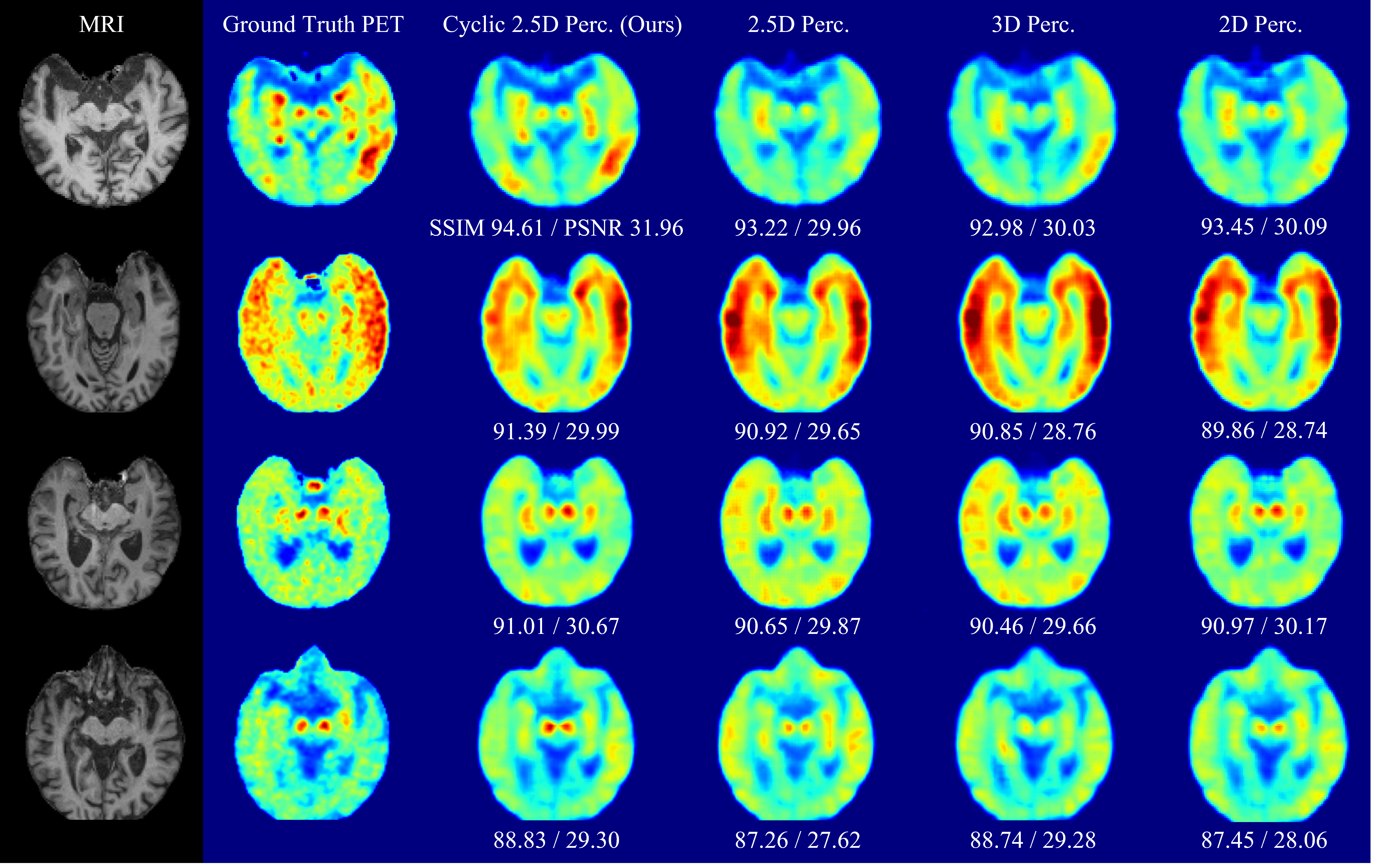}
  \caption{Qualitative comparison of images generated by the 3D U-Net with different perceptual loss functions. The first column presents the input T1w MR images, the second illustrates the ground truth tau PET images, and the subsequent columns show the generated tau PET images. The SSIM and PSNR values relative to the ground truth tau PET image and the generated tau PET image are reported from the third column to the final column. {\it Note:} The SSIM or PSNR values represent performance across the 3D volume.}
  \label{fig:benchmark}
\end{figure*}

\begin{figure*}[t]
  \centering
  \includegraphics[width=\linewidth]{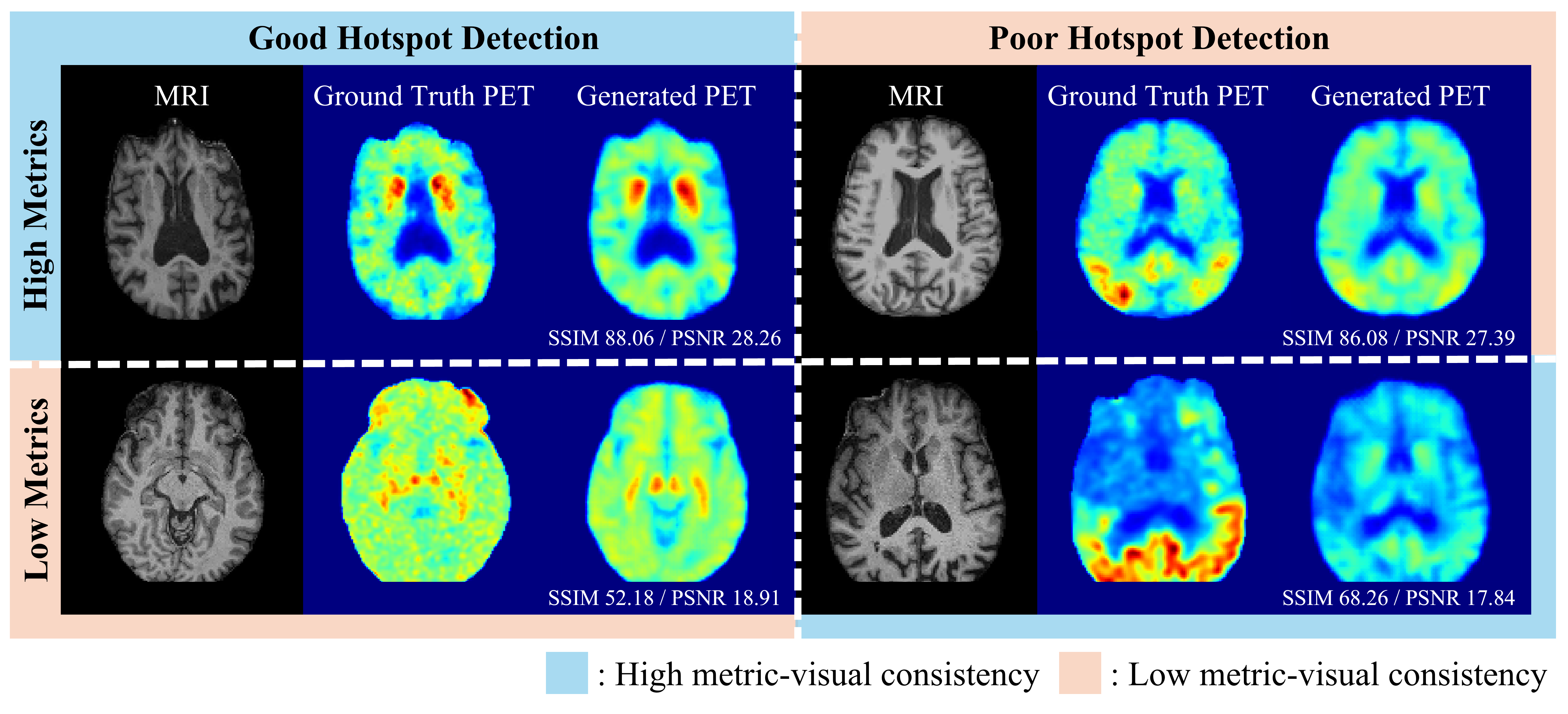}
  \caption{Visualization of multiple pairs comprising an input MR image, the corresponding ground truth tau PET image, and the generated tau PET image. The top-left and bottom-right pairs demonstrate a consistent alignment between the quantitative metrics and the detection of visual hot spots, defined as regions with high SUVRs in the ground truth tau PET images. However, the top-right and bottom-left pairs indicate cases where the metrics do not adequately reflect the visual hot spot detection. {\it Note:} The SSIM or PSNR values are calculated for individual 2D axial slices.}
  \label{fig:metric_consistency}
\end{figure*}

\section{Result}
\label{sec:results}

\subsection{Benchmark Evaluation}
\label{ssec:results}
\textbf{Quantitative comparison}. \Cref{tab:benchmark} presents the quantitative results, demonstrating that the proposed cyclic 2.5D perceptual loss outperforms other methods across most evaluation metrics. The proposed cyclic 2.5D perceptual loss achieves the highest mean SSIM and PSNR in most configurations and remains competitive even when it is not the top performer. Importantly, the SSIM improvements are consistent across all backbones and are evident in both whole‑volume (3D) and plane‑wise (2D) evaluations. The combined loss function incorporating the cyclic 2.5D perceptual loss exhibits superior performance under equivalent conditions compared with the two state-of-the-art open-source frameworks for synthesizing PET from MRI: the PASTA and the ShareGAN with joint learning. The preprocessing method that standardizes PET SUVRs according to the manufacturer ($\dagger$) outperforms the min-max normalization approach, which merely rescales SUVRs to a predetermined range, as evidenced by the results of the U-Net trained with the by-manufacturer standardization (U-Net$^\dagger$) compared to those trained with the min--max normalization (U-Net). Furthermore, the cyclic 2.5D perceptual loss achieves better results across various generative models compared to the 2D, 3D, and 2.5D perceptual losses. Pairwise testing (asterisks in the table) shows that, for the U-Net$^\dagger$, the cyclic 2.5D perceptual loss significantly surpasses the 2D, 3D, and 2.5D perceptual losses across all metrics ($p < 0.05$ throughout). For the transformer backbones (UNETR$^\dagger$ and SwinUNETR$^\dagger$), the cyclic 2.5D method yields clear SSIM gains that are mostly significant, while PSNR differences are small and not consistently significant. Among the GAN-based models, the Pix2Pix$^\dagger$ benefits in both SSIM and PSNR, with all comparisons reaching significance. For the CycleGAN$^\dagger$, the cyclic 2.5D approach achieves the best SSIM, and its PSNR is comparable to the strongest baselines (2D and 2.5D) rather than uniformly higher; it remains significantly better than training with their own loss term (None) or with the 3D perceptual loss. \blue{Comprehensive experimental results, including MAE, and additional distributional summaries of SSIM and PSNR for the 3D U-Net, are reported in Suppl. Section 2.}

\begin{figure*}[tb]
  \centering
  \includegraphics[width=1.0\linewidth]{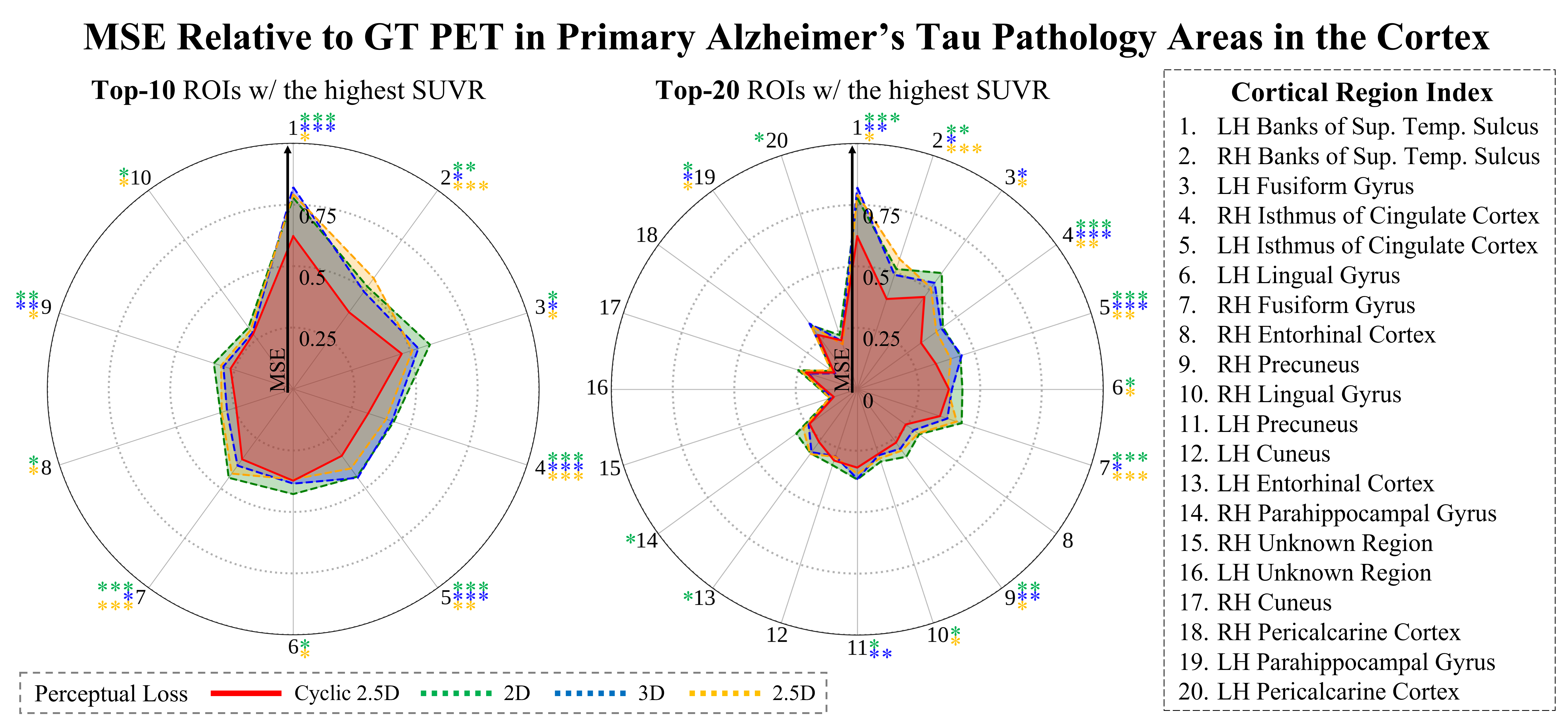}
   \caption{Prediction performance (MSE) of tau PET synthesis across key cortical regions associated with Alzheimer's disease pathology. For each cortical region (ROI), the MSE is computed between the mean SUVR of the actual tau PET and that of the generated tau PET. The ROIs are ranked according to the group-mean ground truth SUVR. The top 10 ROIs with the highest mean SUVR are shown on the left, and the top 20 on the right. The proposed cyclic 2.5D perceptual loss achieves the lowest error among all methods in 9 of the top 10 ROIs and 15 of the top 20 ROIs. Specifically, it produces lower errors than the 2D in all 10 (top 10) and all 20 (top 20) ROIs, lower than the 3D in 10 and 17 ROIs, and lower than the 2.5D in 9 and 16 ROIs. At a significance level of $p<0.05$, the cyclic 2.5D surpasses the 2D in all 10 (top 10) and 13 (top 20) ROIs, exceeds the 3D in 7 and 9 ROIs, and outperforms the 2.5D in all 10 and 10 ROIs. {\it Note:} $^{*}p < 0.05$; $^{**}p < 0.01$; $^{***}p < 0.001.$}
   \label{fig:roi_analysis}
\end{figure*}

\textbf{Qualitative comparison}. The proposed method also produces qualitatively superior results, accurately capturing pathological tau accumulation and NFT distributions. \Cref{fig:benchmark} illustrates the axial slices of PET images generated by the models trained with various configurations of the combined loss, including the cyclic 2.5D, 2D, 3D, and 2.5D perceptual losses. Among these configurations, the model trained with the cyclic 2.5D perceptual loss demonstrates the highest precision in identifying tau tangle hot spots while minimizing the misidentification of neuropathologically spared regions. However, some inconsistencies are observed in certain samples between the SSIM and PSNR values and the visual quality of the generated PET images. These discrepancies are visualized in \Cref{fig:metric_consistency}, which presents axial‑slice comparisons.

\subsection{ROI Analysis}
The ROI analysis demonstrates that the model trained with the proposed cyclic 2.5D perceptual loss generates values that more closely match the actual PET SUVRs in Alzheimer’s tau pathology regions compared with the conventional 2D, 3D, and 2.5D approaches. The primary cortical regions with the highest tau tangle accumulation are identified, and performance is compared in the top 20 (and 10) regions (\Cref{fig:roi_analysis}). The areas are ordered based on the mean PET SUVR across the study participants. Comprehensive ROI analysis results are provided in Suppl. Section 3. Across the top 20 ROIs, the mean squared error in ROI space (MSE$_{\text{ROI}}$) between the generated and ground truth PET images is lower in 15 regions when using the cyclic 2.5D perceptual loss, relative to the 2D, 3D, and 2.5D methods. In detail, the cyclic 2.5D approach achieves lower error than the 2D in 10 (top-10) and 20 (top-20) ROIs, lower error than the 3D in 10 and 17, and lower error than the 2.5D in 9 and 16 ROIs. With statistical significance at $p < 0.05$, the cyclic 2.5D method outperforms 2D in all 10 of the top-10 and 13 of the top-20 ROIs, outperforms 3D in 7 and 9 ROIs, and outperforms 2.5D in all 10 and 10 ROIs, respectively. In specific high-pathology regions, the cyclic 2.5D configuration achieves substantial MSE reductions. In the banks of the superior temporal sulcus, error is reduced by 20.09\% (left hemisphere, LH) and 24.89\% (right hemisphere, RH) relative to the 2D, by 24.32\% (LH) and 21.10\% (RH) relative to the 3D, and by 21.57\% (LH) and 30.66\% (RH) relative to the 2.5D. Reductions in the entorhinal cortex are 17.84\% (LH) and 21.67\% (RH) vs. 2D, 16.36\% (LH) and 13.99\% (RH) vs. 3D, and 18.78\% (LH) and 20.29\% (RH) vs. 2.5D. In the fusiform gyrus, MSE decreased by 20.61\% (LH) and 20.79\% (RH) compared with 2D, by 13.27\% (LH) and 8.21\% (RH) compared with 3D, and by 8.92\% (LH) and 16.87\% (RH) compared with 2.5D. Other primary tau pathology areas showing MSE reduction are the parahippocampal gyrus, precuneus, and isthmus cingulate.

\blue{
\subsection{Diagnosis- and A$\beta$/Tau-Stratified Result}
\textbf{Diagnosis-stratified performance.}}
The proposed cyclic 2.5D perceptual loss significantly outperforms alternative approaches across most diagnostic groups, as summarized in \Cref{tab:diagnosis_stratified}. In particular, the cyclic 2.5D method consistently outperforms the conventional 2D, 3D, and 2.5D perceptual losses across all diagnostic categories, including AD, LMCI, MCI, and EMCI. These improvements are observed in both 3D and 2D analyses. Statistical analyses further reveal that, for the AD, MCI, and EMCI cohorts, the performance gains of the cyclic 2.5D loss over other methods are significant ($p < 0.05$). For the LMCI group, the same performance trend is observed, with statistical significance achieved relative to the 2.5D method for all metrics and to the 2D method for a subset of metrics, but not relative to the 3D method. 

\blue{
When the perceptual loss is fixed to the cyclic 2.5D formulation, synthesis fidelity varies across diagnostic groups. Under the cyclic 2.5D, LMCI exhibits the highest whole-volume SSIM and PSNR, whereas EMCI shows the lowest whole-volume SSIM, and AD shows the lowest whole-volume PSNR. Plane-wise SSIM follows the same ordering (LMCI highest; EMCI lowest) in the axial, coronal, and sagittal planes. Plane-wise PSNR is consistently lowest in AD across all three planes.

\begin{table*}[!t]
\centering
\caption{Quantitative performance stratified by diagnostic group. The results are obtained from a 3D U-Net model trained on datasets preprocessed with by-manufacturer PET SUVR standardization. \blue{The values are reported as mean ± standard deviation across subjects. ``Overall (3D)'' refers to the 3D SSIM or PSNR computed on the entire volume; ``Axial/Coronal/Sagittal'' corresponds to the slice-wise 2D SSIM or PSNR averaged over slices 20–90 in each plane}; $^{*}p < 0.05$; $^{**}p < 0.01.$ The best-performing value in each column is shown in \textbf{bold}.\label{tab:diagnosis_stratified}} 
\resizebox{\textwidth}{!}{%
\begin{tabular}{@{}cccccccccc@{}}
\toprule
\multirow{2}{*}{\textBF{Diagnosis}} & \multirow{2}{*}{\textBF{Perceptual Loss}} & \multicolumn{4}{c}{\textBF{SSIM(\%)↑}} & \multicolumn{4}{c}{\textBF{PSNR(dB)↑}} \\ 
\cmidrule(lr){3-6} \cmidrule(lr){7-10}
  &  & \textBF{Overall (3D)} & \textBF{Axial} & \textBF{Coronal} & \textBF{Sagittal} & \textBF{Overall (3D)} & \textBF{Axial} & \textBF{Coronal} & \textBF{Sagittal} \\ \midrule
\multirow{4}{*}{EMCI}
  & 2D                    & \sig{89.69±2.78}{**} & \sig{85.38±4.27}{**} & \sig{85.44±3.52}{**} & \sig{84.18±3.50}{**} & \sig{27.77±2.96}{*}  & \sig{28.21±4.18}{*}  & \sig{26.27±3.04}{*}  & \sig{26.68±2.96}{*}  \\
  & 3D                    & \sig{89.37±3.35}{**} & \sig{85.08±4.79}{**} & \sig{84.99±4.18}{**} & \sig{83.77±4.21}{**} & \sig{27.80±2.87}{**} & \sig{28.12±4.10}{**} & \sig{26.36±2.87}{**} & \sig{26.71±2.89}{**} \\
  & 2.5D                  & \sig{89.87±2.97}{**} & \sig{85.61±4.43}{**} & \sig{85.64±3.70}{**} & \sig{84.41±3.74}{**} & \sig{27.90±2.48}{**} & \sig{28.37±3.94}{**} & \sig{26.39±2.46}{**} & \sig{26.82±2.51}{**} \\
  & Cyclic 2.5D (Ours)    & \textBF{90.32±2.85}  & \textBF{86.30±4.20}  & \textBF{86.32±3.48}  & \textBF{85.08±3.55}  & \textBF{28.82±2.04}  & \textBF{29.08±3.40}  & \textBF{27.38±2.07}  & \textBF{27.67±2.08}  \\ \midrule
\multirow{4}{*}{MCI}
  & 2D                    & \sig{90.01±3.81}{**} & \sig{86.22±4.40}{**} & \sig{86.15±4.64}{**} & \sig{84.87±4.73}{**} & \sig{28.73±2.53}{**} & \sig{30.11±3.67}{**} & \sig{27.46±2.50}{*}  & \sig{27.89±2.64}{**} \\
  & 3D                    & \sig{89.76±3.98}{**} & \sig{85.94±4.54}{**} & \sig{85.79±4.79}{**} & \sig{84.55±4.89}{**} & \sig{28.67±2.45}{**} & \sig{29.89±3.45}{**} & \sig{27.36±2.44}{**} & \sig{27.86±2.58}{**} \\
  & 2.5D                  & \sig{89.98±3.88}{**} & \sig{86.23±4.52}{**} & \sig{86.10±4.70}{**} & \sig{84.89±4.78}{**} & \sig{28.75±2.73}{*}  & \sig{30.20±3.81}{*}  & \sig{27.49±2.75}{*}  & \sig{27.96±2.88}{**} \\
  & Cyclic 2.5D (Ours)    & \textBF{90.37±3.91}  & \textBF{86.74±4.38}  & \textBF{86.78±4.27}  & \textBF{85.57±4.57}  & \textBF{29.19±2.63}  & \textBF{30.56±3.66}  & \textBF{27.86±2.63}  & \textBF{28.42±2.73}  \\ \midrule
\multirow{4}{*}{LMCI}
  & 2D                    & \sig{91.06±3.86}{*}  & 86.65±5.84            & \sig{87.63±4.73}{*}  & 86.12±6.16            & 28.69±4.09            & 28.71±6.21            & \sig{27.60±3.84}{*}  & \sig{27.90±4.13}{*}  \\
  & 3D                    & 91.28±3.96           & 86.91±5.78            & 87.86±4.90           & 86.31±6.31            & 28.89±3.35            & 28.92±5.66            & 28.04±3.07           & 28.17±3.53           \\
  & 2.5D                  & \sig{91.09±3.59}{*}  & \sig{86.69±5.50}{*}   & \sig{87.60±4.40}{*}  & \sig{86.21±5.88}{*}   & \sig{28.25±4.22}{*}   & \sig{28.34±6.26}{*}   & \sig{27.15±3.79}{*}  & \sig{27.62±4.12}{*}  \\
  & Cyclic 2.5D (Ours)    & \textBF{91.74±3.87}  & \textBF{87.62±5.88}   & \textBF{88.59±4.68}  & \textBF{87.26±6.06}   & \textBF{29.87±4.20}   & \textBF{29.72±6.06}   & \textBF{28.92±3.77}  & \textBF{29.30±4.30}  \\ \midrule
\multirow{4}{*}{AD}
  & 2D                    & \sig{89.71±3.65}{**} & \sig{85.63±4.12}{**} & \sig{85.97±3.82}{**} & \sig{84.89±4.54}{**} & \sig{27.01±2.76}{**} & \sig{27.48±3.28}{**} & \sig{26.10±2.24}{**} & \sig{26.33±3.08}{**} \\
  & 3D                    & \sig{89.73±3.95}{*}  & \sig{85.64±4.52}{*}  & \sig{85.90±4.34}{*}  & \sig{84.96±4.96}{*}  & \sig{27.30±3.07}{*}  & \sig{27.69±3.58}{*}  & \sig{26.35±2.58}{*}  & \sig{26.65±3.21}{*}  \\
  & 2.5D                  & \sig{89.95±3.61}{*}  & \sig{85.91±4.15}{*}  & \sig{86.17±3.89}{*}  & \sig{85.17±4.61}{*}  & \sig{27.17±2.59}{*}  & \sig{27.58±3.33}{*}  & \sig{26.22±2.11}{*}  & \sig{26.51±2.92}{*}  \\
  & Cyclic 2.5D (Ours)    & \textBF{90.49±3.76}  & \textBF{86.73±4.22}  & \textBF{87.01±3.96}  & \textBF{85.98±4.71}  & \textBF{28.14±3.23}  & \textBF{28.38±3.34}  & \textBF{27.21±2.54}  & \textBF{27.33±3.44}  \\ \bottomrule
\end{tabular}
}
\end{table*}
\begin{table*}[!t]
\begingroup\blue 
\centering
\caption{\blue{Quantitative performance stratified by A$\beta$/Tau biomarker status.} The results are obtained from a 3D U-Net model trained on datasets preprocessed with by-manufacturer PET SUVR standardization. The values are reported as mean ± standard deviation across subjects. ``Overall (3D)'' refers to the 3D SSIM or PSNR computed on the entire volume; ``Axial/Coronal/Sagittal'' corresponds to the slice-wise 2D SSIM or PSNR averaged over slices 20–90 in each plane; $^{*}p < 0.05$; $^{**}p < 0.01.$ The best-performing value in each column is shown in \textbf{bold}.\label{tab:abtau_stratified}} 
\resizebox{\textwidth}{!}{%
\begin{tabular}{@{}cccccccccc@{}}
\toprule
\multirow{2}{*}{\makecell{\textBF{A$\beta$/Tau} \\ \textBF{group}}} & \multirow{2}{*}{\textBF{Perceptual Loss}} & \multicolumn{4}{c}{\textBF{SSIM(\%)↑}} & \multicolumn{4}{c}{\textBF{PSNR(dB)↑}} \\ 
\cmidrule(lr){3-6} \cmidrule(lr){7-10}
  &  & \textBF{Overall (3D)} & \textBF{Axial} & \textBF{Coronal} & \textBF{Sagittal} & \textBF{Overall (3D)} & \textBF{Axial} & \textBF{Coronal} & \textBF{Sagittal} \\ \midrule

\multirow{4}{*}{A$\beta$--T--}
  & 2D                 & \sig{89.03±3.89}{*}   & \sig{84.68±5.07}{**}  & \sig{84.90±4.55}{*}   & \sig{83.55±4.70}{**}  & \sig{27.67±2.90}{**}  & \sig{28.59±4.34}{**}  & \sig{26.27±2.89}{*}   & \sig{26.54±2.85}{**}   \\
  & 3D                 & \sig{88.98±4.28}{**}  & \sig{84.62±5.43}{**}  & \sig{84.76±5.02}{**}  & \sig{83.48±5.22}{**}  & \sig{27.89±2.71}{**}  & \sig{28.67±4.01}{**}  & \sig{26.51±2.66}{*}   & \sig{26.78±2.68}{*}    \\
  & 2.5D               & \sig{89.20±4.10}{*}   & \sig{84.94±5.30}{**}  & \sig{85.09±4.80}{*}   & \sig{83.85±4.97}{**}  & \sig{27.84±2.74}{**}  & \sig{28.86±4.28}{**}  & \sig{26.45±2.70}{**}  & \sig{26.77±2.73}{**}   \\
  & Cyclic 2.5D (Ours) & \textBF{89.50±4.07}   & \textBF{85.41±5.15}   & \textBF{85.55±4.64}   & \textBF{84.30±4.85}   & \textBF{28.64±2.33}   & \textBF{29.43±3.88}   & \textBF{27.27±2.23}   & \textBF{27.51±2.29}    \\ \midrule
\multirow{4}{*}{A$\beta$--T+}
  & 2D                 & \sig{90.36±3.77}{*}   & \sig{86.15±4.43}{*}   & \sig{86.63±4.62}{*}   & \sig{85.59±4.38}{*}   & 29.13±3.34            & 28.41±3.46            & 27.84±3.25            & \sig{28.50±3.57}{*}    \\
  & 3D                 & 90.41±4.06            & 86.20±4.60            & 86.62±4.82            & 85.62±4.64            & 29.44±3.06            & 28.56±3.21            & 28.11±3.07            & 28.79±3.34             \\
  & 2.5D               & \sig{90.63±3.81}{*}   & 86.58±4.41            & \sig{86.97±4.64}{*}   & \sig{85.96±4.45}{*}   & 29.54±3.19            & 28.81±3.29            & 28.22±3.26            & 28.99±3.53             \\
  & Cyclic 2.5D (Ours) & \textBF{90.86±3.64}   & \textBF{86.89±4.25}   & \textBF{87.36±4.30}   & \textBF{86.27±4.31}   & \textBF{29.70±2.89}   & \textBF{28.91±3.26}   & \textBF{28.41±2.89}   & \textBF{29.12±3.23}    \\ \midrule 
\multirow{4}{*}{A$\beta$+T--}
  & 2D                 & \sig{90.43±2.74}{*}   & \sig{86.49±3.93}{*}   & \sig{86.88±3.58}{*}   & \sig{85.42±3.78}{**}  & \sig{28.90±3.16}{*}   & \sig{29.55±4.61}{*}   & \sig{27.55±3.20}{*}   & \sig{27.89±3.29}{*}    \\
  & 3D                 & \sig{90.11±3.15}{**}  & \sig{86.06±4.42}{**}  & \sig{86.44±4.11}{*}   & \sig{85.00±4.24}{**}  & \sig{28.46±3.35}{*}   & \sig{29.12±4.78}{**}  & \sig{27.23±3.32}{*}   & \sig{27.56±3.30}{**}   \\
  & 2.5D               & \sig{90.41±2.78}{*}   & \sig{86.44±4.02}{*}   & \sig{86.80±3.60}{*}   & \sig{85.46±3.82}{**}  & \sig{28.59±2.98}{*}   & \sig{29.32±4.58}{*}   & \sig{27.24±2.95}{*}   & \sig{27.70±2.99}{*}    \\
  & Cyclic 2.5D (Ours) & \textBF{91.09±2.77}   & \textBF{87.44±3.77}   & \textBF{87.72±3.49}   & \textBF{86.40±3.83}   & \textBF{29.77±3.01}   & \textBF{30.38±4.28}   & \textBF{28.43±3.03}   & \textBF{28.90±3.44}    \\ \midrule
\multirow{4}{*}{A$\beta$+T+}
  & 2D                 & \sig{90.61±3.26}{**} & \sig{86.90±3.68}{**} & \sig{86.82±3.78}{**} & \sig{85.63±4.24}{**} & \sig{28.45±2.48}{**} & \sig{29.55±3.83}{**} & \sig{27.33±2.16}{**} & \sig{27.78±2.55}{**}  \\
  & 3D                 & \sig{90.34±3.50}{**} & \sig{86.64±3.91}{**} & \sig{86.44±4.09}{**} & \sig{85.31±4.51}{**} & \sig{28.45±2.46}{**} & \sig{29.44±3.75}{**} & \sig{27.30±2.20}{**} & \sig{27.80±2.56}{**}  \\
  & 2.5D               & \sig{90.61±3.22}{**} & \sig{86.88±3.72}{**} & \sig{86.77±3.75}{**} & \sig{85.63±4.22}{**} & \sig{28.40±2.61}{**}  & \sig{29.50±4.06}{**} & \sig{27.27±2.33}{**}  & \sig{27.75±2.73}{**}   \\
  & Cyclic 2.5D (Ours) & \textBF{91.17±3.31}   & \textBF{87.65±3.73}   & \textBF{87.53±3.85}   & \textBF{86.43±4.30}   & \textBF{29.16±2.75}   & \textBF{30.13±3.78}   & \textBF{27.98±2.41}   & \textBF{28.48±2.74}    \\ \bottomrule
\end{tabular}
}
\endgroup
\end{table*}

\textbf{A$\beta$/Tau-stratified performance.}
As summarized in \Cref{tab:abtau_stratified}, the proposed cyclic 2.5D perceptual loss yields the highest mean SSIM and PSNR within each A$\beta$/Tau stratum (A$\beta$--T--, A$\beta$--T+, A$\beta$+T--, and A$\beta$+T+) for both whole-volume and plane-wise evaluations. Pairwise tests show that the cyclic 2.5D significantly outperforms the conventional 2D, 3D, and 2.5D perceptual losses across the reported metrics in the A$\beta$--T--, A$\beta$+T--, and A$\beta$+T+ strata ($p<0.05$). In the A$\beta$--T+ stratum, cyclic 2.5D shows the same performance ordering; however, significance is reached versus the 2D and 2.5D losses only for a subset of metrics, while comparisons with the 3D loss do not reach significance.

Beyond between-loss comparisons, performance under cyclic 2.5D shows systematic variation across A$\beta$/Tau strata. Whole-volume SSIM is lowest in the A$\beta$--T-- group and highest in the A$\beta$+T+ group, with the discordant strata (A$\beta$--T+ and A$\beta$+T--) exhibiting intermediate values. Whole-volume PSNR is also lowest in A$\beta$--T--, whereas the highest PSNR is observed in A$\beta$+T--. When stratified by amyloid status, tau positivity is associated with higher whole-volume SSIM in both A$\beta$-- and A$\beta$+ participants. PSNR exhibits an amyloid-dependent pattern: tau positivity is associated with higher whole-volume PSNR in A$\beta$-- participants (A$\beta$--T+ $>$ A$\beta$--T--), whereas in A$\beta$+ participants, tau positivity is associated with lower PSNR (A$\beta$+T+ $<$ A$\beta$+T--).
}

\blue{
\subsection{Site-Held-Out Evaluation}
On the ADNI site-held-out split, the cyclic 2.5D perceptual loss consistently yields the highest SSIM and PSNR for the whole-volume and for slice-wise evaluations in all three planes (\Cref{tab:site_heldout}). Pairwise tests show that cyclic 2.5D yields significantly higher SSIM than each conventional perceptual loss (2D, 3D, and 2.5D) for the whole-volume metric and for all three planes. For PSNR, cyclic 2.5D differs significantly from the 3D perceptual loss in all comparisons, whereas significance relative to the 2D and baseline 2.5D losses is observed only in some plane-wise evaluations.
}

\blue{
\subsection{Independent-Cohort Replication in SCAN}
The cyclic 2.5D perceptual loss consistently achieves the highest mean SSIM and PSNR at both the whole-volume and plane-wise levels in the SCAN cohort, as summarized in \Cref{tab:scan_res}. Pairwise comparisons show a consistent SSIM advantage, with the cyclic 2.5D method outperforming the 2D, 3D, and 2.5D losses across all views; all differences are statistically significant ($p < 0.01$). For PSNR, the cyclic 2.5D approach also yields the highest mean values, although the strength of evidence varies across planes. Improvements are significant in the axial and sagittal planes relative to all conventional losses, and whole-volume PSNR shows a significant gain compared with the 2.5D perceptual loss. In contrast, PSNR differences in the coronal plane are not statistically significant.

}
\begin{table*}[!t]
\begingroup\blue 
\centering
\caption{\blue{Quantitative performance on the ADNI site-held-out split.} The results are obtained from a 3D U-Net model trained on data preprocessed with the by-manufacturer PET SUVR standardization. \blue{The values are reported as mean ± standard deviation across subjects. ``Overall (3D)'' refers to the 3D SSIM or PSNR computed on the entire volume; ``Axial/Coronal/Sagittal'' corresponds to the slice-wise 2D SSIM or PSNR averaged over slices 20–90 in each plane}; $^{*}p < 0.05$; $^{**}p < 0.01.$ The best-performing value in each column is shown in \textbf{bold}.\label{tab:site_heldout}} 
\resizebox{\textwidth}{!}{%
\begin{tabular}{@{}ccccccccc@{}}
\toprule
\multirow{2}{*}{\textBF{Perceptual Loss}} & \multicolumn{4}{c}{\textBF{SSIM(\%)↑}} & \multicolumn{4}{c}{\textBF{PSNR(dB)↑}} \\ 
\cmidrule(lr){2-5} \cmidrule(lr){6-9}
  & \textBF{Overall (3D)} & \textBF{Axial} & \textBF{Coronal} & \textBF{Sagittal} & \textBF{Overall (3D)} & \textBF{Axial} & \textBF{Coronal} & \textBF{Sagittal} \\ \midrule
2D                    & \sig{91.36±1.66}{**} & \sig{87.07±2.43}{**} & \sig{87.52±2.21}{**} & \sig{86.40±2.33}{**} & 29.09±2.24 & \sig{29.41±3.46}{**} & 27.68±2.20 & \sig{28.51±2.27}{*} \\
3D                    & \sig{90.96±1.84}{**} & \sig{86.58±2.58}{**} & \sig{86.96±2.38}{**} & \sig{85.96±2.49}{**} & \sig{28.68±2.27}{**} & \sig{28.90±3.29}{**} & \sig{27.26±2.26}{**} & \sig{28.09±2.28}{**} \\
2.5D                  & \sig{91.41±1.72}{**} & \sig{87.17±2.44}{**} & \sig{87.57±2.26}{**} & \sig{86.52±2.37}{**} & 28.93±2.43 & \sig{29.34±3.65}{**} & 27.51±2.38 & \sig{28.40±2.41}{*} \\
Cyclic 2.5D (Ours)    & \textBF{91.70±1.70}   & \textBF{87.62±2.50}   & \textBF{87.99±2.29}   & \textBF{86.95±2.40}   & \textBF{29.18±2.60}   & \textBF{29.78±3.99}   & \textBF{27.76±2.53}   & \textBF{28.74±2.59} \\ \bottomrule
\end{tabular}
}
\endgroup
\end{table*}

\begin{table*}[!t]
\begingroup\blue 
\centering
\caption{\blue{Independent-cohort replication performance on the SCAN cohort.} The results are obtained from a 3D U-Net model trained on data preprocessed with the by-manufacturer PET SUVR standardization. \blue{The values are reported as mean ± standard deviation across subjects. ``Overall (3D)'' refers to the 3D SSIM or PSNR computed on the entire volume; ``Axial/Coronal/Sagittal'' corresponds to the slice-wise 2D SSIM or PSNR averaged over slices 20–90 in each plane}; $^{*}p < 0.05$; $^{**}p < 0.01.$ The best-performing value in each column is shown in \textbf{bold}.\label{tab:scan_res}} 
\resizebox{\textwidth}{!}{%
\begin{tabular}{@{}ccccccccc@{}}
\toprule
\multirow{2}{*}{\textBF{Perceptual Loss}} & \multicolumn{4}{c}{\textBF{SSIM(\%)↑}} & \multicolumn{4}{c}{\textBF{PSNR(dB)↑}} \\ 
\cmidrule(lr){2-5} \cmidrule(lr){6-9}
  & \textBF{Overall (3D)} & \textBF{Axial} & \textBF{Coronal} & \textBF{Sagittal} & \textBF{Overall (3D)} & \textBF{Axial} & \textBF{Coronal} & \textBF{Sagittal} \\ \midrule
2D                    & \sig{80.40±3.62}{**} & \sig{73.47±6.02}{**} & \sig{73.85±4.70}{**} & \sig{72.07±5.02}{**} & 29.58±3.15 & \sig{29.44±2.95}{**} & 28.28±3.21 & \sig{28.87±2.84}{**} \\
3D                    & \sig{81.68±3.54}{**} & \sig{74.34±6.01}{**} & \sig{74.71±4.70}{**} & \sig{72.88±5.04}{**} & 29.62±3.22 & \sig{29.54±3.05}{**} & 28.27±3.25 & \sig{29.13±2.84}{**} \\
2.5D                  & \sig{82.40±3.44}{**} & \sig{75.28±5.76}{**} & \sig{75.70±4.54}{**} & \sig{73.92±4.63}{**} & \sig{29.41±3.10}{*} & \sig{29.72±3.07}{**} & 28.27±3.22 & \sig{28.69±2.78}{**} \\
Cyclic 2.5D (Ours)    & \textBF{84.08±3.86}  & \textBF{77.56±6.35}  & \textBF{77.92±5.19}  & \textBF{76.58±5.26}  & \textBF{30.06±3.69}  & \textBF{30.53±3.67}  & \textBF{28.80±3.73}  & \textBF{30.02±3.21}  \\ \bottomrule
\end{tabular}
}
\endgroup
\end{table*}

\blue{
\subsection{Effect of By-Manufacturer SUVR Standardization}
\Cref{fig:pet_suvr_distribution} demonstrates that the expected disease-associated pattern, higher cortical tau uptake in AD and most apparent in the upper tail of the distribution, is preserved across all manufacturers after by-manufacturer standardization. In both the SUVR space (pre-standardization) and the manufacturer-wise $z$-scored space (post-standardization), the AD group consistently exhibits higher p75, p90, and p99 values than the LMCI/MCI/EMCI group for Siemens, GE, and Philips.

The by-manufacturer standardization also reduces scanner- and manufacturer-related distributional differences in cortical tau PET SUVR summary metrics within each diagnostic stratum (\Cref{tab:vendor_effect_prepost}). Before standardization, omnibus KW tests across manufacturers are significant for all subject-level metrics (median, p75, p90) in both AD ($p=0.008$--$0.014$) and LMCI/MCI/EMCI (all $p<0.001$). After standardization, the omnibus manufacturer effect is not significant at $\alpha=0.05$ for any metric in either stratum (AD: $p=0.077$--$0.248$; LMCI/MCI/EMCI: $p=0.191$--$0.654$), with only a trend-level residual effect for AD p90 ($p=0.077$).

These omnibus findings are supported by pairwise MWU tests with BH adjustment. After standardization, no pairwise manufacturer contrasts reach significance in either diagnostic stratum for any metric (all adjusted $p\ge 0.106$). In the pre-standardization SUVR space, Siemens vs.\ GE differs significantly across all three metrics in AD (adjusted $p=0.007$--$0.020$), whereas Siemens vs.\ Philips and GE vs.\ Philips do not differ significantly for any metric (all adjusted $p\ge 0.096$). In LMCI/MCI/EMCI, Siemens vs.\ GE is significant for the median and p75 (adjusted $p<0.001$) and for p90 (adjusted $p=0.008$), and Siemens vs.\ Philips is significant across all three metrics (adjusted $p=0.003$--$0.008$), while GE vs.\ Philips remains non-significant for all metrics (all adjusted $p\ge 0.177$).

\begin{figure*}[t]
  \centering
  \includegraphics[width=\linewidth]{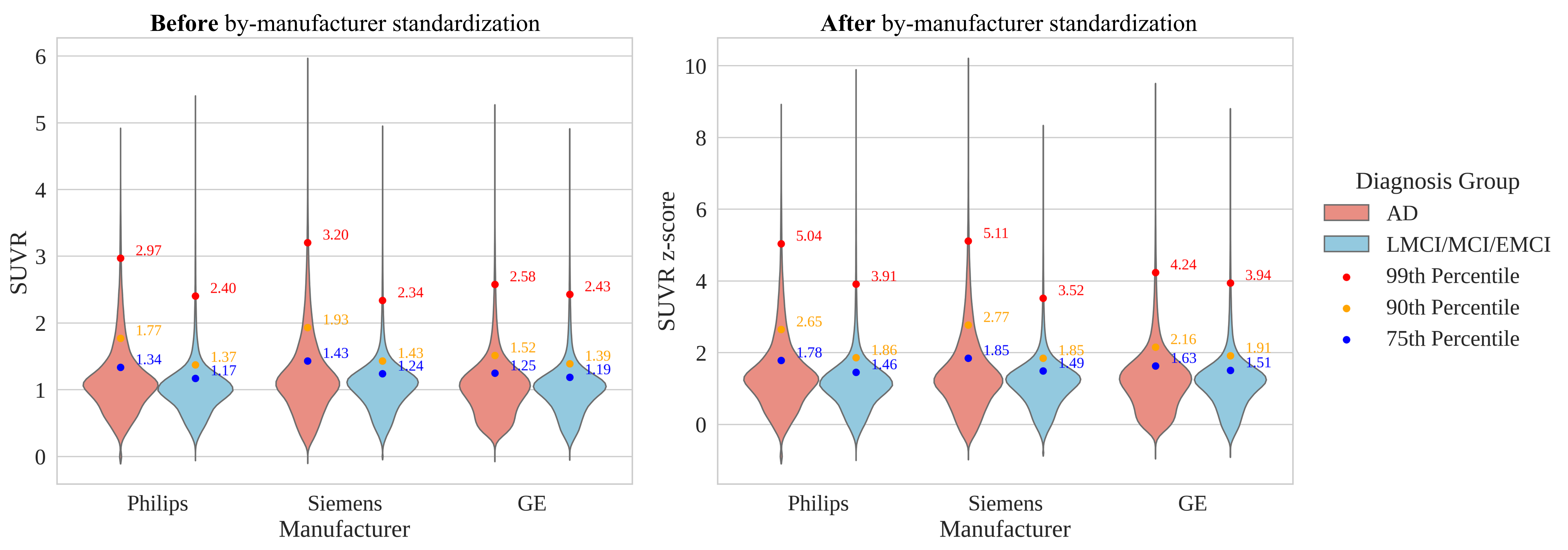}
  \caption{\blue{Cortical tau PET uptake distributions stratified by scanner manufacturer before (left) and after (right) by-manufacturer standardization. The left panel presents cortical tau PET values in SUVR units, and the right panel presents the corresponding manufacturer-wise z-standardized values (SUVR z-scores). For each manufacturer (Philips, Siemens, and GE), violin plots depict the distributions of cortical voxel values for the AD and LMCI/MCI/EMCI groups. Colored markers denote the 75th (blue), 90th (orange), and 99th (red) percentiles, with the corresponding percentile values annotated. Across manufacturers, the AD group shows consistently higher upper-tail uptake than the LMCI/MCI/EMCI group, as indicated by higher 75th, 90th, and 99th percentiles in both the SUVR and standardized z-score panels. Cortical voxels are defined using the Desikan--Killiany atlas.}}
  \label{fig:pet_suvr_distribution}
\end{figure*}

\begin{table}[!t]
\centering
\caption{\blue{Scanner-manufacturer differences in subject-level cortical tau PET SUVR summary metrics before (Pre) and after (Post) by-manufacturer standardization. For each participant, cortical tau PET SUVR values within the Desikan--Killiany cortical mask are summarized using three subject-level quantiles: Median (50th percentile), p75 (75th percentile), and p90 (90th percentile). Here, p75 and p90 refer to percentiles (not statistical $p$-values). Within each diagnostic group, manufacturer effects are evaluated separately for Pre and Post using the KW test (Overall across SI, GE, and PH). Pairwise manufacturer comparisons (SI vs.\ GE, SI vs.\ PH, GE vs.\ PH) are conducted using MWU; reported pairwise $p$-values are BH-adjusted across the three manufacturer pairs within each metric and within each Pre/Post condition. \textit{Abbreviations:} Overall = omnibus across 3 manufacturers; PH = PHilips; SI = SIemens; KW = Kruskal--Wallis H test; MWU = Mann--Whitney U test; BH = Benjamini--Hochberg.\label{tab:vendor_effect_prepost}}}
\begin{tabular}{ccccccc}
\toprule
\multirow{2}{*}{\textbf{Metric}} & \multirow{2}{*}{\textbf{Test}} & \multirow{2}{*}{\textbf{Comparison}} & \multicolumn{2}{c}{\textbf{AD}} & \multicolumn{2}{c}{\textbf{LMCI/MCI/EMCI}} \\
\cmidrule(lr){4-5}\cmidrule(lr){6-7}
 &  &  & \textbf{Pre} & \textbf{Post} & \textbf{Pre} & \textbf{Post} \\
\midrule
\multirow{4}{*}{Median} & KW & Overall & 0.014 & 0.248 & $<$ 0.001 & 0.654 \\ [0.1cm]
 & \multirow{3}{*}{MWU} & SI vs. GE  & 0.020 & 0.266 & $<$ 0.001 & 0.854 \\
 &  & SI vs. PH & 0.521 & 0.945 & 0.008 & 0.705 \\
 &  & GE vs. PH  & 0.096 & 0.266 & 0.877 & 0.705 \\
\midrule
\multirow{4}{*}{p75} & KW & Overall & 0.010 & 0.151 & $<$ 0.001 & 0.367 \\ [0.1cm]
 & \multirow{3}{*}{MWU} & SI vs. GE  & 0.009 & 0.199 & $<$ 0.001 & 0.556 \\
 &  & SI vs. PH & 0.424 & 0.820 & 0.003 & 0.411 \\
 &  & GE vs. PH  & 0.192 & 0.350 & 0.450 & 0.411 \\
\midrule
\multirow{4}{*}{p90} & KW & Overall & 0.008 & 0.077 & $<$ 0.001 & 0.191 \\ [0.1cm]
 & \multirow{3}{*}{MWU} & SI vs. GE  & 0.007 & 0.106 & 0.008 & 0.447 \\
 &  & SI vs. PH & 0.508 & 0.945 & 0.003 & 0.232 \\
 &  & GE vs. PH  & 0.155 & 0.212 & 0.177 & 0.232 \\
\bottomrule
\end{tabular}%
\end{table}

}

\subsection{Ablation Study}
\label{ssec:ablation}
We conduct ablation studies to evaluate the effectiveness of each component comprising the proposed combined loss function. Our method, which employs a combined loss integrating $L^{\phi,j}_{\mathrm{cyc2.5D}}$, $L_{\text{voxel}}$, and $L_{\text{SSIM}}$, outperforms both the individual and combined configurations of $L_{\text{voxel}}$ and $L_{\text{SSIM}}$. The results presented in \Cref{tab:ablation_combined_loss} confirm the effectiveness of the proposed combined loss function. Specifically, we test with $L_{\text{voxel}}$ only, $L_{\text{SSIM}}$ only, combined $L_{\text{voxel}}$ and $L_{\text{SSIM}}$, combined $L_{\text{voxel}}$ and $L^{\phi,j}_{\mathrm{cyc2.5D}}$, and combined $L_{\text{SSIM}}$ and $L^{\phi,j}_{\mathrm{cyc2.5D}}$ to examine performance variations associated with each component within the combined loss $L_{\text{combined}}$. \blue{Additionally, we conduct further ablations to quantify the effects of individual steps in the preprocessing pipeline and the influence of noise augmentation applied to MR and PET images during training; these results are reported in Suppl. Section 4.}

\begin{table}[!t]
\centering
\caption{Ablation study of the components of the proposed combined loss function. The experiments are conducted using a 3D U-Net on a test dataset preprocessed with the by-manufacturer PET SUVR standardization. \blue{The values are reported as mean ± standard deviation across subjects.} The SSIM, PSNR, or MAE results represent the entire 3D volume. The best-performing value in each column is shown in \textbf{bold}.\label{tab:ablation_combined_loss}} 
\setlength{\tabcolsep}{3.5pt}
\begin{tabular}{@{}cccc@{}}
\toprule
\textBF{Loss Function} & \textBF{SSIM(\%)↑} & \textBF{PSNR(dB)↑} & \textBF{MAE↓} \\ 
\midrule
$L_{\text{voxel}}$ & 89.37±3.61 & 28.50±2.64 & 10.37±3.80 \\
$L_{\text{SSIM}}$ & 90.26±3.68 & 28.27±2.47 & 9.93±4.12 \\ 
$L_{\text{voxel}}+L_{\text{SSIM}}$ & 90.32±3.59 & 28.51±2.76 & 9.75±4.09 \\
$L_{\text{voxel}}+L^{\phi,j}_{\mathrm{cyc2.5D}}$ & 89.77±3.38 & 28.57±2.60 & 9.80±3.74 \\
$L_{\text{SSIM}}+L^{\phi,j}_{\mathrm{cyc2.5D}}$ & 90.31±3.62 & 28.47±2.59 & 9.50±4.17 \\
$L_{\text{voxel}}+L_{\text{SSIM}}+L^{\phi,j}_{\mathrm{cyc2.5D}}$ (Ours) & \textBF{90.47±3.60} & \textBF{28.95±2.72} & \textBF{9.11±4.09} \\
\bottomrule
\end{tabular}
\end{table}

\section{Discussion}
\label{sec:discussion}
This study provides proof of concept that T1w MRI can be used to predict regional tau depositions by synthesizing [\textsuperscript{18}F]flortaucipir SUVR maps. We introduce a cyclic 2.5D perceptual loss function that sequentially optimizes three orthogonal planes, improving volumetric fidelity and regional correspondence. In addition, we z-standardize PET SUVRs according to the scanner manufacturers, reducing scanner-induced SUVR heterogeneity and enhancing the preservation of regions with high tau uptake. These methodological advances improve synthesis quality at both the brain-wide level and in AD tau-burden ROIs, yielding closer agreement with the actual PET signal. Importantly, our approach is architecture-agnostic and can be readily integrated into diverse image synthesis frameworks. 

The proposed cyclic 2.5D perceptual loss demonstrates improved performance compared to competing perceptual losses and existing methods for PET image synthesis. It exhibits quantitatively superior performance compared to the two publicly available approaches to synthesize PET images from MRI. It also mostly outperforms conventional 2D, 2.5D, and 3D perceptual losses when applied in 3D U-Net, its variants such as UNETR and SwinUNETR, as well as GAN-based models, including CycleGAN and Pix2Pix. This broad applicability indicates that the cyclic 2.5D perceptual loss can be effectively integrated not only into U-Net-based models but also into more diverse image synthesis paradigms, including adversarial training strategies. Importantly, the cyclic 2.5D perceptual loss achieves the highest SSIM among all evaluated methods. This result suggests that the proposed approach preserves both anatomical integrity and tracer distribution more faithfully than alternative techniques, because SSIM is designed to reflect structural similarity aligned with human visual perception. Moreover, the observed improvements are consistent across volumetric 3D evaluations and 2D analyses along each of the three orthogonal planes, underscoring its ability to optimize performance in a balanced manner across orientations. This closer alignment with the ground-truth PET images highlights its robustness in maintaining both global and local structural fidelity. Taken together, these findings emphasize the technical advantages of the cyclic 2.5D perceptual loss and support its potential as a generalizable tool for cross-modal image synthesis.

The performance improvements achieved by the cyclic 2.5D perceptual loss can be attributed to its unique optimization strategy, which sequentially and evenly refines the neuroanatomical planes. Specifically, the method operates by periodically switching the slicing plane used for 2D perceptual loss calculations during training. This cyclic alternation promotes balanced optimization across all three orthogonal planes, resulting in consistently strong performance in both volumetric 3D metrics and planar 2D metrics. Conceptually, this mechanism can be likened to shaping a clay cube by alternately patting each side, thereby achieving uniform refinement. While initial changes in the slicing plane may lead to transient spikes in validation loss, these diminish over time as the model adapts. A key advantage of the cyclic strategy is its ability to mitigate overfitting by treating all planes equivalently over the same number of epochs. This approach overcomes a limitation of traditional 2.5D perceptual loss, which aggregates plane-wise losses and often disproportionately emphasizes the ``easier” planes. The continuous alternation of the slicing plane further enhances robustness by avoiding excessive reliance on any neuroanatomical plane. This strategy is analogous to the cyclic learning rate approach \citep{cycliclearning}, which enables models to explore broader parameter space and escape saddle points by cyclically varying the learning rate. Similarly, in the cyclic 2.5D perceptual loss, the model undergoes high-learning-rate-like adjustments immediately following a plane change, optimizing perceptual quality on the new plane while fine-tuning during subsequent epochs. The model performs increasingly finer adjustments across all planes as the cycle interval shortens, ultimately achieving a more balanced and optimized synthesis. Through this process, the cyclic 2.5D perceptual loss achieves a more balanced and robust optimization across orientations, enhancing both the perceptual fidelity and structural consistency of synthesized PET images. These properties emphasize its utility for 3D medical image translation, where faithful anatomical preservation is critical.

Across all evaluated architectures and training frameworks, the proposed method achieves the highest SSIM and a competitive overall PSNR. However, PSNR does not consistently reach peak values in certain models. Specifically, in the UNETR, SwinUNETR, and CycleGAN, the cyclic 2.5D perceptual loss yields lower PSNR values compared to the 2D, 2.5D, or 3D perceptual losses for certain planes or full (3D) view. As transformer-CNN hybrid models, UNETR and SwinUNETR utilize transformer encoders \citep{vaswani2017attention} to learn patch-wise feature representations. Since the cyclic 2.5D perceptual loss evaluates feature similarity independently across different planes, it could plausibly be misaligned with the global feature aggregation mechanism inherent to transformer-based architectures. This potential misalignment could hinder the transformer’s ability to integrate contextual information across 3D space, resulting in suboptimal pixel-wise precision (as measured by PSNR) compared to conventional perceptual losses. In the case of the CycleGAN, which employs an unpaired image-to-image translation framework that requires bidirectional modality transformation, the training process involves two generators and two discriminators operating in an adversarial setting. Maintaining cycle consistency is crucial for optimizing the CycleGAN. Since the cyclic 2.5D perceptual loss assesses perceptual similarity independently for each plane, it may conflict with the CycleGAN’s requirement to preserve 3D spatial consistency. However, it is important to note that while higher PSNR values generally indicate lower pixel-wise discrepancies, they do not necessarily guarantee the preservation of structural or pathological features. Notably, our method outperforms all comparison methods in terms of SSIM, the primary metric for translating T1w MRI to tau PET, indicating superior perceptual fidelity in the synthesized images. When stratified by diagnostic subgroup, the LMCI group exhibits statistically significant improvements relative to the 2.5D method across all metrics and relative to the 2D method for a subset of metrics. In contrast, no significant difference is observed relative to the 3D method, likely due to the very small number of LMCI subjects (n = 5) in the test set, which limits statistical power to detect meaningful effects.

Most importantly, our training pipeline synthesizes PET images that more accurately preserve regional tau pathology than those generated using the conventional 2D, 3D, and 2.5D perceptual losses. ROI analysis demonstrates that PET images generated by our method achieve the highest fidelity to the reference PET images, particularly in clinically key regions characterized by high tau burden. This advantage is most pronounced in regions severely affected by tau deposition, which aligns with the areas delineated by Braak staging \citep{braak1991neuropathological,braak2006staging}. The strongest PET signal is observed in the banks of the superior temporal sulcus, consistent with prior studies reporting early elevations in tau PET uptake in this region \citep{insel2020neuroanatomical}. Our method exhibits enhanced performance in this region, further emphasizing its ability to represent the regional tau burden more accurately. Similar improvements are observed in the entorhinal cortex, the initial site of tau tangle formation corresponding to Braak stages I-II \citep{van1991entorhinal,igarashi2023entorhinal}, and in the parahippocampal gyrus, which is associated with early AD and MCI pathology spanning Braak stages I-IV \citep{mitchell2002parahippocampal}. The isthmus cingulate, increasingly affected during Braak stages III-IV \citep{maass2017comparison}, and the fusiform gyrus, linked to cognitive deficits in AD and corresponding to Braak stages III-IV \citep{seemiller2021indication,lowe2018widespread}, also demonstrate improvements. Furthermore, enhanced performance is observed in the precuneus, a region prominently affected in advanced Braak stages V-VI \citep{chen2021staging,pascoal2021microglial}. These findings emphasize the capability of the cyclic 2.5D method to accurately capture tau pathology across brain regions encompassing a wide range of Braak stages, from early (I-II) to advanced (V-VI). Crucially, strong SUVR signals are detected in regions typically associated with early Alzheimer’s pathology, aligning with the ADNI cohort's characteristics, which predominantly includes individuals in the prodromal or early symptomatic stages of AD. 

The qualitative visual analysis confirms that our approach more effectively represents the spatial and pathological characteristics of tau accumulation. These characteristics are essential for accurately modeling the progressive stages of tau-related neuropathology in AD. The proposed method achieves qualitatively superior performance, demonstrating its ability to capture and reflect tau pathology with greater accuracy compared to the existing methods. These results suggest that the cyclic 2.5D approach enables generative models to learn better mappings that retain tau-related pathological information than those trained with traditional perceptual loss configurations. This enhanced capability emphasizes the superiority of our plane-wise cyclic 2.5D perceptual loss in preserving critical pathological information, particularly in the context of AD-related neurodegenerative disorders. Consequently, this approach could improve the preservation of regionally specific patterns relevant to pathological staging in AD and may enhance the consistency of tau PET-like synthesis in research settings.

\blue{
The diagnosis- and A$\beta$/Tau-stratified results suggest that synthesis fidelity as measured by SSIM may track the degree to which tau-related neurodegenerative signatures are expressed on structural MRI. This pattern supports the interpretation that cross-modal translation becomes more learnable when disease-related anatomical alterations are more salient and spatially consistent. In the A$\beta$/Tau analysis, whole-volume SSIM is lowest in A$\beta$--T-- and highest in A$\beta$+T+, and tau positivity corresponds to higher SSIM within both amyloid-negative and amyloid-positive participants. Within the canonical amyloid-first biomarker cascade and the A/T/N framework, the A$\beta$+T-- profile is commonly interpreted as an earlier AD continuum state than A$\beta$+T+. In contrast, an amyloid-negative, tau-positive profile (A$\beta$--T+) does not map straightforwardly onto an intermediate AD stage and is often discussed in the context of age-related or non-AD tau processes, including primary age-related tauopathy \citep{jack2010hypothetical,jack2018nia,crary2014primary}. This framing is consistent with evidence that tau PET burden shows tighter cross-sectional and longitudinal coupling to cortical atrophy than amyloid PET, such that tau-positive strata are more likely to exhibit concordant structure--pathology relationships on MRI \citep{gordon2018cross}. In the diagnostic analysis, group-level SSIM increases from EMCI to MCI and AD, and is highest in LMCI. Notably, SSIM in AD remains lower than in LMCI, suggesting that synthesis fidelity improves across the prodromal spectrum and may attenuate at the dementia stage; however, the LMCI peak warrants cautious interpretation given the small sample size (n=5). From a modeling perspective, strata with greater tau-linked neurodegeneration are expected to exhibit higher-amplitude and more spatially coherent structural changes (e.g., stereotyped atrophy patterns). Such properties increase the effective signal-to-noise ratio of the MRI input, reduce one-to-many ambiguity in the conditional mapping from MRI to tau PET, and provide more consistent supervision for deterministic generators optimized with SSIM- and perceptual-based objectives. Conversely, earlier or biomarker-discordant strata (e.g., A$\beta$+T--) may present subtle or heterogeneous tau deposition with limited macroscopic atrophy, which increases conditional uncertainty and can promote regression-to-the-mean predictions under pixel- and structure-fidelity losses. Consistent with this interpretation, prior cross-modality deep-learning studies report substantially lower tau PET imputation accuracy from T1w MRI than from PET-based inputs \citep{lee2024synthesizing}. Finally, the non-monotonic PSNR patterns across diagnostic and A$\beta$/Tau strata align with PSNR’s sensitivity to voxel-wise intensity discrepancies, which can increase as tau uptake becomes more heterogeneous, even when structural similarity (SSIM) improves.
}

\blue{
The site-held-out evaluation indicates that the performance advantage of the cyclic 2.5D perceptual loss persists when acquisition sites are strictly disjoint. This finding supports the interpretation that the observed gains primarily reflect a generalizable MRI-to-tau PET relationship, rather than an artifact of inadvertent site overlap under random subject-level splitting. In multi-center neuroimaging, site-specific acquisition and reconstruction characteristics can introduce stable, non-biological signatures \citep{fortin2018harmonization,joshi2009reducing}. When the same site contributes data to multiple splits, high-capacity synthesis models can partially leverage these signatures, inflating apparent fidelity without improving true biological correspondence \citep{dinsdale2021deep,snoek2019control}. Although our in-house preprocessing pipeline does not include an explicit statistical harmonization procedure such as ComBat \citep{johnson2007adjusting} or Box--Cox--based transformations \citep{box1964analysis}, the benefit of the cyclic 2.5D approach under a site-disjoint setting suggests reduced sensitivity to site-dependent signatures and more robust learning of disease-relevant spatial structure.
}
%

\blue{
The independent-cohort replication in the etiologically heterogeneous SCAN cohort demonstrates that the proposed cyclic 2.5D strategy yields consistent improvements over alternative perceptual-loss variants, indicating that its benefits generalize beyond an AD-enriched research setting. Because structural MRI predominantly reflects downstream neurodegeneration rather than upstream molecular pathology \citep{jack2010hypothetical,jack2018nia}, learning the inverse mapping from atrophy patterns to tau tracer uptake is inherently challenging in SCAN. In this cohort, cognitive impairment and MRI-visible neurodegeneration frequently arise from etiologically diverse and sometimes mixed non-AD processes, including VCID \citep{gorelick2011vascular,iadecola2013pathobiology}, FTD-spectrum disorders \citep{bang2015frontotemporal}, and LBD \citep{walker2015lewy}. Moreover, [\textsuperscript{18}F]flortaucipir is optimized for Alzheimer-type paired helical filament tau and therefore shows limited or heterogeneous retention across many non-AD neurodegenerative syndromes, including FTD-spectrum disorders and LBD \citep{mattay2020brain,petersen2022overview,tsai201918f,wolters2020tau}. In this context, the lower SSIM relative to ADNI is consistent with both the smaller paired training set in SCAN and the greater etiologic diversity, which weakens any one-to-one correspondence between MRI-visible neurodegeneration and tracer uptake. Importantly, the cyclic 2.5D strategy still yields the strongest performance among perceptual-loss variants across both volumetric and plane-wise metrics, supporting the interpretation that its gains arise from plane-balanced optimization rather than dependence on ADNI-specific recruitment characteristics.
}
%

\blue{
We find that the by-manufacturer standardization of tau PET SUVRs consistently yields stronger quantitative performance than a global min--max rescaling. In particular, the by-manufacturer standardization more faithfully preserves high-uptake patterns that are pathophysiologically salient for AD while reducing cross-manufacturer distributional mismatch. By contrast, min--max normalization enforces an identical fixed dynamic range across scans and can disproportionately compress the upper SUVR tail, attenuating diagnostically informative right-tail variation. We emphasize that this step functions as an optimization-oriented intensity normalization for learning-based synthesis, introduced to stabilize training and mitigate domain shift, rather than as an additional multicenter PET harmonization procedure (e.g., phantom- or physics-based inter-scanner corrections; \citealt{joshi2009reducing}; \citealt{akamatsu2023review}; \citealt{landau2025positron}). This interpretation aligns with common PET and machine-learning workflows, in which intensity normalization is typically applied after centralized preprocessing to support robust modeling \citep{young2021influence,park2023machine,landau2025positron}.

The by-manufacturer standardization is particularly effective because residual vendor-specific biases often persist even after nominally harmonized preprocessing and manifest primarily as systematic shifts in SUVR scale and dispersion. Despite standardized pipelines in ADNI and SCAN, scanner- and manufacturer-related SUVR biases can remain because quantification depends on reconstruction settings and correction procedures beyond nominal resolution matching \citep{joshi2009reducing}. Moreover, early enforcement of a target FWHM does not guarantee identical effective resolution after subsequent resampling and registrations \citep{carbonell2025novel}. Consistent with this, prior multisite work using [\textsuperscript{18}F]flortaucipir shows that smoothing to a common resolution does not reliably reduce scanner (batch) effects, and that additional correction/harmonization steps improve robustness \citep{luo2025effect}. In our ADNI cohort, we likewise observe manufacturer-dependent differences in the scale and dispersion of cortical tau SUVRs before standardization, which are substantially attenuated after by-manufacturer standardization. This transformation is affine and rank-preserving within each manufacturer, rescaling intensities into a common numeric regime without reordering subjects or exchanging distributions across manufacturers. Importantly, disease-relevant signal remains intact: in both the original SUVR space and the manufacturer-wise $z$-scored space, the AD group consistently shows elevated right-tail uptake (p75, p90, p99) relative to the LMCI/MCI/EMCI group across Siemens, GE, and Philips.
}

A growing body of evidence from in vivo imaging, longitudinal cohort studies, and neuropathological investigations supports the neurobiological basis for approximating [\textsuperscript{18}F]flortaucipir tau PET SUVR from T1w MRI. The regional cortical thickness derived from T1w MRI has been reported to show robust negative correlations with regional [\textsuperscript{18}F]flortaucipir SUVR across 68 cortical regions (e.g., $r \approx -0.82$ to $-0.51$, all $p<0.001$), while no comparable correlation is observed with amyloid PET \citep{xia2017association}. Also, studies have reported that an elevated [\textsuperscript{18}F]flortaucipir binding co-localizes with reduced gray-matter intensity or cortical thickness in the cortices of the temporoparietal and limbic association, in cognitively normal older adults and throughout the continuum of AD \citep{sepulcre2016vivo,lapoint2017association}. Notably, these relationships persist after correction for atrophy-related partial-volume effects and are observed within both A$\beta$--positive and A$\beta$--negative subgroups \citep{lapoint2017association,mak2018vivo}. Consistent with these findings, recent multimodal imaging work in mild AD dementia has shown that tau PET signal strongly predicts local gray-matter volume loss, whereas amyloid burden shows no such regional specificity \citep{iaccarino2018local}. Longitudinal studies further demonstrate that baseline tau PET intensity and spatial distribution predict the subsequent rate and topography of cortical atrophy over 12--18 months, independent of baseline thickness, consistent with tau accumulation contributing to the neurodegeneration captured by structural MRI \citep{fyfe2020imaging,gordon2018cross}. These imaging results accord with the AD biomarker cascade, in which A$\beta$ deposition precedes changes in soluble tau, followed by PET-detectable tau pathology, and ultimately widespread neurodegeneration with MRI-visible atrophy \citep{jack2010hypothetical,guo2021characterization}. Neuropathological staging corroborates this temporal ordering: tau pathology first appears in the entorhinal cortex (Braak stages I–II) without overt atrophy, extends to limbic structures in stages III–IV, and involves widespread neocortical regions in stages V–VI, in parallel with cognitive decline \citep{braak1991neuropathological,braak2006staging}. By the onset of MCI or early dementia, substantial tau deposition in the limbic and neocortical regions is accompanied by measurable tissue loss in these same areas. Collectively, these converging temporal and spatial patterns suggest that structural MRI captures disease-relevant information about regional tau burden beyond generalized atrophy. This provides a pathophysiological rationale for estimating the tau PET contrast (i.e., ``pseudo tau PET'') from T1w MRI. 

The proposed MRI-based synthesis framework could provide a non-ionizing, lower-cost surrogate of regional tau burden for research and exploratory clinical use. A promising future application is research-oriented referral prioritization or clinical trial enrichment before specialty referral: synthesized tau PET SUVR maps derived from MRI could stratify patients by their likelihood of AD-related pathology \citep{dang2023tau}. Because MRI is already routine in memory clinics and general neurology, patients predicted to have higher tau burden could be prioritized for confirmatory testing (e.g., tau PET or cerebrospinal fluid analysis) or referral to tertiary care. In contrast, those with lower predicted burden could be monitored locally according to clinical judgment. This strategy aligns with efforts to deploy less invasive, more accessible biomarkers to reduce unnecessary confirmatory testing \blue{\citep{jack2020predicting}} and mirrors the growing use of blood-based biomarkers in primary care \citep{pleen2024blood}. An MRI-derived pseudo-SUVR is not intended to replace established biomarkers but to complement them as a triage tool that identifies individuals most likely to benefit from advanced imaging. By enabling earlier recognition of probable AD pathology, it may streamline decisions about confirmatory testing and care pathways---an increasingly important consideration, given the recent approval of disease-modifying monoclonal antibodies, which typically require biomarker confirmation for treatment eligibility \citep{van2023lecanemab,cummings2024anti}. Given the cost and logistical constraints of tau PET, population-level deployment remains impractical \citep{lee2024synthesizing}. If rigorously validated across sites, scanners, and cohorts, with careful consideration of potential biases and the risk of false negatives, MRI-based tau estimation could serve as a scalable, non-invasive adjunct for patient stratification and research prioritization.

Several images with high PSNR or SSIM values exhibit poor visual quality, particularly in PET images with predominantly low SUVRs interspersed with a few prominent tau hot spots. This observation highlights the inherent limitations of relying solely on SSIM or PSNR for evaluating tau PET synthesis tasks and emphasizes the need for region-specific analyses that account for variations across different brain areas. This problem is particularly crucial given that the primary objective of tau PET imaging is to identify the pathological localization of NFTs in the brain. The failure to detect hot spots compromises the clinical utility of synthesized tau PET images, emphasizing the need for an evaluation metric that addresses these shortcomings. Therefore, developing a metric that effectively accounts for these challenges is essential.

Despite its advantages, the cyclic 2.5D perceptual loss also has certain limitations. It requires many epochs due to the current configuration, which activates early stopping in later stages. However, premature termination of training in earlier stages increases the risk of overfitting to a single plane, mainly due to the large number of epochs within the initial cycle durations. Additionally, the method considers all slices, including those with minimal or irrelevant anatomical information. Many peripheral slices contain negligible tissue, for which more straightforward metrics, such as MSE and SSIM, are sufficient. Future work should address these issues to maximize the benefits of cyclic 2.5D perceptual loss while reducing computational costs. Furthermore, the development of a robust evaluation metric tailored for PET image synthesis---particularly one sensitive to regionally focal uptake---is warranted, as traditional metrics often fail to fully capture the visual quality and domain relevance of the generated images. \blue{More broadly, extending MRI-to-tau PET synthesis beyond supervised, paired cohorts and ensuring robustness to cross-cohort distribution shift will require additional multi-cohort datasets and further methodological advances.}

\section{Conclusion}
\label{sec:conclusion}
This study provides a proof-of-concept demonstration of estimating regional tau PET SUVR from structural MRI by generating 3D maps of [\textsuperscript{18}F]flortaucipir SUVR directly from 3D T1w images. We introduce a cyclic 2.5D perceptual loss function that improves spatial fidelity and enables more robust cross-modal 3D image synthesis. We additionally propose a by-manufacturer PET SUVR standardization strategy to reduce variability between scanners. \blue{Experiments on the ADNI and SCAN cohorts support the broader applicability of the proposed approach to Alzheimer’s disease and related dementias.} Our method demonstrates competitive whole-brain tau PET synthesis performance, with significant improvements in many key tau pathology regions associated with Alzheimer’s disease. This approach has the potential to serve as a non-invasive tool for research and early-stage screening for clinical trials. \blue{Future work should focus on further optimizing the loss function, developing PET synthesis–specific evaluation metrics, and assessing robustness under cross-cohort distribution shifts.}

\section*{Author contributions}
\textbf{Moon, J.: }Methodology; Data curation; Formal Analysis; Investigation; Software; Validation; Visualization; Writing – Review \& Editing. \textbf{Kim, S.: }Methodology; Data curation; Formal Analysis; Investigation; Software; Visualization; Writing – original draft; Writing – Review \& Editing. \textbf{Chung, H.: }Funding acquisition; Resources; Writing – Review \& Editing. \textbf{Jang, I.: } Conceptualization; Methodology; Project Administration; Formal Analysis; Funding acquisition; Resources; Supervision; Writing – Review \& Editing.

\section*{Acknowledgments}
The authors appreciate Chanik Kang, Daehwan Kim, and Jiwoong Yang for providing valuable feedback during the preparation of this manuscript. 
This work was supported by the National Institute of Health (NIH) research projects (2024ER040700, 2025ER040300), the National Research Foundation of Korea (NRF) grants funded by the Ministry of Science and ICT (MSIT) (RS-2024-00455720, RS-2024-00338048, RS-2024-00414119), the National Supercomputing Center with supercomputing resources including technical support (KSC-2024-CRE-0021, KSC-2025-CRE-0065), the High-Performance Computing Support project (RQT-25-070083) funded by MSIT, a grant of the Korea Health Technology R\&D Project through the Korea Health Industry Development Institute (KHIDI), funded by the Ministry of Health \& Welfare (RS-2025-02220534), the Technology Innovation Program (or Industrial Strategic Technology Development Program - Biotechnology)(RS-2025-13002970) funded By the Ministry of Trade Industry \& Energy (MOTIE, Korea), and Hankuk University of Foreign Studies Research Fund of 2025. 
The work was also supported by Culture, Sports and Tourism R\&D Program through the Korea Creative Content Agency grant funded by the Ministry of Culture, Sports and Tourism (RS-2024-00332210), Artificial Intelligence Graduate School Program (RS-2020-II201373, Hanyang University) supervised by the IITP, and under the artificial intelligence semiconductor support program to nurture the best talents ((IITP-(2025)-RS-2023-00253914) grant funded by the Korea government.

Data collection and sharing for the Alzheimer's Disease Neuroimaging Initiative (ADNI) is funded by the National Institute on Aging (National Institutes of Health Grant U19 AG024904). The grantee organization is the Northern California Institute for Research and Education. In the past, ADNI has also received funding from the National Institute of Biomedical Imaging and Bioengineering, the Canadian Institutes of Health Research, and private sector contributions through the Foundation for the National Institutes of Health (FNIH) including generous contributions from the following: AbbVie, Alzheimer's Association; Alzheimer's Drug Discovery Foundation; Araclon Biotech; BioClinica, Inc.; Biogen; Bristol-Myers Squibb Company; CereSpir, Inc.; Cogstate; Eisai Inc.; Elan Pharmaceuticals, Inc.; Eli Lilly and Company; EuroImmun; F. Hoffmann-La Roche Ltd and its affiliated company Genentech, Inc.; Fujirebio; GE Healthcare; IXICO Ltd.; Janssen Alzheimer Immunotherapy Research \& Development, LLC.; Johnson \& Johnson Pharmaceutical Research \& Development LLC.; Lumosity; Lundbeck; Merck \& Co., Inc.; Meso Scale Diagnostics, LLC.; NeuroRx Research; Neurotrack Technologies; Novartis Pharmaceuticals Corporation; Pfizer Inc.; Piramal Imaging; Servier; Takeda Pharmaceutical Company; and Transition Therapeutics.

The NACC database is funded by NIA/NIH Grant U24 AG072122. NACC data are contributed by the NIA-funded ADRCs: P30 AG062429 (PI James Brewer, MD, PhD), P30 AG066468 (PI Oscar Lopez, MD), P30 AG062421 (PI Bradley Hyman, MD, PhD), P30 AG066509 (PI Thomas Grabowski, MD), P30 AG066514 (PI Mary Sano, PhD), P30 AG066530 (PI Helena Chui, MD), P30 AG066507 (PI Marilyn Albert, PhD), P30 AG066444 (PI David Holtzman, MD), P30 AG066518 (PI Lisa Silbert, MD, MCR), P30 AG066512 (PI Thomas Wisniewski, MD), P30 AG066462 (PI Scott Small, MD), P30 AG072979 (PI David Wolk, MD), P30 AG072972 (PI Charles DeCarli, MD), P30 AG072976 (PI Andrew Saykin, PsyD), P30 AG072975 (PI Julie A. Schneider, MD, MS), P30 AG072978 (PI Ann McKee, MD), P30 AG072977 (PI Robert Vassar, PhD), P30 AG066519 (PI Frank LaFerla, PhD), P30 AG062677 (PI Ronald Petersen, MD, PhD), P30 AG079280 (PI Jessica Langbaum, PhD), P30 AG062422 (PI Gil Rabinovici, MD), P30 AG066511 (PI Allan Levey, MD, PhD), P30 AG072946 (PI Linda Van Eldik, PhD), P30 AG062715 (PI Sanjay Asthana, MD, FRCP), P30 AG072973 (PI Russell Swerdlow, MD), P30 AG066506 (PI Glenn Smith, PhD, ABPP), P30 AG066508 (PI Stephen Strittmatter, MD, PhD), P30 AG066515 (PI Victor Henderson, MD, MS), P30 AG072947 (PI Suzanne Craft, PhD), P30 AG072931 (PI Henry Paulson, MD, PhD), P30 AG066546 (PI Sudha Seshadri, MD), P30 AG086401 (PI Erik Roberson, MD, PhD), P30 AG086404 (PI Gary Rosenberg, MD), P20 AG068082 (PI Angela Jefferson, PhD), P30 AG072958 (PI Heather Whitson, MD), P30 AG072959 (PI James Leverenz, MD). SCAN is a multi-institutional project that was funded as a U24 grant (AG067418) by the National Institute on Aging in May 2020. Data collected by SCAN and shared by NACC are contributed by the NIA-funded ADRCs as follows: Arizona Alzheimer’s Center - P30 AG072980 (PI: Eric Reiman, MD); R01 AG069453 (PI: Eric Reiman (contact), MD); P30 AG019610 (PI: Eric Reiman, MD); and the State of Arizona which provided additional funding supporting our center; Boston University - P30 AG013846 (PI Neil Kowall MD); Cleveland ADRC - P30 AG062428 (James Leverenz, MD); Cleveland Clinic, Las Vegas – P20AG068053; Columbia - P50 AG008702 (PI Scott Small MD); Duke/UNC ADRC – P30 AG072958; Emory University - P30AG066511 (PI Levey Allan, MD, PhD); Indiana University - R01 AG19771 (PI Andrew Saykin, PsyD); P30 AG10133 (PI Andrew Saykin, PsyD); P30 AG072976 (PI Andrew Saykin, PsyD); R01 AG061788 (PI Shannon Risacher, PhD); R01 AG053993 (PI Yu-Chien Wu, MD, PhD); U01 AG057195 (PI Liana Apostolova, MD); U19 AG063911 (PI Bradley Boeve, MD); and the Indiana University Department of Radiology and Imaging Sciences; Johns Hopkins - P30 AG066507 (PI Marilyn Albert, Phd.); Mayo Clinic - P50 AG016574 (PI Ronald Petersen MD PhD); Mount Sinai - P30 AG066514 (PI Mary Sano, PhD); R01 AG054110 (PI Trey Hedden, PhD); R01 AG053509 (PI Trey Hedden, PhD); New York University - P30AG066512-01S2 (PI Thomas Wisniewski, MD); R01AG056031 (PI Ricardo Osorio, MD); R01AG056531 (PIs Ricardo Osorio, MD; Girardin Jean-Louis, PhD); Northwestern University - P30 AG013854 (PI Robert Vassar PhD); R01 AG045571 (PI Emily Rogalski, PhD); R56 AG045571, (PI Emily Rogalski, PhD); R01 AG067781, (PI Emily Rogalski, PhD); U19 AG073153, (PI Emily Rogalski, PhD); R01 DC008552, (M.-Marsel Mesulam, MD); R01 AG077444, (PIs M.-Marsel Mesulam, MD, Emily Rogalski, PhD); R01 NS075075 (PI Emily Rogalski, PhD); R01 AG056258 (PI Emily Rogalski, PhD); Oregon Health \& Science University - P30 AG066518 (PI Lisa Silbert, MD, MCR); Rush University - P30 AG010161 (PI David Bennett MD); Stanford – P30AG066515; P50 AG047366 (PI Victor Henderson MD MS); University of Alabama, Birmingham – P20; University of California, Davis - P30 AG10129 (PI Charles DeCarli, MD); P30 AG072972 (PI Charles DeCarli, MD); University of California, Irvine - P50 AG016573 (PI Frank LaFerla PhD); University of California, San Diego - P30AG062429 (PI James Brewer, MD, PhD); University of California, San Francisco - P30 AG062422 (Rabinovici, Gil D., MD); University of Kansas - P30 AG035982 (Russell Swerdlow, MD); University of Kentucky - P30 AG028283-15S1 (PIs Linda Van Eldik, PhD and Brian Gold, PhD); University of Michigan ADRC - P30AG053760 (PI Henry Paulson, MD, PhD) P30AG072931 (PI Henry Paulson, MD, PhD) Cure Alzheimer's Fund 200775 - (PI Henry Paulson, MD, PhD) U19 NS120384 (PI Charles DeCarli, MD, University of Michigan Site PI Henry Paulson, MD, PhD) R01 AG068338 (MPI Bruno Giordani, PhD, Carol Persad, PhD, Yi Murphey, PhD) S10OD026738-01 (PI Douglas Noll, PhD) R01 AG058724 (PI Benjamin Hampstead, PhD) R35 AG072262 (PI Benjamin Hampstead, PhD) W81XWH2110743 (PI Benjamin Hampstead, PhD) R01 AG073235 (PI Nancy Chiaravalloti, University of Michigan Site PI Benjamin Hampstead, PhD) 1I01RX001534 (PI Benjamin Hampstead, PhD) IRX001381 (PI Benjamin Hampstead, PhD); University of New Mexico - P20 AG068077 (Gary Rosenberg, MD); University of Pennsylvania - State of PA project 2019NF4100087335 (PI David Wolk, MD); Rooney Family Research Fund (PI David Wolk, MD); R01 AG055005 (PI David Wolk, MD); University of Pittsburgh - P50 AG005133 (PI Oscar Lopez MD); University of Southern California - P50 AG005142 (PI Helena Chui MD); University of Washington - P50 AG005136 (PI Thomas Grabowski MD); University of Wisconsin - P50 AG033514 (PI Sanjay Asthana MD FRCP); Vanderbilt University – P20 AG068082; Wake Forest - P30AG072947 (PI Suzanne Craft, PhD); Washington University, St. Louis - P01 AG03991 (PI John Morris MD); P01 AG026276 (PI John Morris MD); P20 MH071616 (PI Dan Marcus); P30 AG066444 (PI John Morris MD); P30 NS098577 (PI Dan Marcus); R01 AG021910 (PI Randy Buckner); R01 AG043434 (PI Catherine Roe); R01 EB009352 (PI Dan Marcus); UL1 TR000448 (PI Brad Evanoff); U24 RR021382 (PI Bruce Rosen); Avid Radiopharmaceuticals / Eli Lilly; Yale - P50 AG047270 (PI Stephen Strittmatter MD PhD); R01AG052560 (MPI: Christopher van Dyck, MD; Richard Carson, PhD); R01AG062276 (PI: Christopher van Dyck, MD); 1Florida - P30AG066506-03 (PI Glenn Smith, PhD); P50 AG047266 (PI Todd Golde MD PhD)

\section*{Data and code availability statements}
The code for the proposed method is available at \url{https://github.com/labhai/Cyclic-2.5D-Perceptual-Loss}.

Data used in preparation of this article were obtained from the Alzheimer’s Disease Neuroimaging Initiative (ADNI) database (adni.loni.usc.edu). As such, the investigators within the ADNI contributed to the design and implementation of ADNI and/or provided data but did not participate in analysis or writing of this report. A complete listing of ADNI investigators can be found at: \url{http://adni.loni.usc.edu/wp-content/uploads/how_to_apply/ADNI_Acknowledgement_List.pdf}.

Data used in the preparation of this article were also obtained from the National Alzheimer’s Coordinating Center (NACC) database, including MRI and/or PET data from the Standardized Centralized Alzheimer’s and Related Dementias Neuroimaging (SCAN) Initiative. NACC/SCAN data are available to qualified researchers via the NACC Data Request Process and are subject to the NACC Data Use Agreement; the authors do not have permission to share NACC/SCAN data directly.

\bibliographystyle{plainnat}
\bibliography{references}

\end{document}


\maketitle

\setcounter{tocdepth}{1}
\tableofcontents
\bigskip

\section{Cyclic 2.5D Perceptual Loss Model Selection}
\Cref{tab:supp_perceptual_loss_model} presents the overall experimental results for selecting the perceptual loss model $\phi$ and weight $\lambda$. The cyclic 2.5D perceptual loss is tested with $\lambda$ values of \{0.1, 0.5, 1, 3, 5\} across all the perceptual loss models used in these experiments. The highest performance is achieved when utilizing features extracted from the 23rd layer (10th ReLU) of the VGG-16 model \citep{vgg} for perceptual loss calculation. Based on these findings, the perceptual loss model $\phi$ is fixed as VGG-16, the feature extraction layer $j$ as 23 (10th ReLU), and the weight $\lambda$ as 0.5 for all subsequent experiments. In all experiments, image translation is performed using a 3D U-Net \citep{cciccek20163d} model.

\begin{table*}[t]
\caption{Overall experimental results for selecting the cyclic 2.5D perceptual loss model $\phi$ and optimal weight $\lambda$. Evaluations are conducted using the validation set. The perceptual loss weight $\lambda$ is systematically tested with values of \{0.1, 0.5, 1, 3, 5\} for each model. The feature extraction layers are selected while maintaining comparable parameter counts across the model. \blue{The values are reported as mean ± standard deviation across subjects.} The SSIM, PSNR, or MAE values represent performance across the 3D volume.\label{tab:supp_perceptual_loss_model}} 
\resizebox{\textwidth}{!}{%
\begin{tabular}{@{}cccccccc@{}}
\toprule
Dataset & \begin{tabular}[c]{@{}c@{}}Perceptual Loss\\Model ($\phi$)\end{tabular} & \begin{tabular}[c]{@{}c@{}}Feature Extraction\\Layer ($j$)\end{tabular} & $n_{\text{params}}$ & \begin{tabular}[c]{@{}c@{}}Perceptual Loss\\Weight ($\lambda$)\end{tabular} & SSIM(\%)↑ & PSNR(dB)↑ & MAE↓ \\
\midrule
\multirow{10}{*}{RadImageNet} & \multirow{5}{*}{ResNet-50} & \multirow{5}{*}{3rd Block Stack} & \multirow{5}{*}{8.53M} & 0.1 & 90.30±3.74 & 28.94±3.17 & 10.30±4.66 \\
& & & & 0.5 & 90.37±3.63 & 28.99±3.06 & 10.09±4.62 \\
& & & & 1.0 & 90.37±3.58 & 29.03±3.09 & 10.05±4.85 \\
& & & & 3.0 & 89.97±3.54 & 28.94±2.96 & 10.02±4.59 \\
& & & & 5.0 & 89.71±3.57 & 28.68±2.89 & 10.37±4.94 \\ \cmidrule(lr){2-8}

& \multirow{5}{*}{InceptionNet v3} & \multirow{5}{*}{8th Inception Module} & \multirow{5}{*}{8.97M} & 0.1 & 90.45±3.68 & 29.06±3.17 & 10.11±4.80 \\
& & & & 0.5 & 90.48±3.67 & 29.11±3.18 & 10.03±4.79 \\
& & & & 1.0 & 90.45±3.64 & 29.08±3.10 & 10.05±4.79 \\
& & & & 3.0 & 90.08±3.62 & 28.87±2.97 & 10.34±4.71 \\
& & & & 5.0 & 89.99±3.67 & 28.90±2.99 & 10.24±4.72 \\
\midrule

\multirow{15}{*}{ImageNet-1K} & \multirow{5}{*}{ResNet-50} & \multirow{5}{*}{3rd Block Stack} & \multirow{5}{*}{8.53M} & 0.1 & 90.48±3.65 & 29.03±3.12 & 10.05±4.87 \\
& & & & 0.5 & 90.45±3.68 & 29.06±3.16 & 9.99±4.97 \\
& & & & 1.0 & 90.31±3.72 & 29.01±3.12 & 10.00±4.98\\
& & & & 3.0 & 89.94±3.70 & 28.86±3.03 & 10.17±4.83 \\
& & & & 5.0 & 89.41±3.84 & 28.64±3.15 & 10.49±5.19 \\ \cmidrule(lr){2-8}

& \multirow{5}{*}{InceptionNet v3} & \multirow{5}{*}{8th Inception Module} & \multirow{5}{*}{8.97M} & 0.1 & 90.20±3.77 & 28.72±3.11 & 10.46±4.60 \\
& & & & 0.5 & 90.65±3.57 & 29.30±3.09 & 9.79±4.65 \\
& & & & 1.0 & 90.51±3.59 & 29.20±3.04 & 9.85±4.76 \\
& & & & 3.0 & 90.34±3.56 & 28.99±2.90 & 10.04±4.63 \\
& & & & 5.0 & 90.43±3.67 & 29.16±3.04 & 9.86±4.64 \\ \cmidrule(lr){2-8}

& \multirow{5}{*}{VGG-16} & \multirow{5}{*}{10th ReLU} & \multirow{5}{*}{7.63M} & 0.1 & 90.41±3.71 & 28.95±3.20 & 10.22±4.56 \\
& & & & 0.5 & \textBF{90.69±3.59} & \textBF{29.31±3.19} & \textBF{9.73±4.83} \\
& & & & 1.0 & 90.43±3.67 & 29.18±3.19 & 9.86±4.90 \\
& & & & 3.0 & 89.92±3.71 & 28.82±3.16 & 10.22±4.97  \\
& & & & 5.0 & 89.86±3.64 & 28.81±2.95 & 10.17±4.72 \\
\bottomrule
\end{tabular}
}
\end{table*}

\section{Benchmark Results}
\Cref{tab:supp_benchmark} provides the comprehensive experimental results comparing the cyclic 2.5D perceptual loss with other methods. The table reports quantitative metrics, including structural similarity index measure (SSIM), peak signal-to-noise ratio (PSNR), and normalized mean absolute error (NMAE), across overall (3D) performance and individual 2D views in the axial, coronal, and sagittal planes. Rows marked with a dagger symbol ($\dagger$) correspond to experiments conducted on PET data preprocessed via by-manufacturer standardization. In contrast, rows without the dagger indicate results from PET data preprocessed with min--max normalization. Under by-manufacturer standardization, PET SUVRs are not scaled to a fixed range (e.g., $-1$ to 1), distinguishing them from min--max normalization. Thus, the mean absolute error (MAE) is replaced with the NMAE to ensure a fair comparison across preprocessing methods. The NMAE is computed by normalizing the MAE with respect to the range of the ground truth PET SUVRs ($y$), as defined below:
\begin{equation}
\text{NMAE}(\hat{y}, y) = \frac{\text{MAE}}{\max(y)-\min(y)}.
\label{eq:SSIM}
\end{equation}

\begin{sidewaystable*}
\centering
\caption{Comprehensive evaluation of the cyclic 2.5D perceptual loss. Performance is assessed across multiple state-of-the-art MRI-to-PET synthesis frameworks and representative perceptual loss variants. Models marked with a dagger symbol ($\dagger$) are trained on datasets preprocessed using by-manufacturer standardization. By contrast, models without the dagger symbol are trained on datasets preprocessed with min–max normalization. Under min–max normalization, scaling procedures differ by framework: PASTA (row 1) \citep{li2024pasta} scales SUVRs to the range $(0, 1)$, while ShareGAN (row 2) \citep{wang2024joint} and U-Net (row 3) scale them to $(-1, 1)$. \blue{The values are reported as mean ± standard deviation across subjects.} All NMAE values are scaled by a factor of 100 for readability. \blue{``Overall (3D)'' refers to the 3D SSIM or PSNR computed on the entire volume; ``Axial/Coronal/Sagittal'' corresponds to the slice-wise 2D SSIM or PSNR averaged over slices 20–90 in each plane}; $^{*}p < 0.05$; $^{**}p < 0.01.$\label{tab:supp_benchmark}} 
\resizebox{\textheight}{!}{%
\begin{tabular}{@{}cccccccccccccc@{}}
\toprule
\multirow{2}{*}{Model} & \multirow{2}{*}{Perceptual Loss} & \multicolumn{4}{c}{SSIM(\%)↑} & \multicolumn{4}{c}{PSNR(dB)↑} & \multicolumn{4}{c}{NMAE$(\times 100)$↓} \\ 
 \cmidrule(lr){3-6} \cmidrule(lr){7-10} \cmidrule(lr){11-14}
                                      &                                   & \multicolumn{1}{c}{Overall (3D)} & \multicolumn{1}{c}{Axial} & \multicolumn{1}{c}{Coronal} & \multicolumn{1}{c}{Sagittal} & \multicolumn{1}{c}{Overall (3D)} & \multicolumn{1}{c}{Axial} & \multicolumn{1}{c}{Coronal} & \multicolumn{1}{c}{Sagittal} & \multicolumn{1}{c}{Overall (3D)} & \multicolumn{1}{c}{Axial} & \multicolumn{1}{c}{Coronal} & \multicolumn{1}{c}{Sagittal} \\ \midrule
\begin{tabular}[c]{@{}c@{}}PASTA\\{\small w/ MRI only}\end{tabular} & None & 85.32±4.49 & 80.79±5.28 & 80.26±4.81 & 79.10±5.12 & 24.98±2.88 & 25.98±3.52 & 23.55±2.87 & 24.22±3.07 & 2.30±1.00 & 3.30±1.30 & 3.40±1.30 & 3.60±1.40 \\ \midrule 
\begin{tabular}[c]{@{}c@{}}ShareGAN\\{\small w/ AD vs. MCI}\end{tabular} & None & 85.49±4.59 & 79.98±6.06 & 79.41±5.85 & 78.02±6.53 & 25.01±2.57 & 26.19±4.32 & 23.57±2.48 & 24.24±2.86 & 4.30±1.70 & 2.40±1.70 & 6.10±2.60 & 3.50±1.70 \\ \midrule 

\multirow{4}{*}{U-Net}
  & 2D
    & \sig{87.67±5.50}{**} & \sig{82.63±6.69}{*} & 82.66±7.04 & 81.43±7.58
    & 26.46±2.43 & 26.16±2.34 & \textBF{25.41±2.57} & 25.41±2.59
    & \sig{5.70±1.90}{**} & \sig{7.00±2.30}{*} & 7.00±2.50 & 7.30±2.60 \\

  & 3D
    & \sig{87.18±5.47}{**} & \sig{82.23±6.71}{**} & \sig{82.12±7.03}{**} & \sig{80.89±7.62}{**}
    & \sig{25.51±1.99}{**} & \sig{25.01±1.80}{**} & \sig{24.48±2.12}{**} & \sig{24.44±2.16}{**}
    & \sig{7.80±2.30}{**} & \sig{8.90±1.80}{**} & \sig{9.00±2.00}{**} & \sig{9.30±2.10}{**} \\

  & 2.5D
    & \sig{85.33±5.35}{**} & \sig{80.40±6.48}{**} & \sig{80.18±6.80}{**} & \sig{79.03±7.37}{**}
    & \sig{23.09±1.47}{**} & \sig{22.74±1.44}{**} & \sig{22.55±1.66}{**} & \sig{22.52±1.70}{**}
    & \sig{12.10±3.10}{**} & \sig{12.90±2.70}{**} & \sig{13.00±2.90}{**} & \sig{13.20±3.10}{**} \\

  & Cyclic 2.5D (Ours)
    & \textBF{88.03±5.89} & \textBF{82.78±7.59} & \textBF{82.75±7.59} & \textBF{81.49±8.26}
    & \textBF{26.71±2.81} & \textBF{26.56±2.68} & 25.32±2.83 & \textBF{25.48±2.87}
    & \textBF{5.10±1.80} & \textBF{6.60±2.30} & \textBF{6.70±2.40} & \textBF{7.00±2.60} \\ \midrule

\multirow{4}{*}{U-Net$^\dagger$}
  & 2D
    & \sig{89.95±3.53}{**} & \sig{85.95±4.35}{**} & \sig{86.05±4.22}{**} & \sig{84.80±4.48}{**}
    & \sig{28.19±2.81}{**} & \sig{29.10±3.98}{**} & \sig{26.95±2.69}{**} & \sig{27.33±2.92}{**}
    & \sig{1.49±0.51}{**} & \sig{2.12±0.76}{**} & \sig{2.13±0.69}{**} & \sig{2.28±0.77}{**} \\

  & 3D
    & \sig{89.77±3.80}{**} & \sig{85.76±4.60}{**} & \sig{85.77±4.54}{**} & \sig{84.57±4.80}{**}
    & \sig{28.23±2.73}{**} & \sig{29.02±3.82}{**} & \sig{26.99±2.61}{**} & \sig{27.40±2.84}{**}
    & \sig{1.50±0.55}{**} & \sig{2.11±0.79}{**} & \sig{2.11±0.73}{**} & \sig{2.25±0.79}{**} \\

  & 2.5D
    & \sig{90.02±3.57}{**} & \sig{86.06±4.41}{**} & \sig{86.11±4.27}{**} & \sig{84.93±4.54}{**}
    & \sig{28.23±2.77}{**} & \sig{29.18±4.01}{**} & \sig{26.98±2.66}{**} & \sig{27.41±2.91}{**}
    & \sig{1.47±0.56}{**} & \sig{2.10±0.80}{**} & \sig{2.10±0.72}{**} & \sig{2.20±0.78}{**} \\

  & Cyclic 2.5D (Ours)
    & \textBF{90.47±3.62} & \textBF{86.70±4.38} & \textBF{86.74±4.27} & \textBF{85.57±4.57}
    & \textBF{28.95±2.73} & \textBF{29.77±3.76} & \textBF{27.70±2.56} & \textBF{28.10±2.86}
    & \textBF{1.38±0.51} & \textBF{1.96±0.72} & \textBF{1.96±0.64} & \textBF{2.07±0.70} \\ \midrule

\multirow{4}{*}{UNETR$^\dagger$}
  & 2D & 89.84±3.54 & 85.94±4.32 & \sig{85.97±4.19}{*} & \sig{84.60±4.48}{*} & 28.32±2.45 & 29.22±3.53 & 27.10±2.30 & 27.50±2.55 & \sig{1.52±0.50}{*} & 2.11±0.66 & 2.09±0.59 & \sig{2.22±0.66}{*} \\
  & 3D & \sig{89.68±3.76}{*} & 85.80±4.46 & \sig{85.80±4.39}{*} & \sig{84.54±4.73}{**} & 28.25±2.41 & \sig{28.82±3.26}{**} & \textBF{27.16±2.22} & 27.40±2.48 & \sig{1.51±0.50}{*} & \sig{2.10±0.66}{**} & \sig{2.08±0.59}{*} & \sig{2.21±0.66}{**} \\ 
  & 2.5D & \sig{89.68±3.66}{**} & \sig{85.77±4.30}{**} & \sig{85.76±4.22}{**} & \sig{84.58±4.50}{**} & \textBF{28.34±2.29} & 29.16±3.38 & 27.11±2.03 & \textBF{27.57±2.44} & \sig{1.44±0.46}{**} & \sig{2.01±0.61}{**} & \sig{2.03±0.54}{**} & \sig{2.14±0.61}{**} \\ 
  & Cyclic 2.5D (Ours) & \textBF{89.93±3.29} & \textBF{86.03±3.94} & \textBF{86.07±3.92} & \textBF{84.84±4.21} & 28.34±2.52 & \textBF{29.32±3.67} & 27.13±2.30 & 27.57±2.65 & \textBF{1.44±0.52} & \textBF{2.04±0.71} & \textBF{2.03±0.63} & \textBF{2.16±0.70} \\ \midrule

\multirow{4}{*}{SwinUNETR$^\dagger$}
  & 2D & \sig{90.16±3.55}{*} & \sig{86.35±4.33}{**} & \sig{86.46±4.13}{*} & \sig{85.14±4.43}{**} & 28.53±2.55 & \textBF{29.49±3.68} & 27.30±2.36 & 27.79±2.72 & \sig{1.41±0.52}{*} & \sig{1.98±0.73}{**} & \sig{1.99±0.66}{*} & \sig{2.10±0.71}{**} \\
  & 3D & \sig{90.20±3.68}{*} & \sig{86.42±4.44}{*} & \sig{86.45±4.29}{**} & \sig{85.26±4.64}{**} & \sig{28.28±2.64}{*} & \sig{29.08±3.72}{*} & \sig{27.12±2.38}{**} & \sig{27.46±2.76}{*} & \sig{1.49±0.56}{*} & \sig{2.11±0.83}{*} & \sig{2.08±0.69}{**} & \sig{2.21±0.78}{*} \\ 
  & 2.5D & 90.23±3.55 & \sig{86.42±4.35}{*} & \sig{86.51±4.19}{*} & \sig{85.30±4.53}{*} & \textBF{28.60±2.58} & 29.47±3.61 & 27.37±2.41 & \textBF{27.86±2.72} & 1.40±0.52 & \sig{1.97±0.76}{*} & \sig{1.98±0.66}{*} & \sig{2.09±0.71}{*} \\ 
  & Cyclic 2.5D (Ours) & \textBF{90.29±3.40} & \textBF{86.57±4.11} & \textBF{86.69±4.01} & \textBF{85.36±4.38} & 28.52±2.59 & 29.47±3.59 & \textBF{27.41±2.35} & 27.72±2.71 & \textBF{1.39±0.53} & \textBF{1.97±0.76} & \textBF{1.95±0.64} & \textBF{2.08±0.72} \\ \midrule

\multirow{5}{*}{CycleGAN$^\dagger$} 
  & None
    &  \sig{84.69±4.81}{**} & \sig{79.19±5.58}{**} & \sig{78.49±5.40}{**} & \sig{76.94±5.75}{**}
    & \sig{26.12±2.39}{**} & \sig{26.70±3.16}{**} & \sig{24.75±2.22}{**} & \sig{25.25±2.51}{**}
    & \sig{1.97±0.66}{**} & \sig{2.76±0.86}{**} & \sig{2.81±0.80}{**} & \sig{2.96±0.89}{**} \\
  & 2D
    & \sig{86.93±4.42}{**} & \sig{81.82±5.19}{**} & \sig{81.68±5.05}{**} & \sig{80.03±5.38}{**}
    & 27.15±2.32 & \textBF{27.83±3.33} & 25.85±2.11 & 26.20±2.44
    & 1.73±0.58 & \sig{2.45±0.78}{*} & 2.44±0.69 & \sig{2.58±0.76}{*} \\
  & 3D
    & \sig{85.63±5.03}{**} & \sig{80.58±5.67}{**} & \sig{79.84±5.65}{**} & \sig{78.39±5.98}{**}
    & \sig{26.34±2.38}{**} & \sig{26.87±3.17}{**} & \sig{24.99±2.20}{**} & \sig{25.45±2.55}{**}
    & \sig{1.94±0.70}{**} & \sig{2.71±0.88}{**} & \sig{2.76±0.83}{**} & \sig{2.92±0.93}{**} \\
  & 2.5D
    & \sig{87.15±4.32}{**} & \sig{82.28±5.01}{**} & \sig{82.16±4.89}{**} & \sig{80.75±5.14}{**}
    & \textBF{27.18±2.45} & 27.71±3.26 & \textBF{25.91±2.24} & \textBF{26.27±2.51}
    & \textBF{1.69±0.58} & \textBF{2.39±0.77} & \textBF{2.38±0.68} & \textBF{2.51±0.76} \\
  & Cyclic 2.5D (Ours)
    & \textBF{87.69±3.97} & \textBF{82.82±4.78} & \textBF{82.78±4.62} & \textBF{81.27±4.95}
    & 27.16±2.32 & 27.62±3.21 & 25.90±2.19 & 26.24±2.43
    & 1.78±0.56 & 2.46±0.76 & 2.47±0.69 & 2.60±0.74 \\ \midrule

\multirow{5}{*}{Pix2Pix$^\dagger$} 
  & None
    & \sig{82.69±4.39}{**} & \sig{76.73±5.05}{**} & \sig{76.35±4.83}{**} & \sig{74.56±5.20}{**}
    & \sig{26.48±2.01}{**} & \sig{26.56±2.48}{**} & \sig{25.14±1.82}{**} & \sig{25.40±2.06}{**}
    & \sig{2.05±0.53}{**} & \sig{2.77±0.67}{**} & \sig{2.80±0.63}{**} & \sig{2.93±0.69}{**} \\
  & 2D
    & \sig{85.70±4.28}{**} & \sig{80.30±4.96}{**} & \sig{80.19±4.76}{**} & \sig{78.48±5.26}{**}
    & \sig{26.70±2.09}{**} & \sig{27.02±2.86}{**} & \sig{25.38±1.95}{**} & \sig{25.50±2.12}{**}
    & \sig{1.86±0.61}{**} & \sig{2.55±0.73}{**} & \sig{2.56±0.70}{**} & \sig{2.68±0.76}{**} \\
  & 3D
    & \sig{84.10±4.72}{**} & \sig{78.02±5.30}{**} & \sig{77.82±5.11}{**} & \sig{75.90±5.49}{**}
    & \sig{26.47±2.14}{**} & \sig{26.48±2.62}{**} & \sig{25.12±1.90}{**} & \sig{25.35±2.20}{**}
    & \sig{2.13±0.71}{**} & \sig{2.84±0.81}{**} & \sig{2.87±0.79}{**} & \sig{3.02±0.86}{**} \\
  & 2.5D
    & \sig{86.66±4.08}{**} & \sig{81.41±4.74}{**} & \sig{81.45±4.54}{**} & \sig{79.72±4.85}{**}
    & \sig{27.06±2.12}{**} & \sig{27.41±2.90}{**} & \sig{25.74±1.95}{**} & \sig{26.02±2.23}{**}
    & \sig{1.83±0.61}{**} & \sig{2.52±0.73}{**} & \sig{2.53±0.70}{**} & \sig{2.65±0.76}{**} \\
  & Cyclic 2.5D (Ours)
    & \textBF{87.63±3.91} & \textBF{82.65±4.63} & \textBF{82.67±4.45} & \textBF{81.08±4.74}
    & \textBF{27.55±2.21} & \textBF{28.18±3.24} & \textBF{26.25±1.99} & \textBF{26.66±2.34}
    & \textBF{1.61±0.52} & \textBF{2.27±0.67} & \textBF{2.28±0.61} & \textBF{2.40±0.68} \\ \bottomrule
\end{tabular}}
\end{sidewaystable*}

\begin{sidewaysfigure*}
  \centering
  \includegraphics[width=\linewidth]{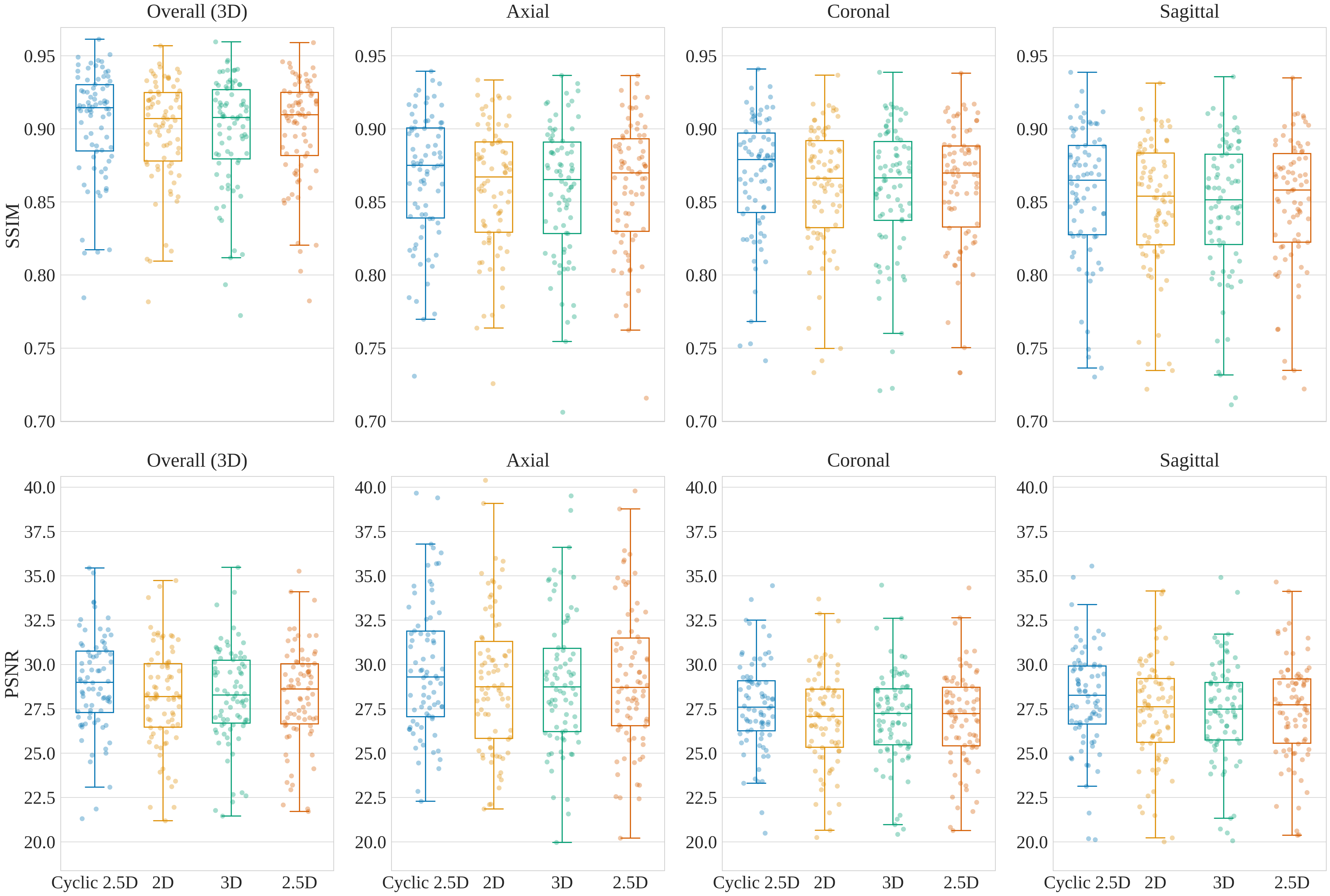}
  \caption{\blue{Distribution of SSIM (top row) and PSNR (bottom row) across subjects for the 3D U-Net with by-manufacturer PET SUVR standardization. Performance is reported for whole-volume (Overall, 3D) evaluation and for slice-wise evaluation in the axial, coronal, and sagittal planes. The plots compare the proposed Cyclic 2.5D approach with 2D, 3D, and conventional 2.5D baselines under perceptual-loss training. Boxes represent the IQR with the median indicated, while whiskers and overlaid points illustrate dispersion and subject-level variability.}}
  \label{fig:supp_dist_SSIM_PSNR}
\end{sidewaysfigure*}

\blue{
\Cref{fig:supp_dist_SSIM_PSNR} summarizes the distributions of SSIM (top row) and PSNR (bottom row) across test subjects for the 3D U-Net trained on datasets preprocessed with by-manufacturer PET SUVR standardization. Results are reported for the full 3D volume (Overall) and for slice-wise evaluations in the axial, coronal, and sagittal planes. The boxplots show the median and interquartile range (IQR), with whiskers indicating the broader spread, and the overlaid markers denoting individual-subject values. Because SSIM is bounded and both metrics may exhibit skewed distributions, these robust summaries complement mean-based statistics by emphasizing central tendency and dispersion. Across all evaluation views, the proposed Cyclic 2.5D method achieves the highest medians and exhibits a consistent upward shift in the distributions relative to the 2D, 3D, and conventional 2.5D baselines, while maintaining comparable IQR and tail behavior. This distributional trend indicates that the performance gains are not driven by a small number of outliers; rather, they reflect broadly improved and orientation-consistent reconstruction fidelity across the cohort.
}

\newsavebox{\abltab} 
\begin{lrbox}{\abltab}
\begin{tabular}{@{}lccc cc@{}} 
\toprule
\multirow{2}{*}{Ablation} & \multicolumn{3}{c}{Preprocessing} & \multirow{2}{*}{SSIM(\%)↑} & \multirow{2}{*}{PSNR(dB)↑} \\
\cmidrule(lr){2-4}
 &  N3Norm   &   FSNorm    &   Skull   &                   &                                      \\
\midrule
w/o N3Norm \& FSNorm &  &  & \cmark & 90.28±3.53 &  \textBF{28.88±2.59}  \\
w/o FSNorm &  & \cmark & \cmark & 90.23±3.55 & 28.81±2.49 \\
w/o Skull & \cmark & \cmark &  & 51.92±9.69 & 18.35±1.92 \\
\textBF{Ours} & \cmark & \cmark & \cmark & \textBF{90.36±3.59} & 28.79±2.58  \\
\bottomrule
\end{tabular}
\end{lrbox}
\begin{table*}[tb]
\centering
\begin{minipage}{\wd\abltab} 
\caption{Quantitative comparisons of tau PET synthesis processed using the original and the modified preprocessing pipelines. Image translation is performed using a 3D U-Net model trained with the proposed combined loss function. Performance evaluation is conducted on the test dataset. \blue{The values are reported as mean ± standard deviation across subjects.} The SSIM and PSNR values represent evaluations of the entire 3D image volume. The abbreviations are defined as follows: w/o N3Norm indicates the absence of N3 normalization and FreeSurfer intensity normalization on MRI; w/o FSNorm refers to the absence of FreeSurfer intensity normalization on MRI; and w/o Skull denotes the absence of skull stripping on both MRI and PET data. \label{tab:ablation_preprocessing}}
\noindent\usebox{\abltab}
\end{minipage}
\end{table*}

\begingroup\blue 
\newsavebox{\augtab} 
\begin{lrbox}{\augtab}
\begin{tabular}{@{}lcccc@{}} 
\toprule
\multirow{2}{*}{Ablation} & \multicolumn{2}{c}{Noise Augmentation} & \multirow{2}{*}{SSIM(\%)↑} & \multirow{2}{*}{PSNR(dB)↑} \\
\cmidrule(lr){2-3}
 &     MRI    &   PET   &                   &                                      \\
\midrule
No Noise Augmentation  &  & & 90.18±3.70 &  \textBF{28.68±2.91}  \\
MRI-Only & \cmark &  & \textBF{90.25±3.66} & 28.64±2.91 \\
MRI+PET & \cmark & \cmark & 89.98±3.72 & 28.48±2.87  \\
\bottomrule
\end{tabular}
\end{lrbox}
\begin{table*}[tb]
\centering
\begin{minipage}{\wd\augtab} 
\caption{\blue{Ablation of noise augmentation strategies during training. Image translation is performed using a 3D U-Net model trained with the proposed combined loss function. Performance evaluation is conducted on the test dataset. The values are reported as mean ± standard deviation across subjects. The SSIM and PSNR values represent evaluations of the entire 3D image volume. The abbreviations are defined as follows: No Noise Augmentation indicates training without noise simulation in either modality; MRI-Only adopts noise simulation only to the MRI input, leaving the PET target unchanged; and MRI+PET applies noise simulation to both the MRI input and the PET target.} \label{tab:ablation_noiseAug}}
\noindent\usebox{\augtab}
\end{minipage}
\end{table*}
\endgroup

\section{Cortical ROI Analysis}
\begin{sidewaysfigure*}
  \centering
  \includegraphics[width=1.0\textheight]{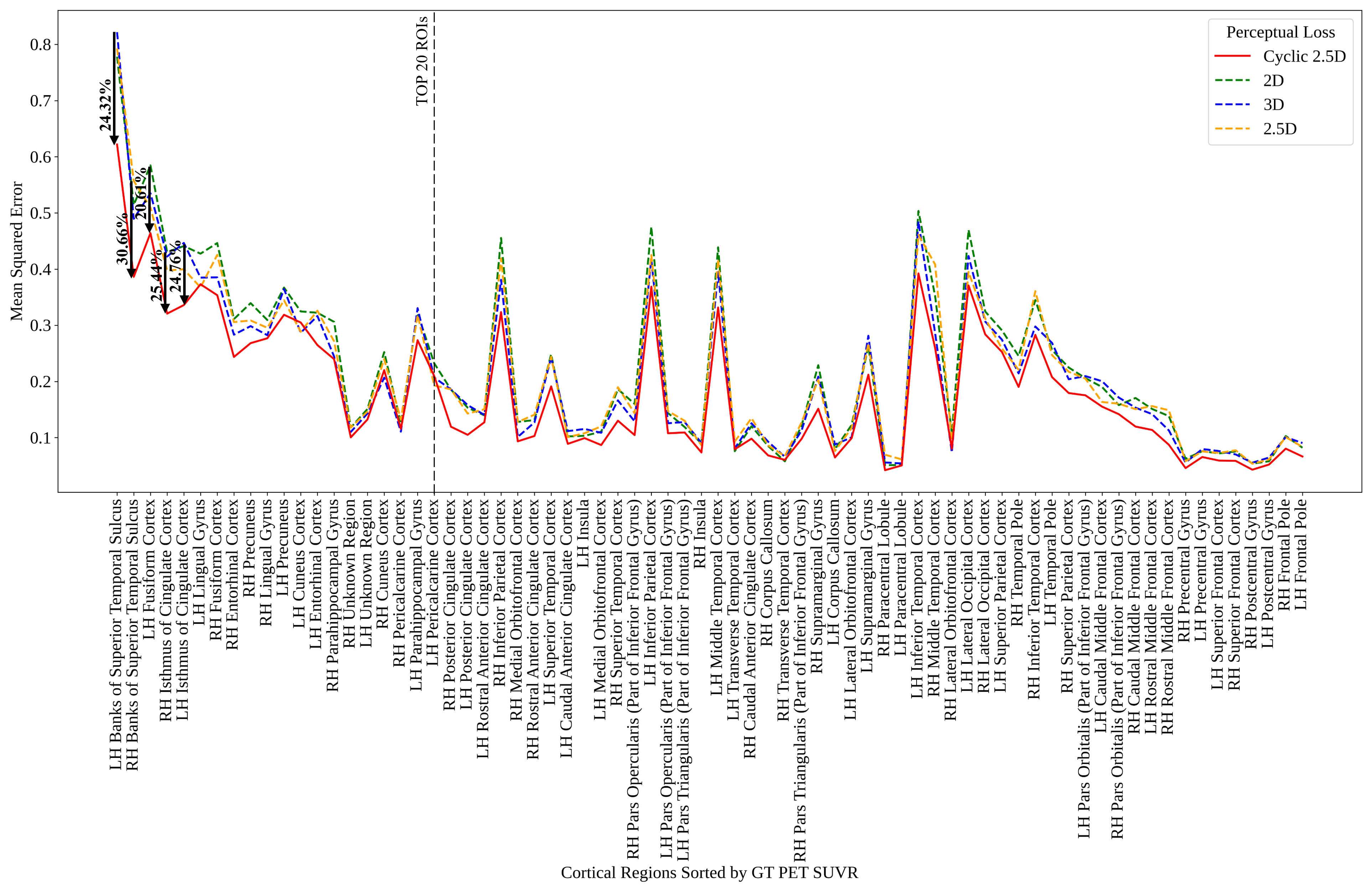}
   \caption{Prediction performance (MSE) across cortical regions. MSE is calculated between the average ground truth PET SUVR and the average generated PET SUVR for each ROI. Cortical regions are ranked in descending order based on the mean ground truth PET SUVRs across all study participants. MSE reduction rates for the five highest-ranked ROIs are highlighted. The cyclic 2.5D perceptual loss achieves lower MSE in 64 of 72 cortical regions compared to conventional perceptual losses. ROI definitions follow the Desikan--Killiany atlas.}
   \label{fig:supp_roi_analysis}
\end{sidewaysfigure*}

\Cref{fig:supp_roi_analysis} illustrates the results of an ROI analysis conducted across all cerebral cortex regions, ranked in descending order based on the mean ground truth PET SUVR across the study participants. The cyclic 2.5D perceptual loss achieves a higher reduction in mean squared error (MSE) in 64 out of the 72 regions when compared to the 2D, 2.5D, and 3D perceptual losses. These findings indicate that the cyclic 2.5D perceptual loss exhibits superior performance in modeling Alzheimer-type tau pathology, which primarily affects cortical regions (gray matter), compared to the conventional approaches.

\section{Additional Ablations}
An ablation study is performed to evaluate the effectiveness of each step in the preprocessing pipeline. Datasets processed with the complete preprocessing pipeline are compared to those processed with certain components removed. In detail, we apply the proposed combined loss function to datasets excluding N3 normalization, initial MRI intensity normalization, and skull stripping. The results are presented in \Cref{tab:ablation_preprocessing}. Based on our assessment that SSIM is a more suitable metric for evaluating tau PET synthesis, as it reflects perceptual similarity aligned with the human visual system more effectively than PSNR, which calculates pixel-level errors, we select SSIM as the primary evaluation metric. The dataset processed using the complete pipeline achieves the highest SSIM score, and this fully processed dataset is utilized for model training and evaluation.

\blue{
Beyond the preprocessing ablations, we assess whether noise simulation should be applied consistently across modalities. \Cref{tab:ablation_noiseAug} compares three configurations: no noise augmentation, MRI-only noise augmentation, and joint MRI and PET noise augmentation. Across all configurations, we use the same 3D U-Net architecture and train it with the proposed combined loss. The MRI-only augmentation yields a modest improvement in SSIM relative to the no-noise baseline. In contrast, augmenting PET with noise reduces both SSIM and PSNR compared with the no-noise baseline. Although the no-noise configuration achieves a slightly higher PSNR than MRI-only augmentation, we prioritize SSIM as the primary metric because it better reflects perceptually meaningful structural agreement. Overall, these results indicate that perturbing the MRI input improves robustness to acquisition-related variability, whereas injecting the PET target effectively introduces label noise and degrades learning stability. Consequently, we adopt MRI-only noise augmentation in all experiments.
}

\bibliographystyle{plainnat}
\bibliography{supp_refs}